\pdfoutput=1
\documentclass{article}
\usepackage{spconf,amsmath,graphicx,subfigure}
\usepackage{tabularx}
\usepackage{color}
\usepackage{dcolumn}
\usepackage{subfig}
\usepackage{tabularx}

\title{CAN PARAMETRIC STATISTICAL METHODS BE TRUSTED \\ FOR FMRI BASED GROUP STUDIES?}
\name{Anders Eklund$^{\: a,b,c}$, Thomas Nichols$^{\: d}$, Hans Knutsson$^{\: a,c}$}
\address{$^a$Division of Medical Informatics, Department of Biomedical Engineering, \\ Link\"{o}ping University, Link\"{o}ping, Sweden \\ $^b$Division of Statistics and Machine Learning, Department of Computer and Information Science, \\ Link\"{o}ping University, Link\"{o}ping, Sweden \\ $^c$Center for Medical Image Science and Visualization (CMIV), \\ Link\"{o}ping University, Link\"{o}ping, Sweden \\ $^d$Department of Statistics, University of Warwick, Coventry, United Kingdom \\   }

\begin{document}
%
\maketitle

\thispagestyle{plain}
\pagestyle{plain}

\begin{abstract}

The most widely used task fMRI analyses use parametric methods that depend on a variety of assumptions. While individual aspects of these fMRI models have been evaluated, they have not been evaluated in a comprehensive manner with empirical data. In this work, a total of 2 million random task fMRI group analyses have been performed using resting state fMRI data, to compute empirical familywise error rates for the software packages SPM, FSL and AFNI, as well as a standard non-parametric permutation method. While there is some variation, for a nominal familywise error rate of 5\% the parametric statistical methods are shown to be conservative for voxel-wise inference and invalid for cluster-wise inference; in particular, cluster size inference with a cluster defining threshold of p = 0.01 generates familywise error rates up to 60\%.  We conduct a number of follow up analyses and investigations that suggest the cause of the invalid cluster inferences is spatial auto correlation functions that do not follow the assumed Gaussian shape. By comparison, the non-parametric permutation test, which is based on a small number of assumptions, is found to produce valid results for voxel as well as cluster wise inference. Using real task data, we compare the results between one parametric method and the permutation test, and find stark differences in the conclusions drawn between the two using cluster inference. These findings speak to the need of validating the statistical methods being used in the neuroimaging field.

\end{abstract}

\section{Introduction}
\label{sec:introduction}
Functional magnetic resonance imaging (fMRI)~\cite{ogawa,fMRI} has since its beginning some 20 years ago been a popular tool for increasing the knowledge about the human brain, with some 28,000 published papers according to PubMed (fMRI in the title or abstract). The first fMRI experiments consisted of simple motor tasks, while more recent examples involve resting state fMRI to study (dynamic) brain connectivity~\cite{biswal,dynamic}. Despite the popularity of fMRI as a tool for studying brain function, the statistical methods used have rarely been validated using real data, likely due to the high cost of fMRI data collection. Validations have instead mainly been performed using simulated data~\cite{welvaert}, but it is obviously very hard to simulate the complex spatiotemporal noise that arises from a living human subject in an MR scanner. 

Through the introduction of international data sharing initiatives in the neuroimaging field~\cite{biswal2,essen,poldrack,openfmri,adni1,adni2,poline}, it has become possible to evaluate the statistical methods using real data. Scarpazza et al.~\cite{scarpazza} for example used freely available anatomical images from 396 healthy controls~\cite{biswal2} to investigate the validity of parametric statistical methods for voxel based morphometry~\cite{vbm}, when comparing a single subject to a group. Silver et al.~\cite{silver} instead used image and genotype data from 181 subjects in the Alzheimer's disease neuroimaging initiative (ADNI)~\cite{adni1,adni2}. The data were used to evaluate statistical methods common in imaging genetics, where the goal is to find genes that can explain variation in brain structure or function. Another example of the use of open data is our previous work~\cite{eklund_ni}, where we investigated the validity of the SPM software for single subject fMRI analysis. A total of 1484 resting state fMRI data sets from the 1000 functional connectomes project~\cite{biswal2} were used as null data, to test how likely it is to find significant brain activity when a subject has not performed any specific task in the MR scanner. The main conclusion was that the noise model in SPM is too simple, yielding a high degree of false positives (up to 70\% incidence of any false positives, compared to the expected 5\% under familywise error control). It was, however, not clear if these problems would propagate to group studies, where inter-subject variability in a per-subject response is less complex than intrasubject time series data. Another unanswered question was the statistical validity of other fMRI software packages. Here, we therefore present a statistical evaluation of the three most common fMRI software packages (SPM~\cite{friston,ashburner}, FSL~\cite{fsl}, AFNI~\cite{afni}) for group inference. Specifically, we evaluate the packages in their entirety, submitting the null data to recommended suite of preprocessing steps integrated into each package.

The main idea of this study is the same for our previous one~\cite{eklund_ni}; to analyze null data and simply count the number analyses that give rise to any false positives. Since two groups of subjects are randomly drawn from a group of healthy controls, the null hypothesis of no group difference in brain activation is true. By performing many random group comparisons, using only healthy controls of similar age, it is possible to compute the empirical false positive rate of a two sample t-test. A similar approach has previously been used to investigate the validity of parametric statistics for voxel based morphometry~\cite{vbm,scarpazza2}. Using resting state fMRI data, which should not contain specific forms of brain activity, it is also possible to compute the empirical false positive rate of a one sample t-test (group activation). 

Briefly, our results show that the parametric statistical methods used in the three most common fMRI software packages are conservative for voxel-wise inference with familywise error control, where each voxel in a volume is independently tested for significance. However, the parametric methods can give a very high degree of false positives (up to 60\%, compared to the nominal 5\%) for cluster-wise inference~\cite{friston_cluster,forman_cluster,woo}, where groups of neighboring voxels are tested simultaneously (to increase the statistical power). By comparison, the non-parametric permutation test~\cite{nichols2002,winkler} is found to produce valid results for both voxel- and cluster-wise inference.

\section{Results}
\label{sec:results}

A total of 1,920,000 random group analyses were performed to compute the empirical false positive rates of SPM, FSL and AFNI (1,000 random analyses repeated for 128 parameter combinations, three thresholding approaches and five functions in the three softwares). The tested parameter combinations, given in Table~\ref{table:levels}, are common in the fMRI field according to a recent review~\cite{carp2}. The following five software tools were tested; SPM OLS, FSL OLS, FSL FLAME1, AFNI OLS (3dttest++) and AFNI 3dMEMA. The OLS (ordinary least squares) functions only use the beta estimates from each subject in the group analysis, while FLAME1 in FSL and 3dMEMA in AFNI also consider the variance of the beta estimates. To compare the parametric statistical methods used by SPM, FSL and AFNI to a non-parametric method, all analyses were also performed using a permutation test~\cite{nichols2002,winkler,broccoli}. 

Resting state fMRI data from 396 healthy controls, downloaded from the 1000 functional connectomes project~\cite{biswal2}, were used for all of the analyses. Resting state data should not contain systematic changes in brain activity, but our previous work~\cite{eklund_ni} showed that the used (pretended) activity paradigm can have a large impact on the degree of false positives. Several different activity paradigms were therefore used; two block based (B1, B2) and two event related (E1, E2), see Table~\ref{table:paradigms}. 

\begin{figure*}
\centering
\subfigure[]{
\includegraphics[scale=0.425]{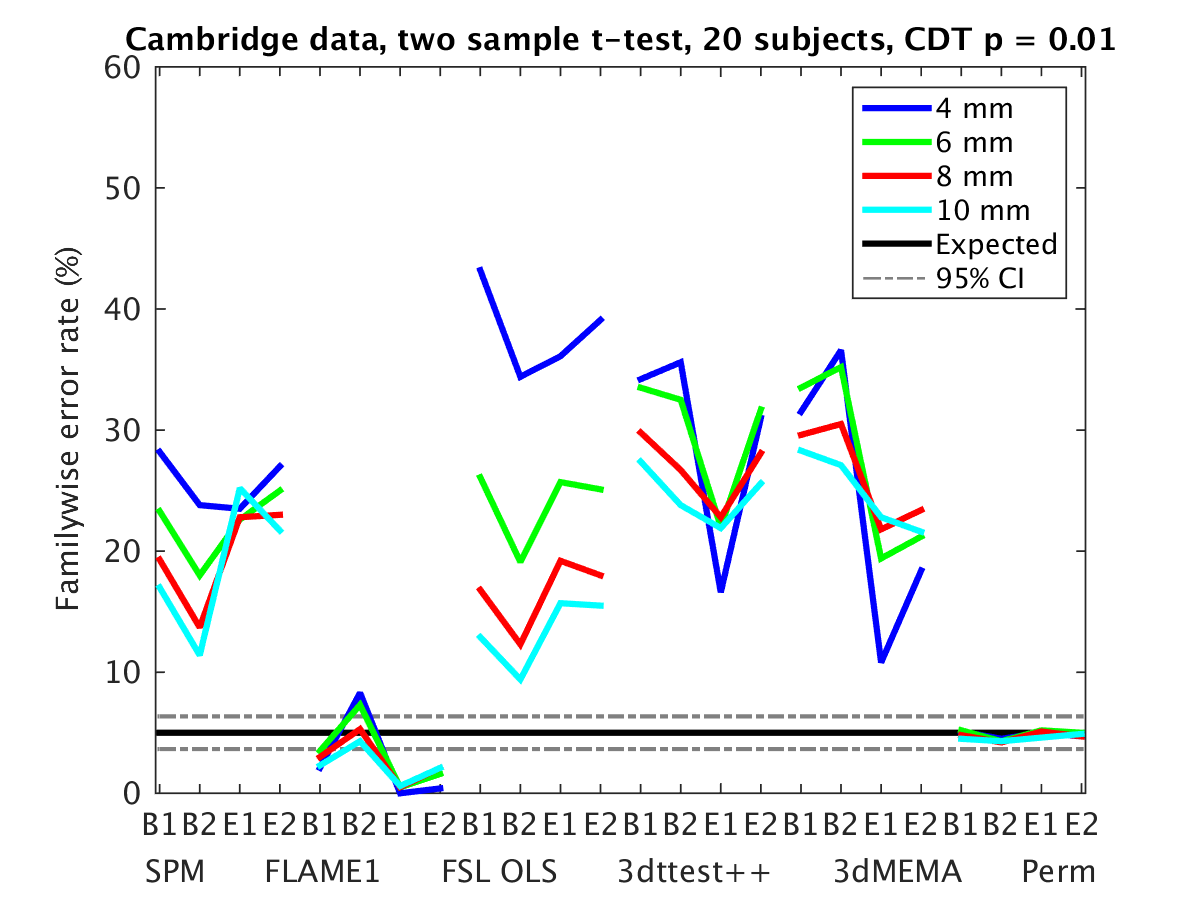}
}
\subfigure[]{
\includegraphics[scale=0.425]{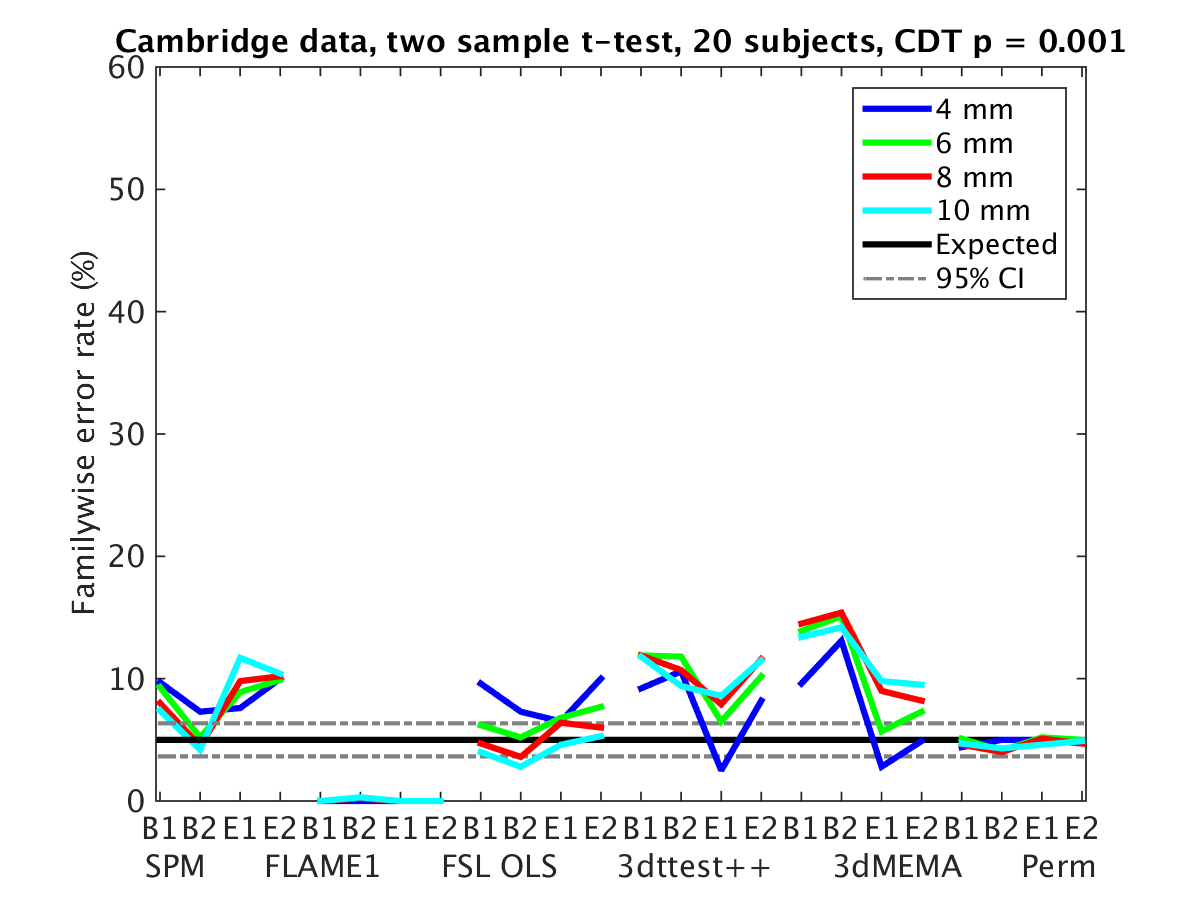}
}
\subfigure[]{
\includegraphics[scale=0.425]{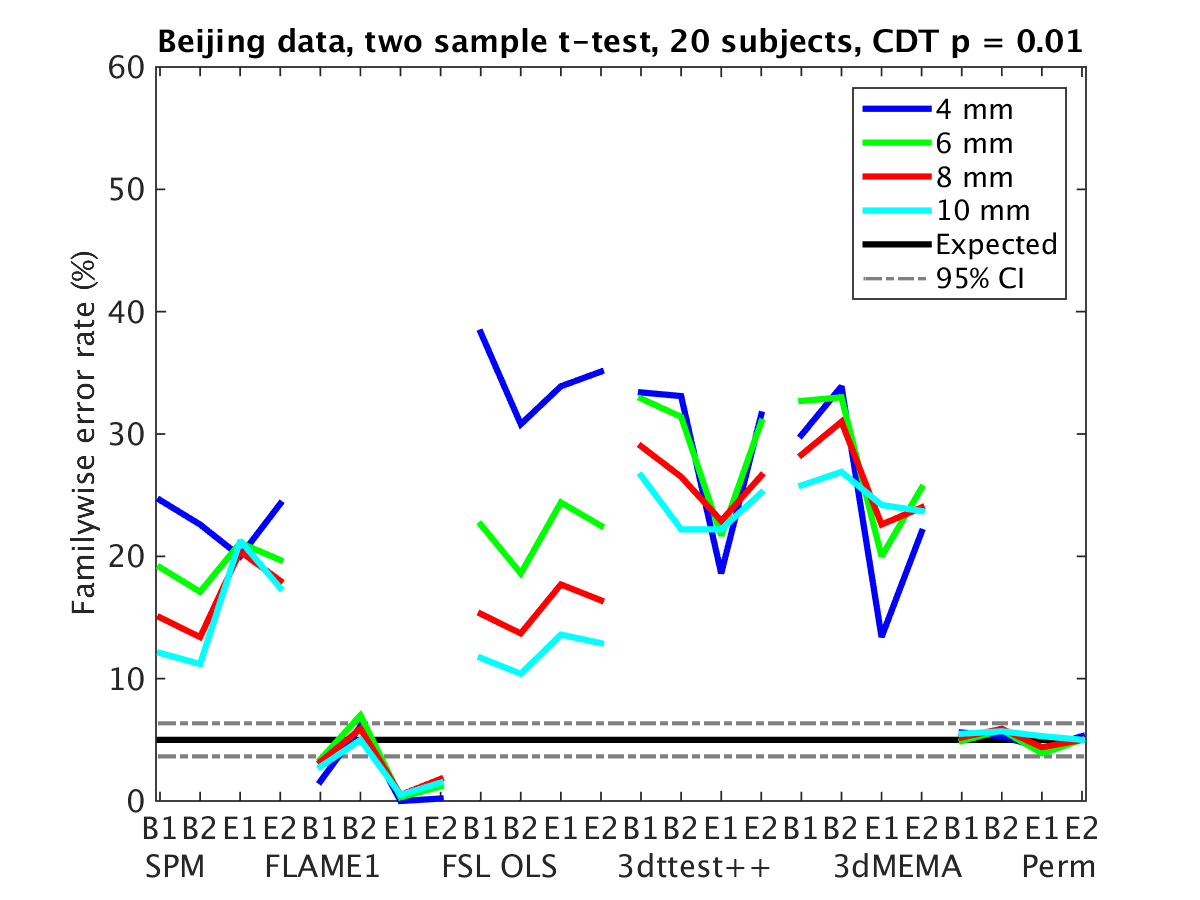}
}
\subfigure[]{
\includegraphics[scale=0.425]{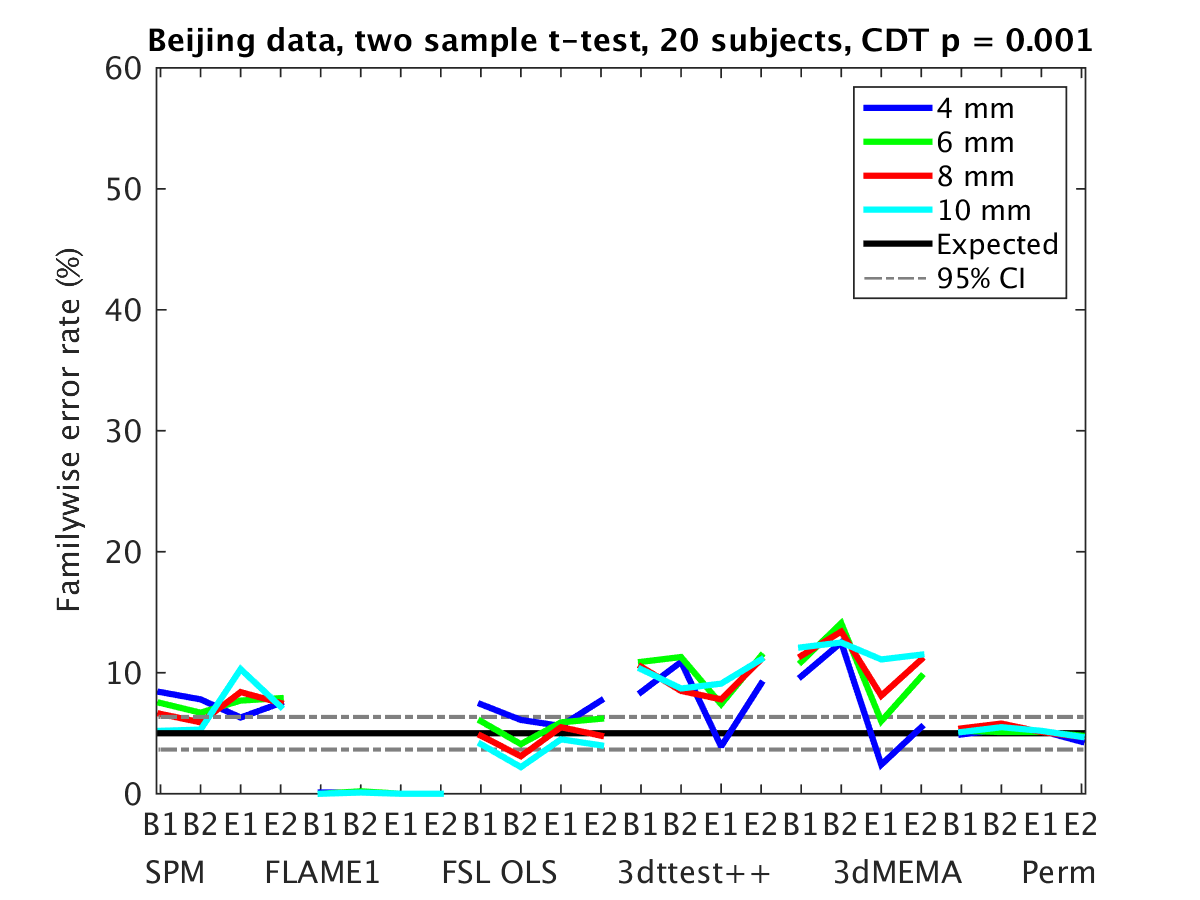}
}
\caption{\emph{Results for two sample t-test and cluster-wise inference, showing estimated familywise error rates for 4-10 mm of smoothing and four different activity paradigms (B1, B2, E1, E2), for SPM, FSL, AFNI and a permutation test. These results are for a group size of 10 (giving a total of 20 subjects). Each statistic map was first thresholded using a cluster defining threshold (CDT) (p = 0.01 or p = 0.001, uncorrected for multiple comparisons), and the surviving clusters were then compared to a FWE-corrected cluster extent threshold, $p_{FWE} = 0.05$. The estimated familywise error rates are simply the number of analyses with any significant group differences divided by the number of analyses (1000). Note that the default CDT is p = 0.001 in SPM and p = 0.01 in FSL (AFNI does not have a default setting). Also note that the default amount of smoothing is 8 mm in SPM, 5 mm in FSL and 4 mm in AFNI.  \textbf{(a)} results for Cambridge data and a CDT of p = 0.01 \textbf{(b)} results for Cambridge data and a CDT of p = 0.001  \textbf{(c)} results for Beijing data and a CDT of p = 0.01 \textbf{(d)} results for Beijing data and a CDT of p = 0.001.}}
\label{fig:fwe_cluster_twosample_20_subjects}
\end{figure*}

\begin{figure*}
\centering
\subfigure[]{
\includegraphics[scale=0.425]{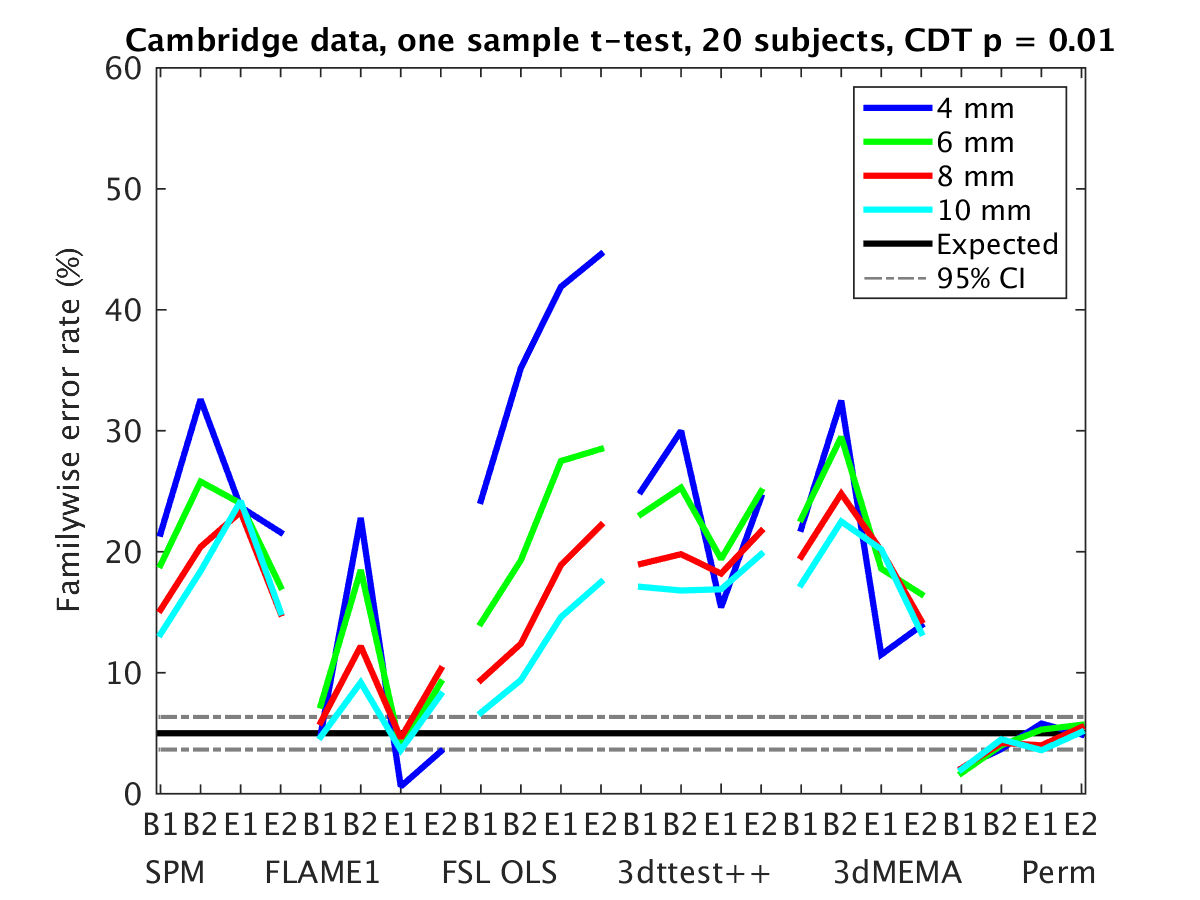}
}
\subfigure[]{
\includegraphics[scale=0.425]{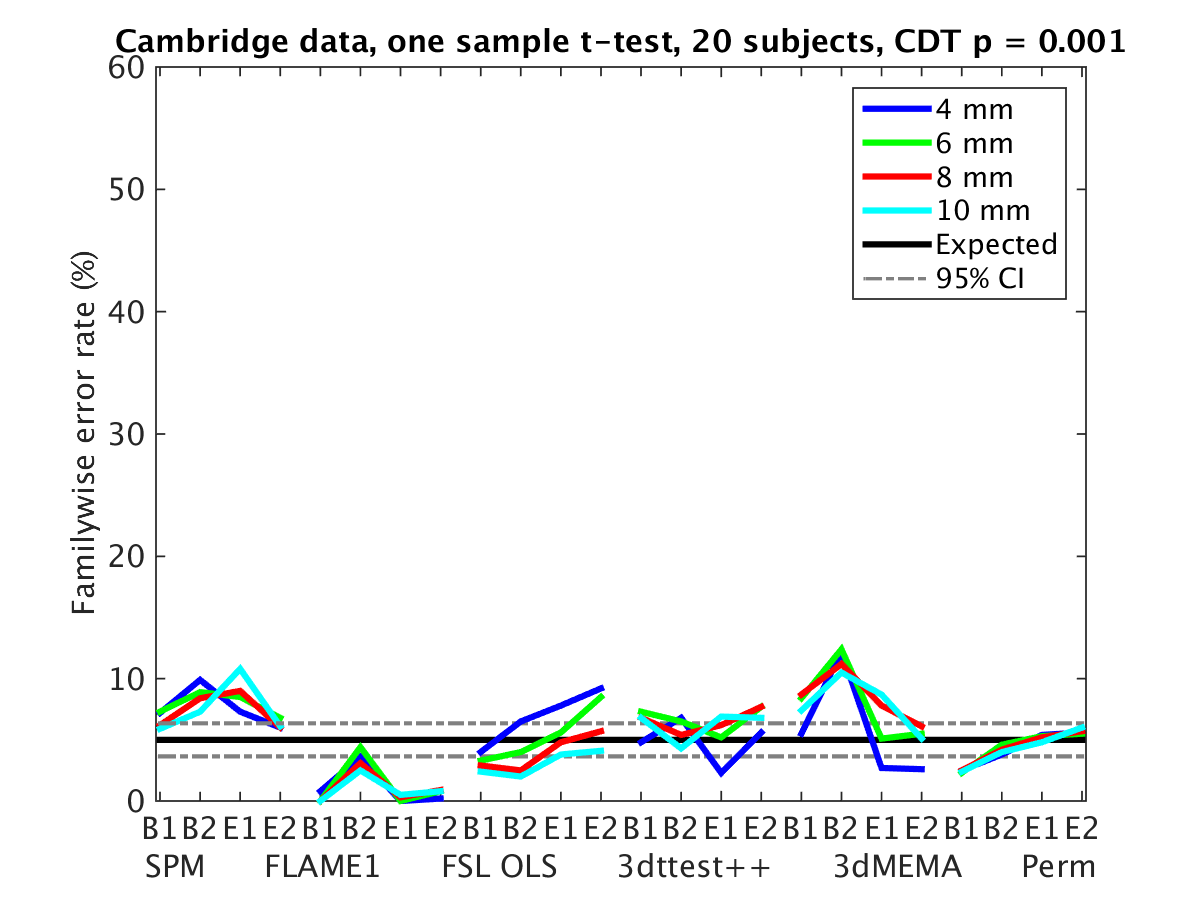}
}
\subfigure[]{
\includegraphics[scale=0.425]{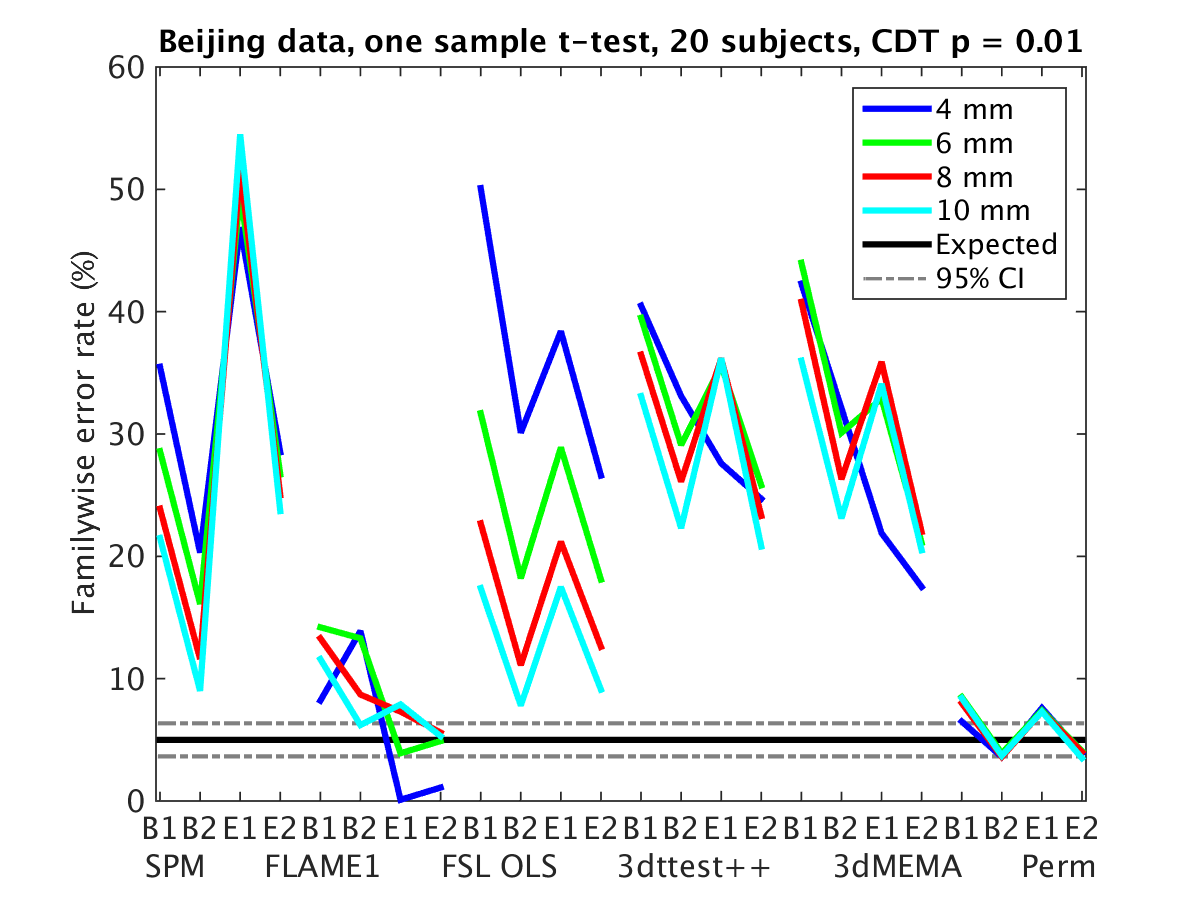}
}
\subfigure[]{
\includegraphics[scale=0.425]{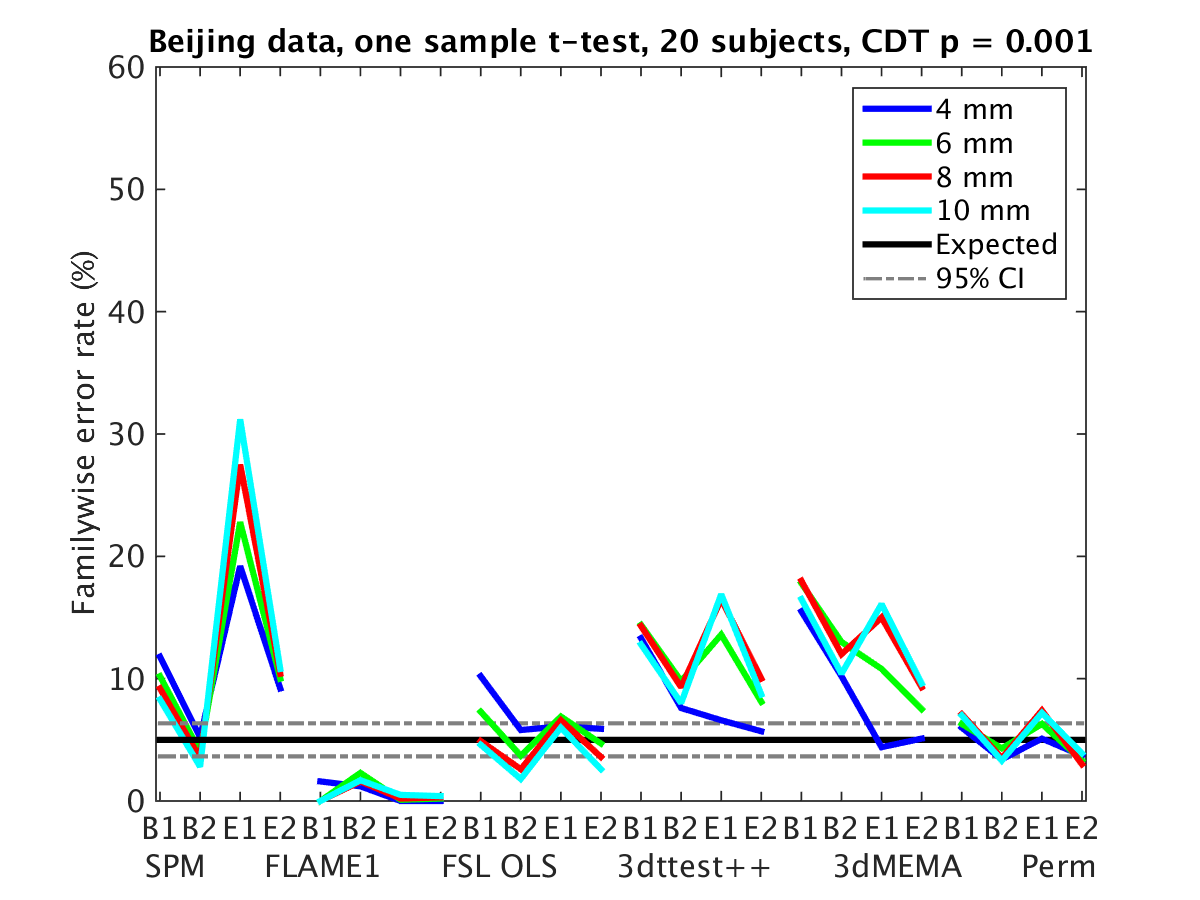}
}
\caption{\emph{Results for one sample t-test and cluster-wise inference, showing estimated familywise error rates for 4-10 mm of smoothing and four different activity paradigms (B1, B2, E1, E2), for SPM, FSL, AFNI and a permutation test. These results are for a group size of 20. Each statistic map was first thresholded using a cluster defining threshold (CDT) (p = 0.01 or p = 0.001, uncorrected for multiple comparisons), and the surviving clusters were then compared to a FWE-corrected cluster extent threshold, $p_{FWE} = 0.05$. The estimated familywise error rates are simply the number of analyses with any significant group activations divided by the number of analyses (1000). Note that the default CDT is p = 0.001 in SPM and p = 0.01 in FSL (AFNI does not have a default setting). Also note that the default amount of smoothing is 8 mm in SPM, 5 mm in FSL and 4 mm in AFNI.  \textbf{(a)} results for Cambridge data and a CDT of p = 0.01 \textbf{(b)} results for Cambridge data and a CDT of p = 0.001  \textbf{(c)} results for Beijing data and a CDT of p = 0.01 \textbf{(d)} results for Beijing data and a CDT of p = 0.001.}}
\label{fig:fwe_cluster_onesample_20_subjects}
\end{figure*}

\begin{figure*}
\centering
\subfigure[]{
\includegraphics[scale=0.425]{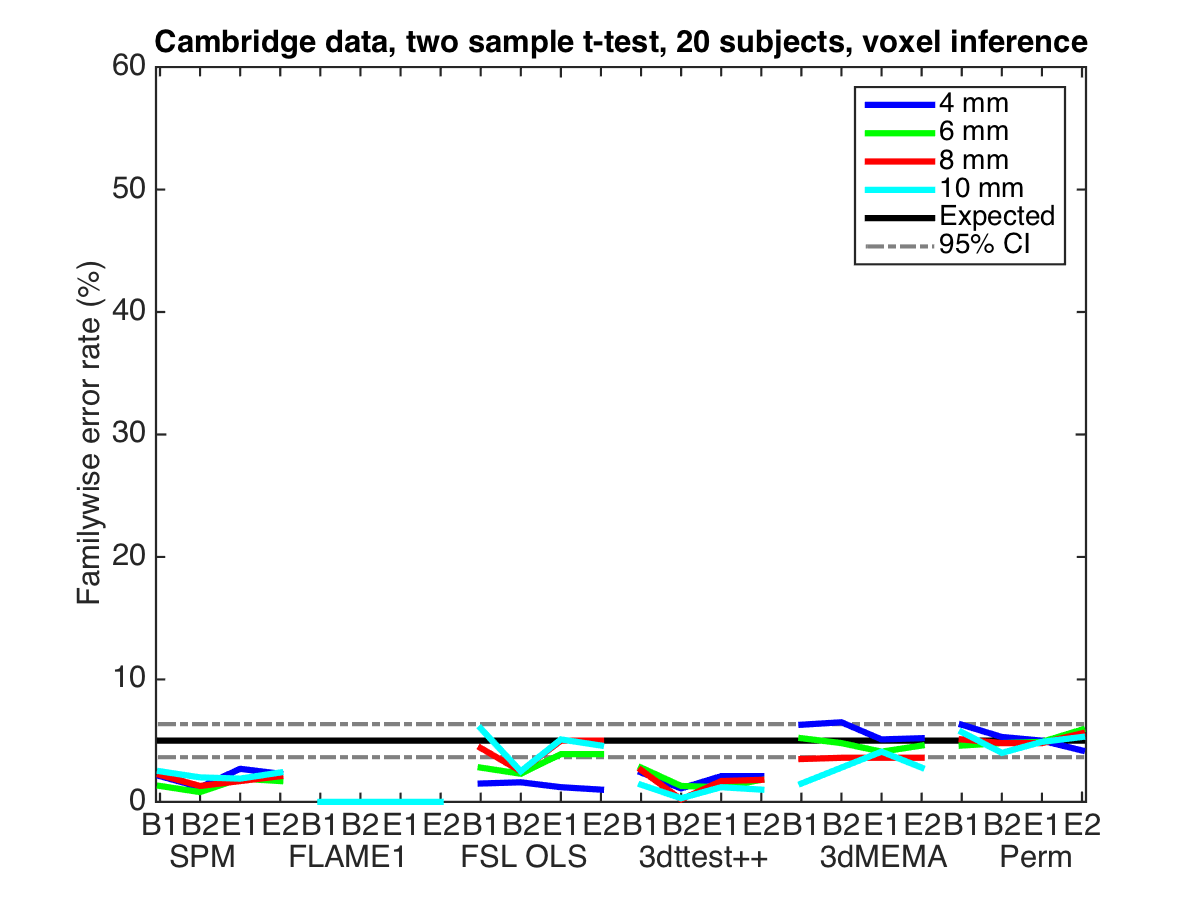}
}
\subfigure[]{
\includegraphics[scale=0.425]{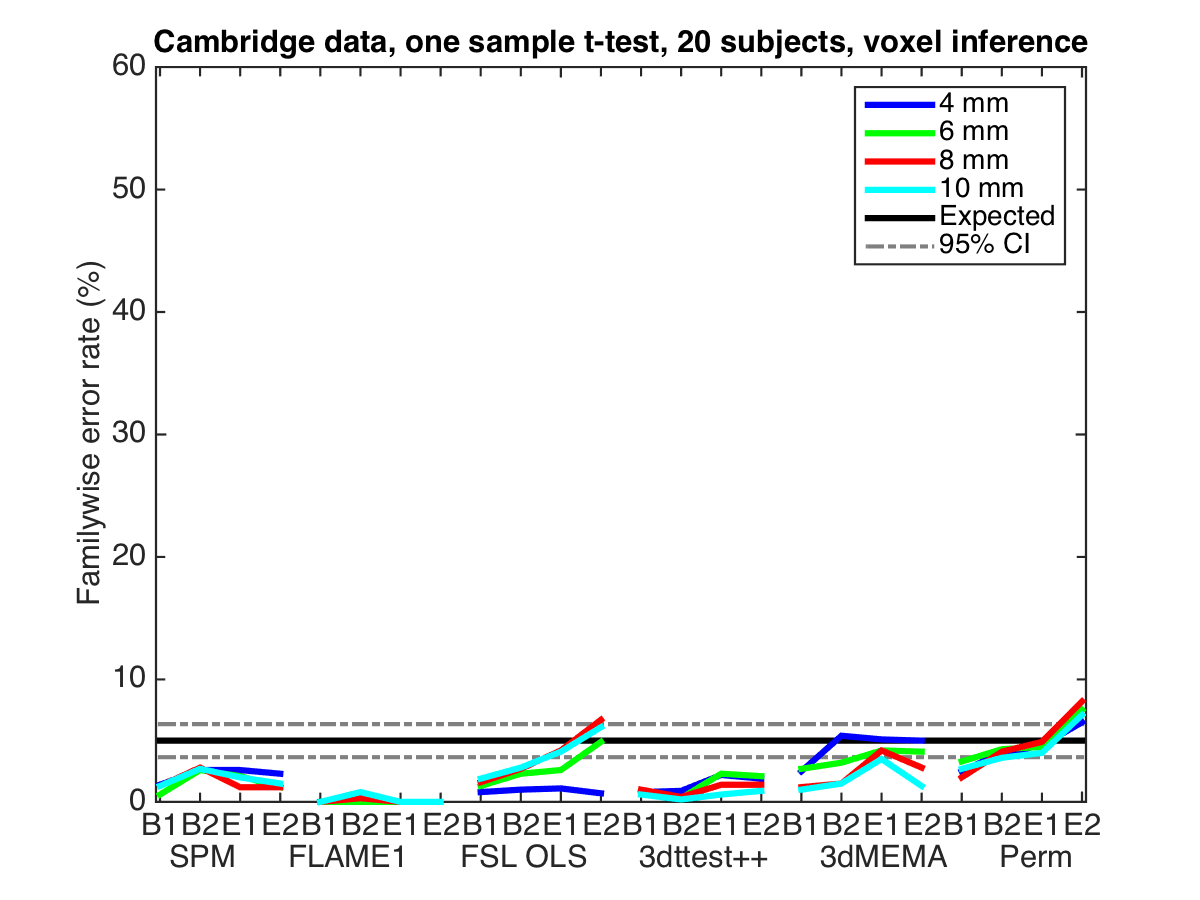}
}
\subfigure[]{
\includegraphics[scale=0.425]{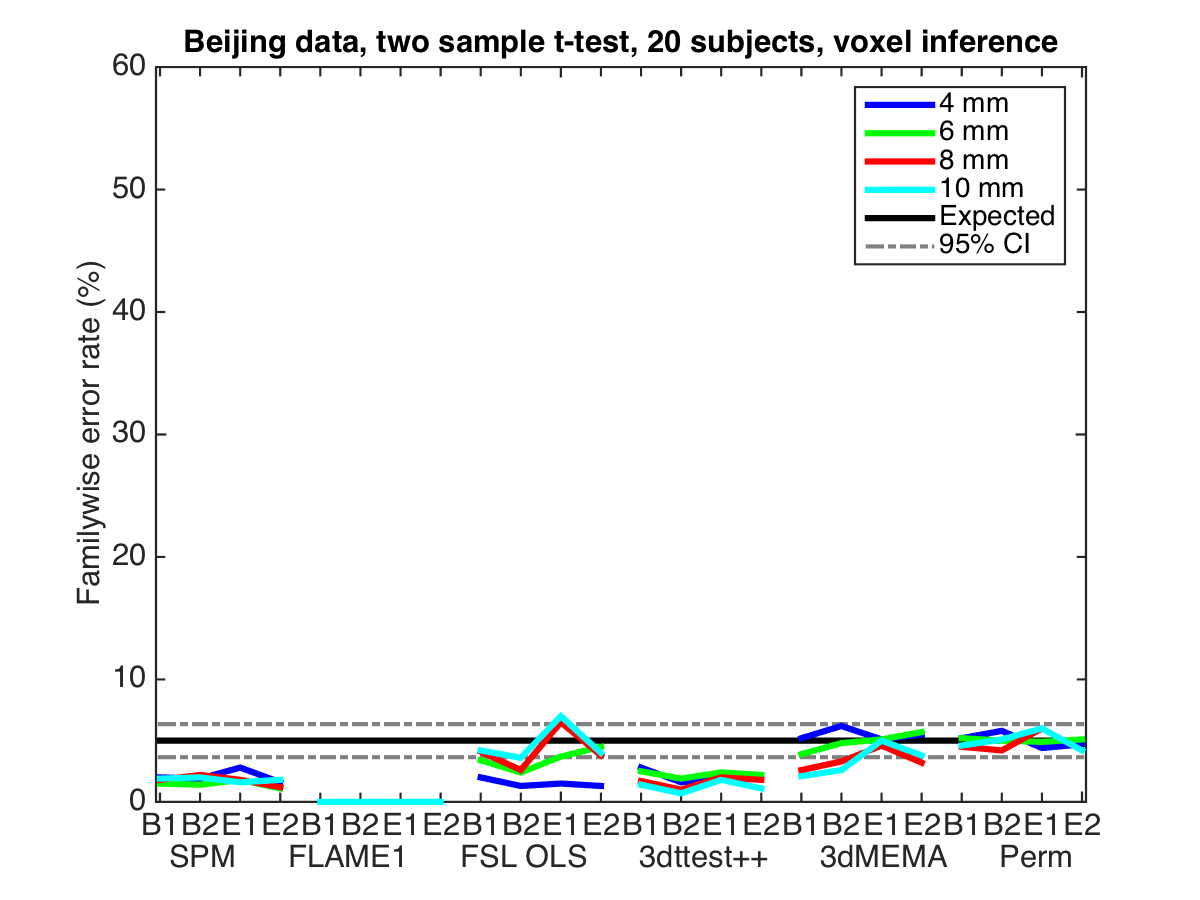}
}
\subfigure[]{
\includegraphics[scale=0.425]{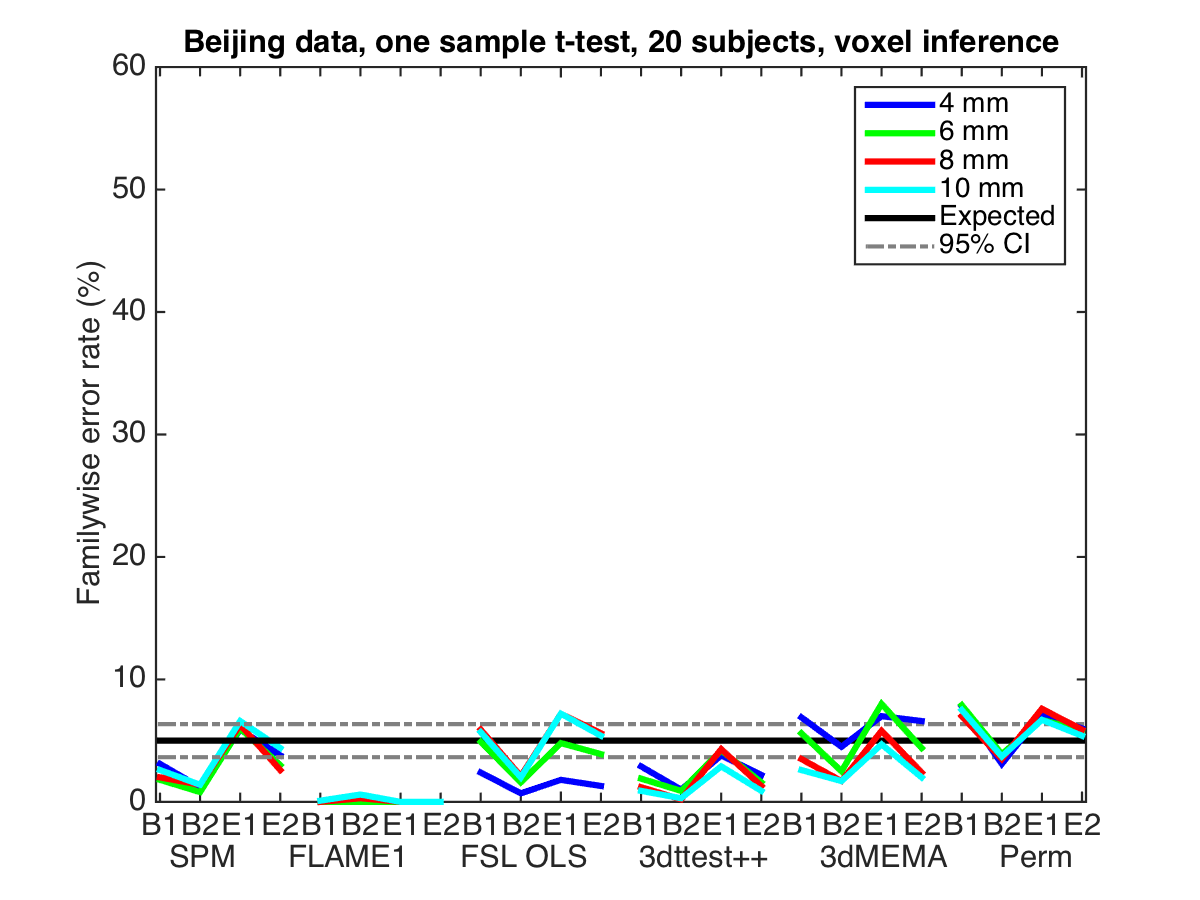}
}
\caption{\emph{Results for two-sample (left) and one-sample (right) t-test and voxel-wise inference, showing estimated familywise error rates for 4-10 mm of smoothing and four different activity paradigms (B1, B2, E1, E2), for SPM, FSL, AFNI and a permutation test. Each statistic map was thresholded using a FWE-corrected voxel-wise threshold of $p_{FWE} = 0.05$. The estimated familywise error rates are simply the number of analyses with any significant results divided by the number of analyses (1000). Note that the default amount of smoothing is 8 mm in SPM, 5 mm in FSL and 4 mm in AFNI. \textbf{(a)} results for two sample t-tests using Cambridge data \textbf{(b)} results for one sample t-tests using Cambridge data \textbf{(c)} results for two sample t-tests using Beijing data \textbf{(d)} results for one sample t-tests using Beijing data.}}
\label{fig:fwe_voxel_20_subjects}
\end{figure*}

\begin{figure*}
\centering
\subfigure[]{
\includegraphics[scale=0.425]{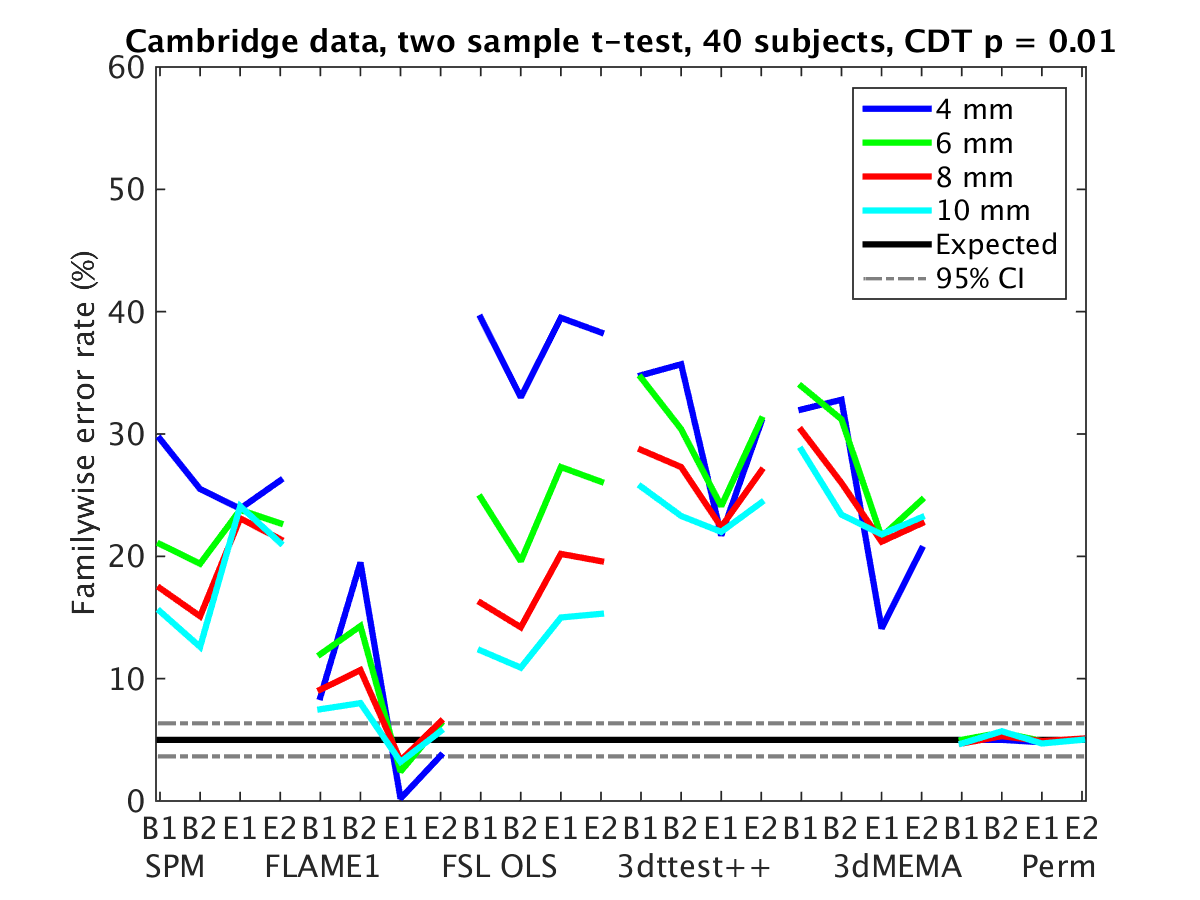}
}
\subfigure[]{
\includegraphics[scale=0.425]{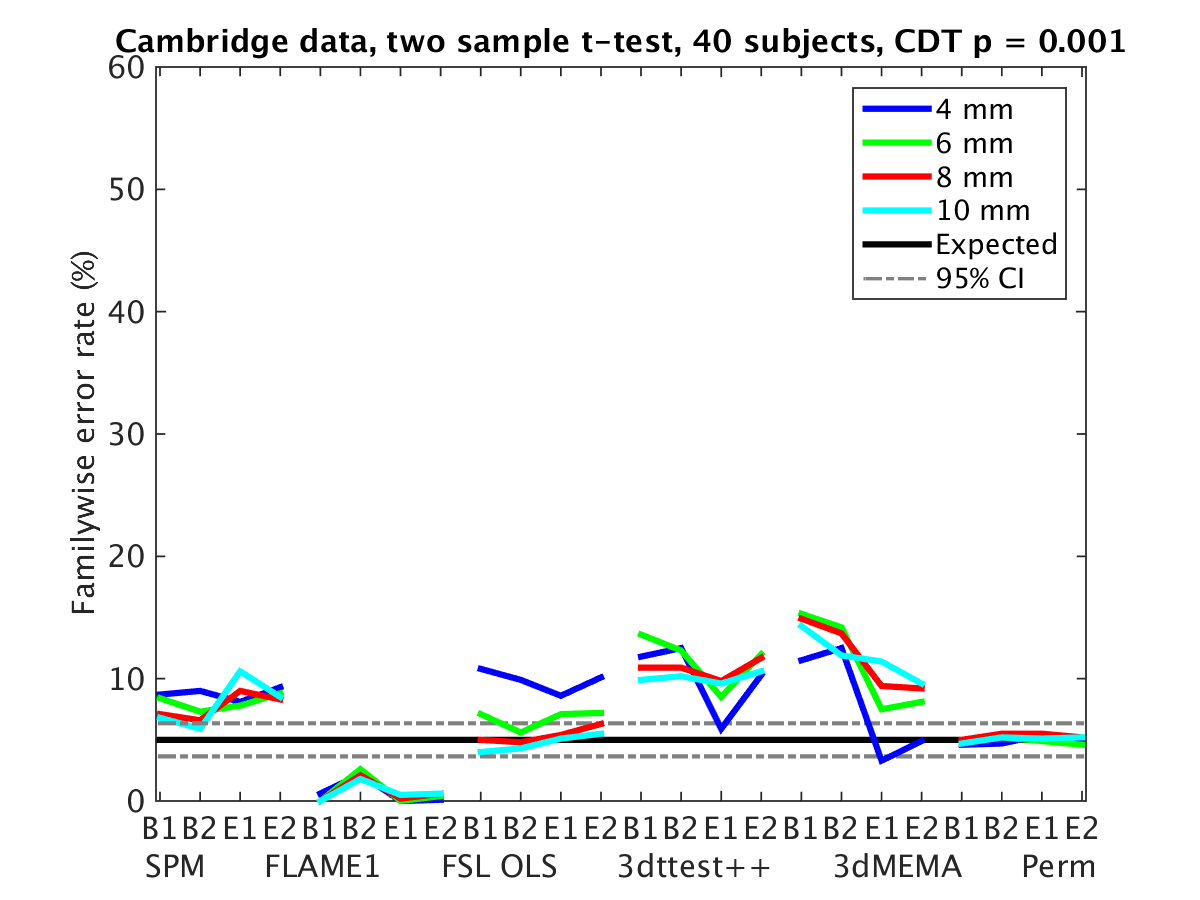}
}
\subfigure[]{
\includegraphics[scale=0.425]{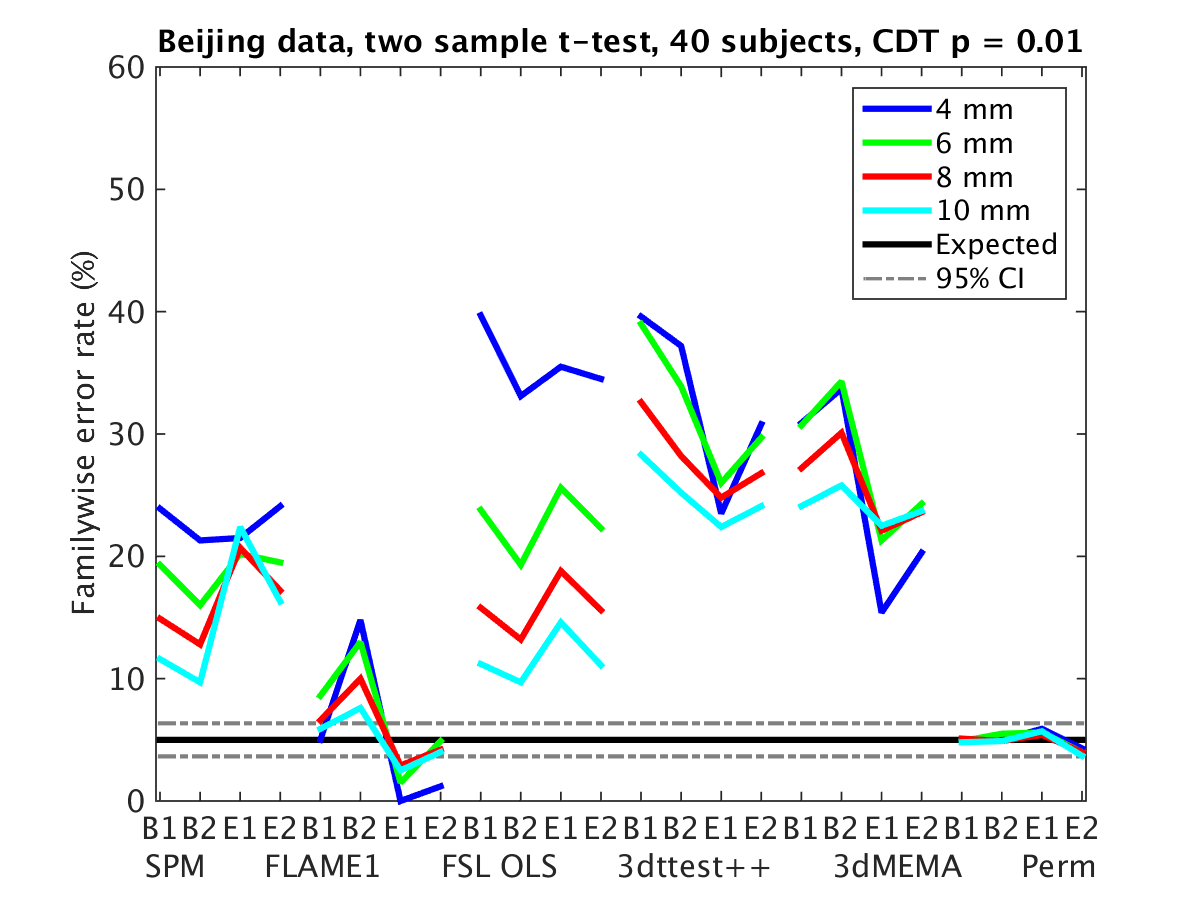}
}
\subfigure[]{
\includegraphics[scale=0.425]{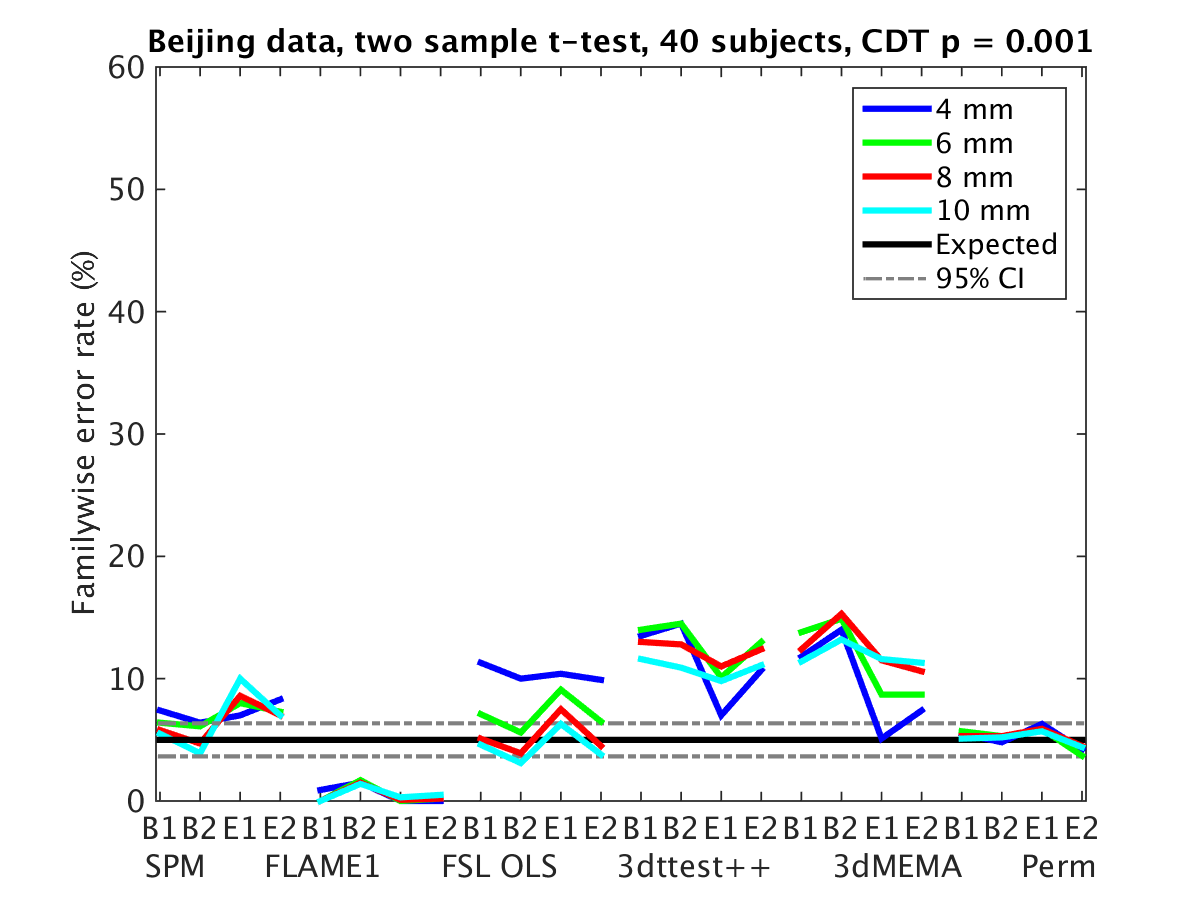}
}
\caption{\emph{Results for two-sample t-test and cluster-wise inference, showing estimated familywise error rates for 4-10 mm of smoothing and four different activity paradigms (B1, B2, E1, E2), for SPM, FSL, AFNI and a permutation test. These results are for a group size of 20 (giving a total of 40 subjects). Each statistic map was first thresholded using a cluster defining threshold (CDT) (p = 0.01 or p = 0.001, uncorrected for multiple comparisons), and the surviving clusters were then compared to a FWE-corrected cluster extent threshold, $p_{FWE} = 0.05$. The estimated familywise error rates are simply the number analyses with any significant group differences divided by the number of analyses (1000). Note that the default CDT is p = 0.001 in SPM and p = 0.01 in FSL (AFNI does not have a default setting). Also note that the default amount of smoothing is 8 mm in SPM, 5 mm in FSL and 4 mm in AFNI.  \textbf{(a)} results for Cambridge data and a CDT of p = 0.01 \textbf{(b)} results for Cambridge data and a CDT of p = 0.001  \textbf{(c)} results for Beijing data and a CDT of p = 0.01 \textbf{(d)} results for Beijing data and a CDT of p = 0.001.}}
\label{fig:fwe_cluster_twosample_40_subjects}
\end{figure*}

\begin{figure*}
\centering
\subfigure[]{
\includegraphics[scale=0.425]{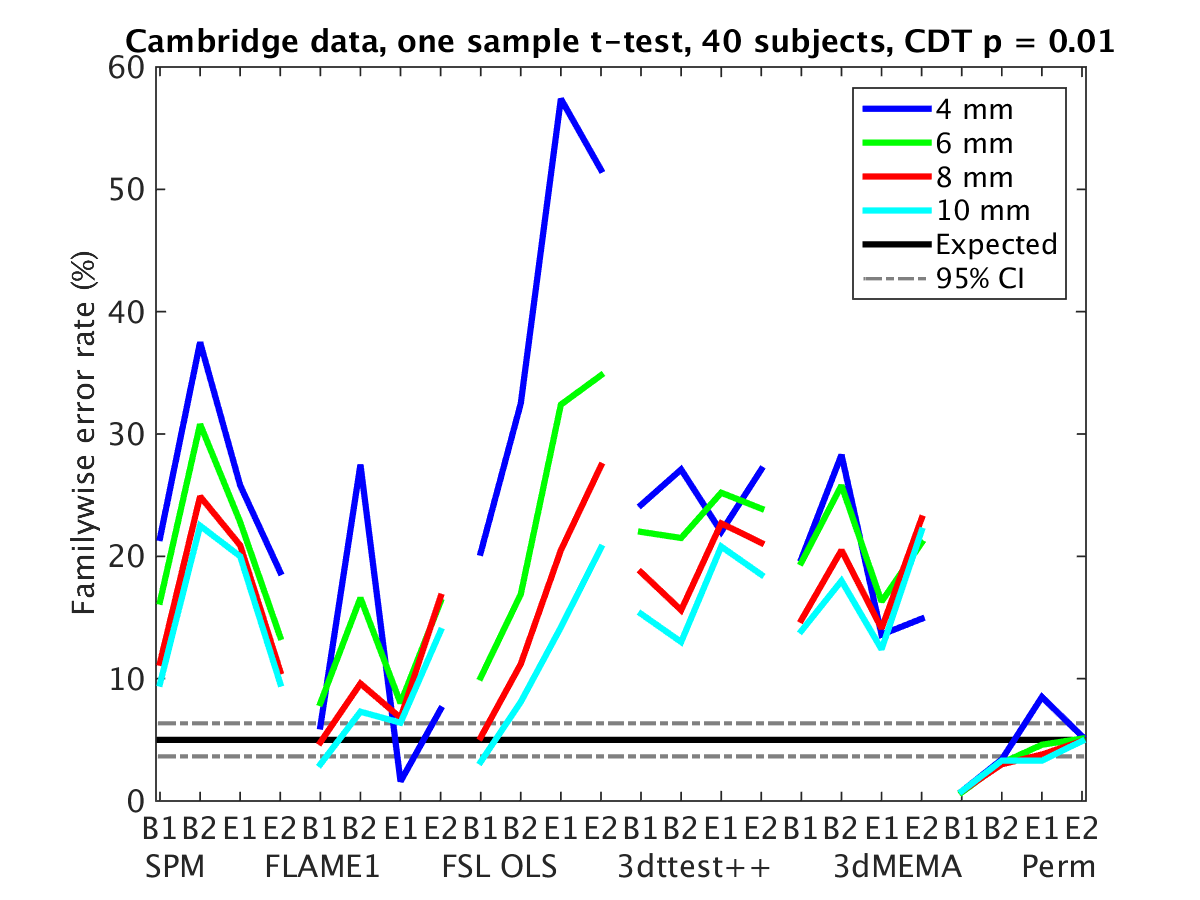}
}
\subfigure[]{
\includegraphics[scale=0.425]{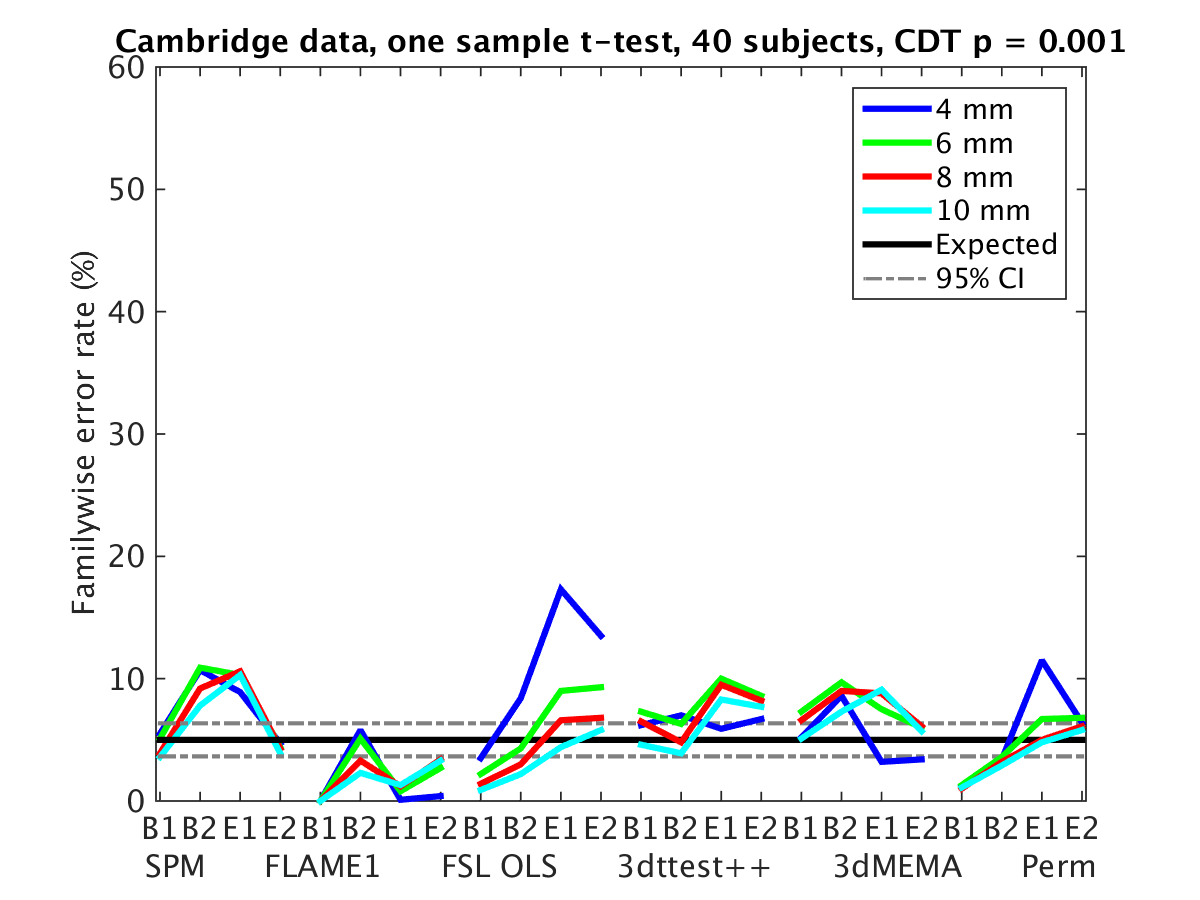}
}
\subfigure[]{
\includegraphics[scale=0.425]{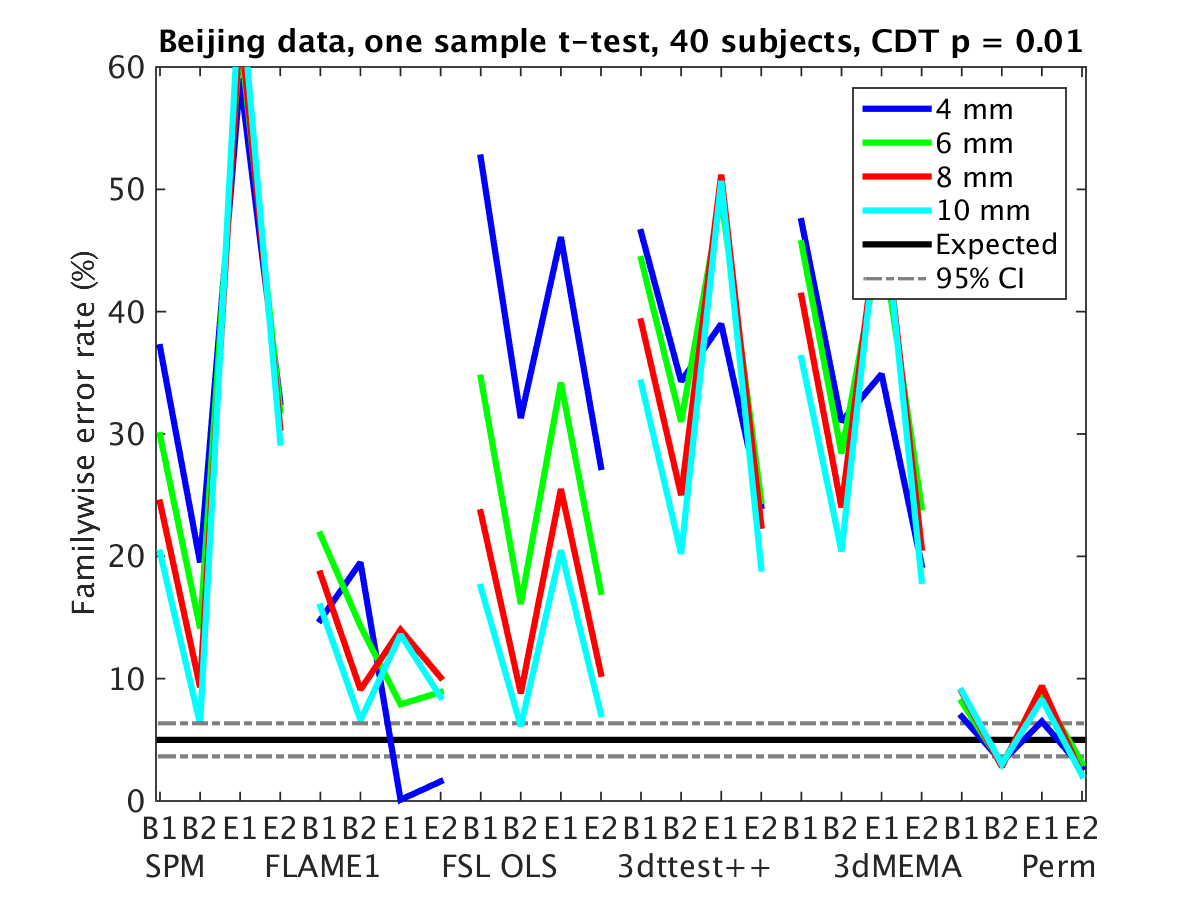}
}
\subfigure[]{
\includegraphics[scale=0.425]{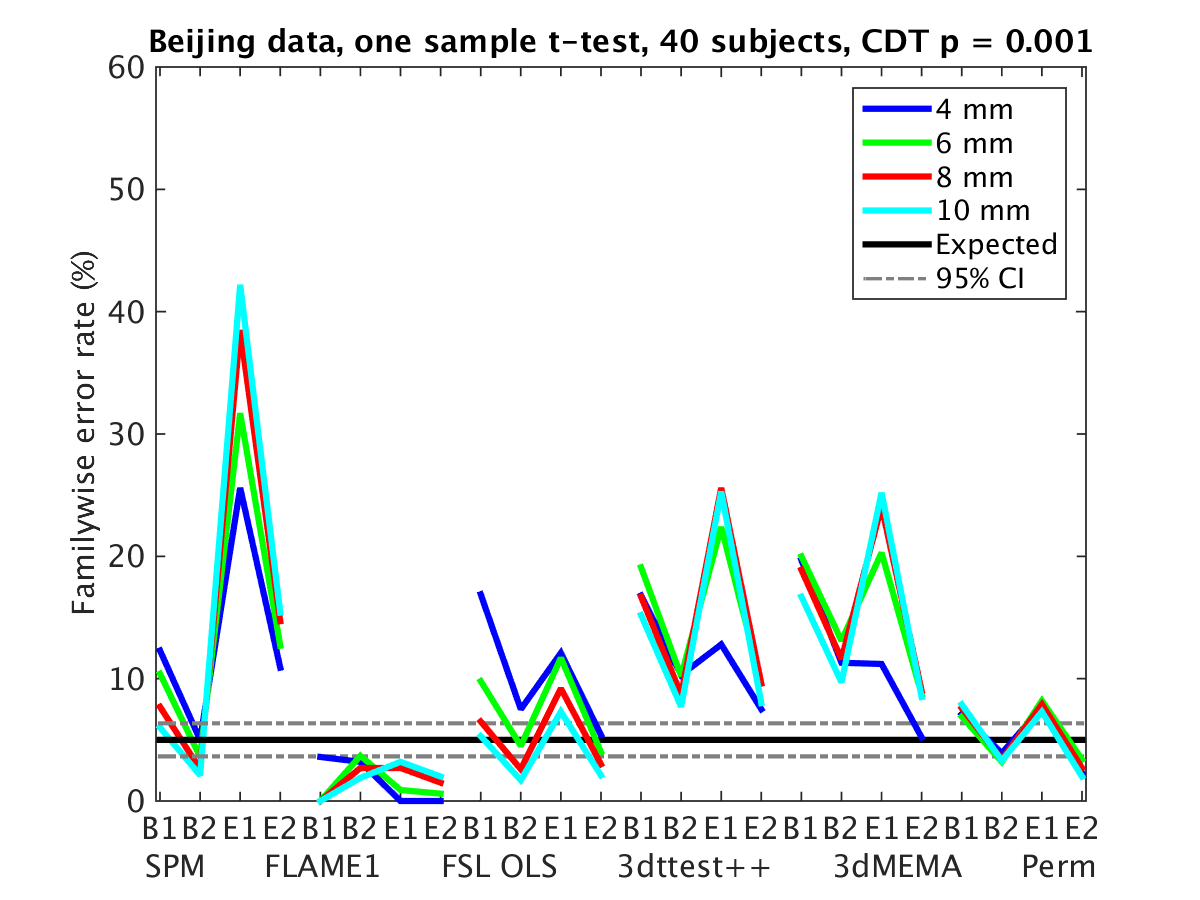}
}
\caption{\emph{Results for one-sample t-test and cluster-wise inference, showing estimated familywise error rates for 4-10 mm of smoothing and four different activity paradigms (B1, B2, E1, E2), for SPM, FSL, AFNI and a permutation test. These results are for a group size of 40. Each statistic map was first thresholded using a cluster defining threshold (CDT) (p = 0.01 or p = 0.001, uncorrected for multiple comparisons), and the surviving clusters were then compared to a FWE-corrected cluster extent threshold, $p_{FWE} = 0.05$. The estimated familywise error rates are simply the number of analyses with significant group activations divided by the number of analyses (1000). Note that the default CDT is p = 0.001 in SPM and p = 0.01 in FSL (AFNI does not have a default setting). Also note that the default amount of smoothing is 8 mm in SPM, 5 mm in FSL and 4 mm in AFNI.  \textbf{(a)} results for Cambridge data and a CDT of p = 0.01 \textbf{(b)} results for Cambridge data and a CDT of p = 0.001  \textbf{(c)} results for Beijing data and a CDT of p = 0.01 \textbf{(d)} results for Beijing data and a CDT of p = 0.001. }}
\label{fig:fwe_cluster_onesample_40_subjects}
\end{figure*}

\begin{figure*}
\centering
\subfigure[]{
\includegraphics[scale=0.425]{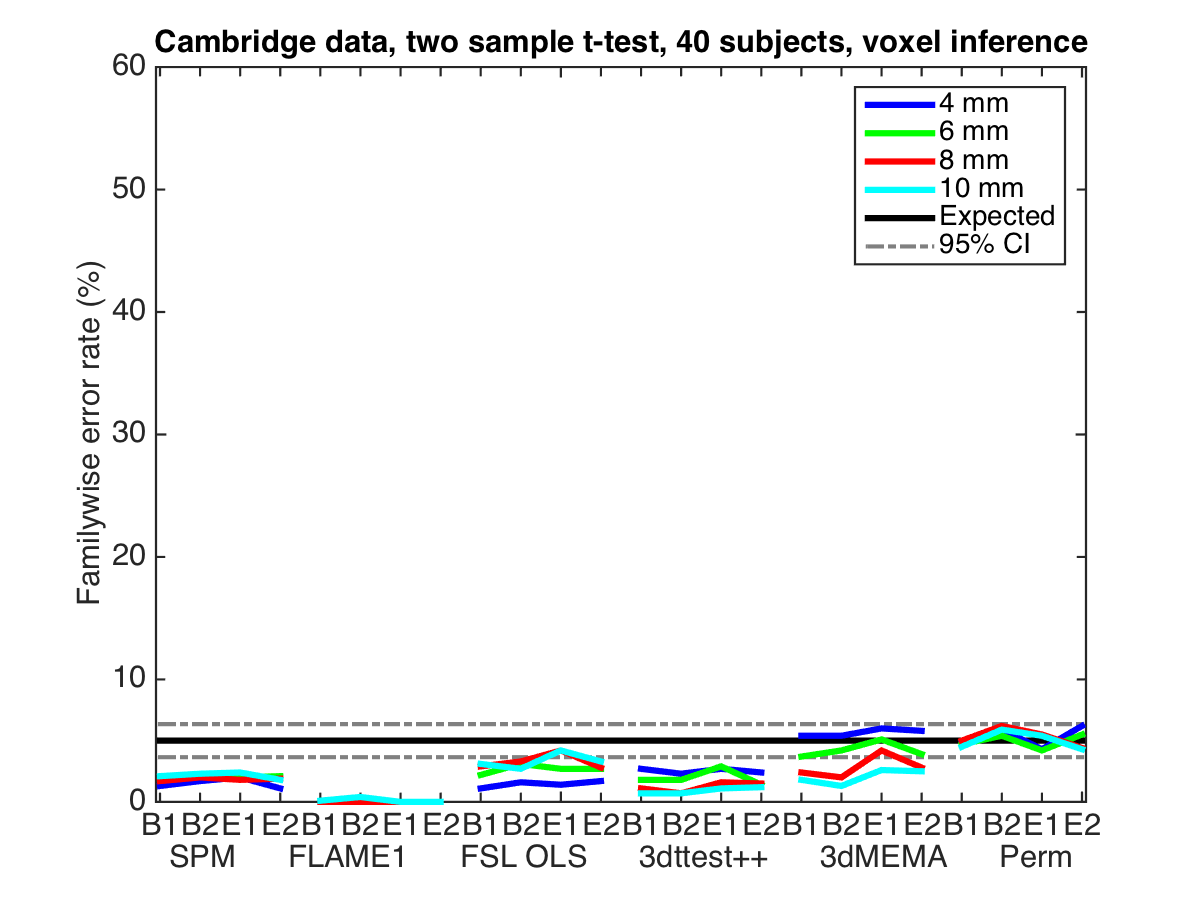}
}
\subfigure[]{
\includegraphics[scale=0.425]{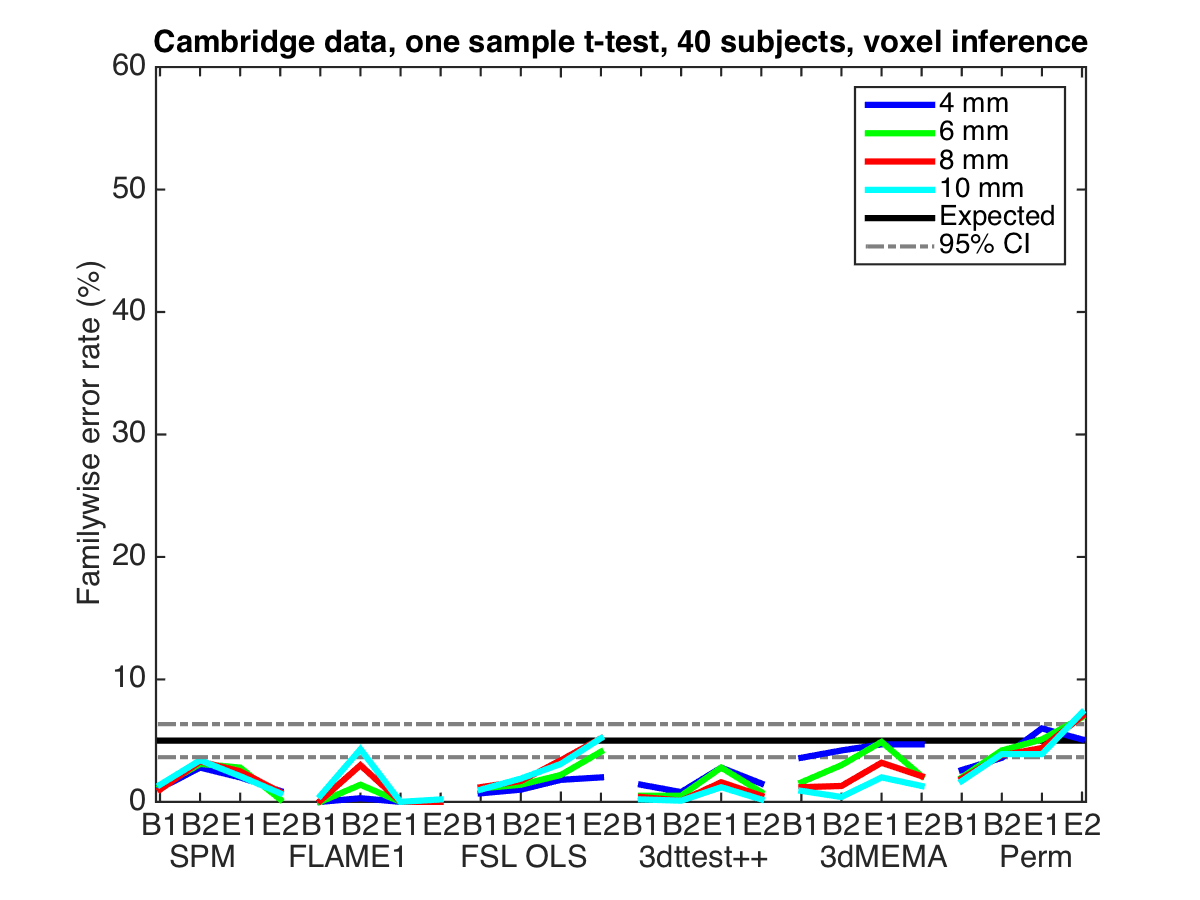}
}
\subfigure[]{
\includegraphics[scale=0.425]{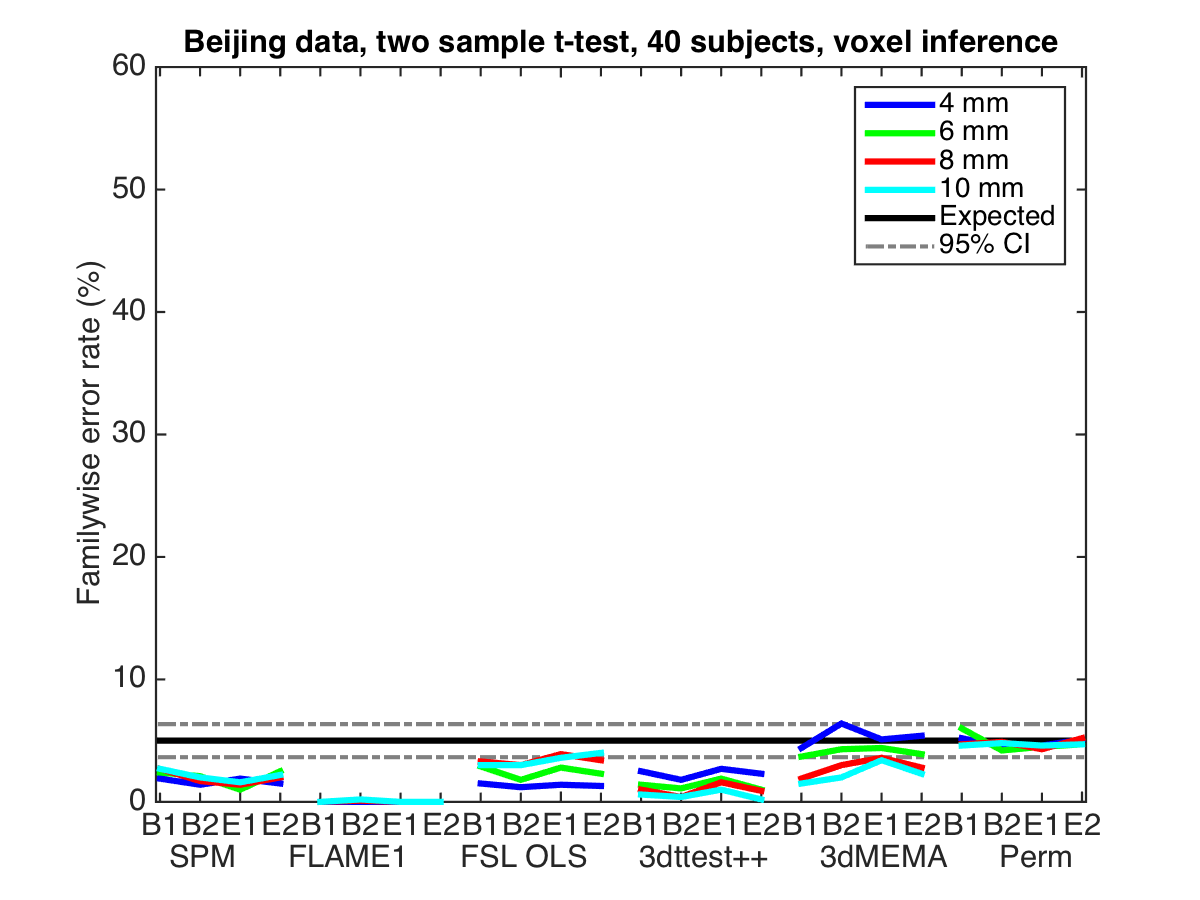}
}
\subfigure[]{
\includegraphics[scale=0.425]{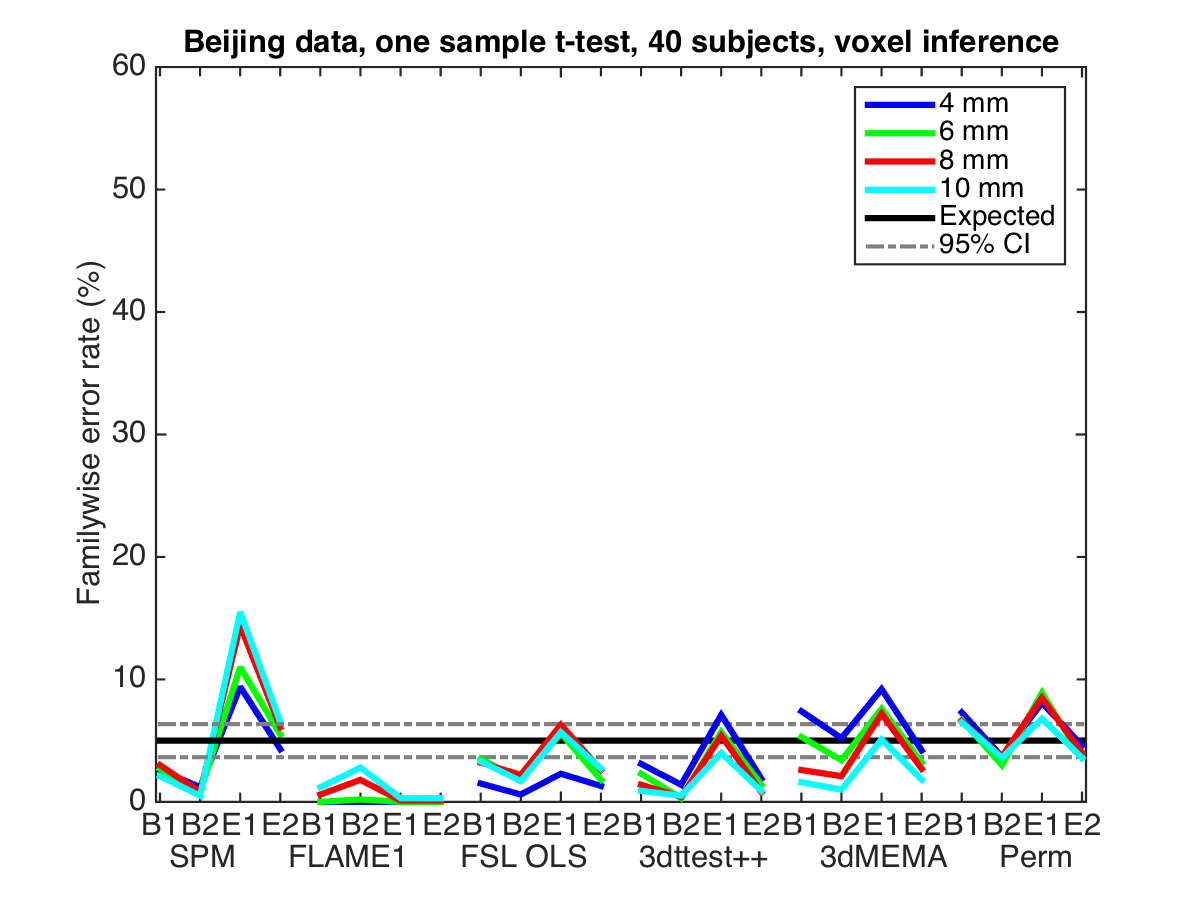}
}
\caption{\emph{Results for two-sample (left) and one-sample (right) t-test and voxel-wise inference, showing estimated familywise error rates for 4-10 mm of smoothing and four different activity paradigms (B1, B2, E1, E2), for SPM, FSL, AFNI and a permutation test. These results are for 40 subjects. Each statistic map was thresholded using a FWE-corrected voxel-wise threshold of $p_{FWE} = 0.05$. The estimated familywise error rates are simply the number of analyses with any significant results divided by the number of analyses (1000). Note that the default amount of smoothing is 8 mm in SPM, 5 mm in FSL and 4 mm in AFNI. \textbf{(a)} results for two sample t-tests using Cambridge data \textbf{(b)} results for one sample t-tests using Cambridge data \textbf{(c)} results for two sample t-tests using Beijing data \textbf{(d)} results for one sample t-tests using Beijing data.}}
\label{fig:fwe_voxel_40_subjects}
\end{figure*}

Figures~\ref{fig:fwe_cluster_twosample_20_subjects} -~\ref{fig:fwe_voxel_20_subjects} present the main findings of the study, showing cluster-wise (Figure~\ref{fig:fwe_cluster_twosample_20_subjects}, two-sample t-test; Figure~\ref{fig:fwe_cluster_onesample_20_subjects}, one-sample t-test) and voxel-wise (Figure~\ref{fig:fwe_voxel_20_subjects}) results for a total sample size of 20 (Figures~\ref{fig:fwe_cluster_twosample_40_subjects} -~\ref{fig:fwe_voxel_40_subjects} show corresponding results for a total sample size of 40). In broad summary, parametric software's familywise error (FWE) rates for cluster-wise inference far exceed their nominal 5\% level, while parametric voxel-wise inferences are valid but often conservative, often falling below 5\%. Permutation false positives are controlled at a nominal 5\% except for cluster-wise inference with a one sample t-test, mainly with the Beijing data with designs B1 and E1. The impact of cluster defining threshold (CDT) was appreciable for the parametric methods, with CDT p = 0.001 having much better FWE control than CDT p = 0.01. Among the parametric software packages, FSL's FLAME1 cluster-wise inference stood out as having much lower FWE, often being valid (under 5\%). But with cluster-wise CDT p = 0.001, and voxel-wise inference, FLAME1 was often very conservative.


\begin{table*}[htb]
\scriptsize
\caption{Parameters tested for the different fMRI software packages, giving a total of 128 parameter combinations and 3 thresholding approaches. One thousand group analyses were performed for each parameter combination.}
\begin{center}
\begin{tabular}{|c|c|}
\hline 
\textbf{\normalsize Parameter}  & \textbf{\normalsize Values used}  \\[0.2ex]
\hline
\normalsize fMRI data & \normalsize Beijing (198 subjects), Cambridge (198 subjects) \\[0.2ex]
\normalsize Activity paradigm & \normalsize Block (B1, B2), event (E1, E2) \\[0.2ex]

\normalsize Smoothing & \normalsize 4, 6, 8, 10 mm FWHM \\[0.2ex]
\normalsize Analysis type & \normalsize One sample t-test (group activation), two sample t-test (group difference) \\[0.2ex]
\normalsize Number of subjects & \normalsize 20, 40 \\[0.2ex]
\normalsize Inference level & \normalsize Voxel, cluster \\[0.2ex]
\normalsize Cluster defining threshold & \normalsize p = 0.01 (z = 2.3), p = 0.001 (z = 3.1) \\[0.2ex]
\hline
\end{tabular}
\end{center}
\label{table:levels}
\end{table*}

\begin{table*}[htb]
\scriptsize
\caption{Length of activity and rest periods for the used (pretended) activity paradigms, R stands for randomized. Number of periods}
\begin{center}
\begin{tabular}{|c|c|c|}
\hline 
\textbf{\normalsize Paradigm}  & \textbf{\normalsize Activity duration (s)} & \textbf{\normalsize Rest duration (s)}  \\[0.2ex]
\hline
\normalsize B1 & \normalsize 10  & \normalsize 10 \\[0.2ex]
\normalsize B2 & \normalsize 30  & \normalsize 30 \\[0.2ex]
\normalsize E1 & \normalsize 2  & \normalsize 6 \\[0.2ex]
\normalsize E2 & \normalsize 1-4 (R) & \normalsize 3-6 (R) \\[0.2ex]
\hline
\end{tabular}
\end{center}
\label{table:paradigms}
\end{table*}

\clearpage
\newpage

\cleardoublepage
\newpage

\cleardoublepage
\newpage

\cleardoublepage
\newpage

\cleardoublepage
\newpage

\subsection{What about other common cluster thresholding approaches?}

\begin{figure}
\center
\includegraphics[scale=0.45]{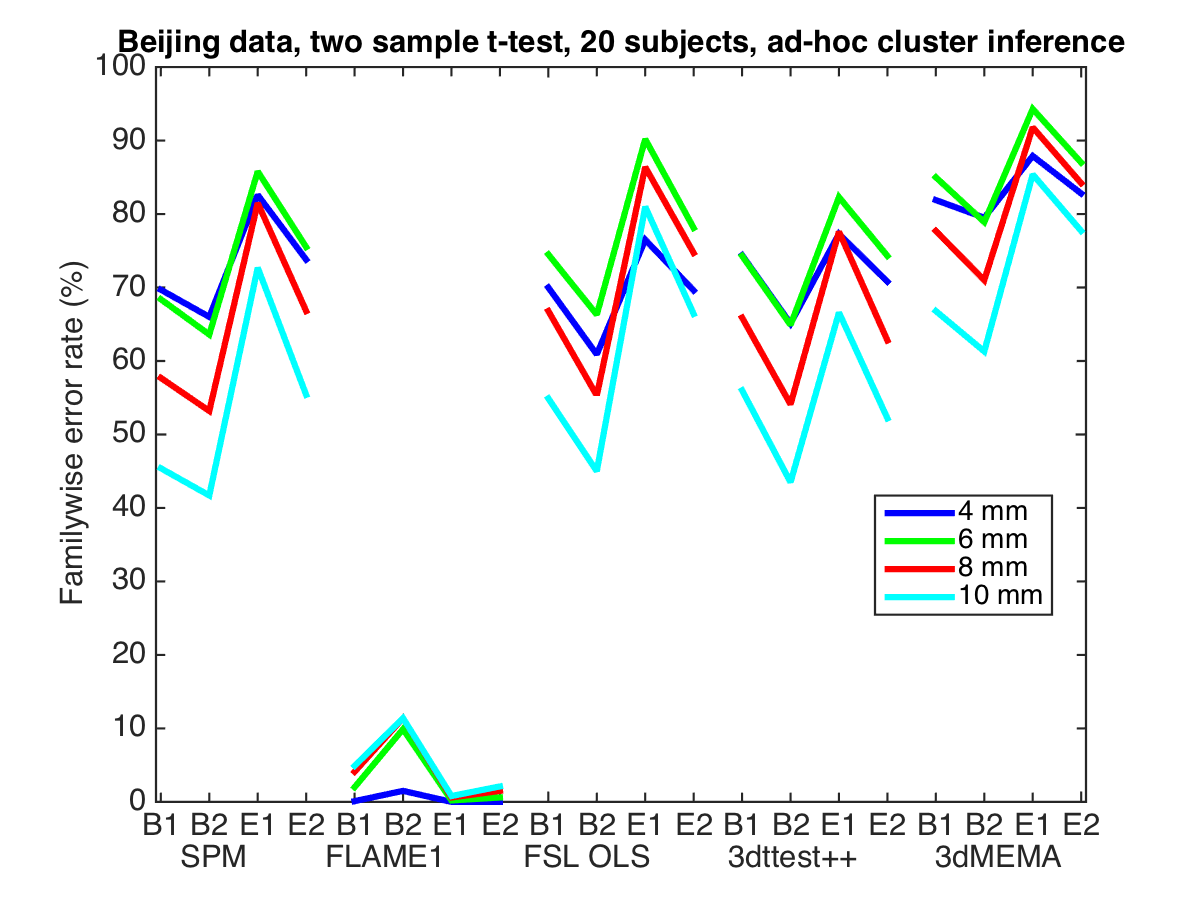}
\caption{\emph{Results for two-sample t-test and ad-hoc cluster-wise inference, showing estimated familywise error rates for 4-10 mm of smoothing and four different activity paradigms (B1, B2, E1, E2), for SPM, FSL and AFNI. These results are for a total of 20 subjects. Each statistic map was first thresholded using a cluster defining threshold of p = 0.001 (uncorrected for multiple comparisons), and the surviving clusters were then compared to a cluster extent threshold of 80 mm$^3$. The estimated familywise error rates are simply the number of analyses with a significant result divided by the number of analyses (1000).}}
\label{fig:fwe_adhoc_cluster}
\end{figure}

A common thresholding approach in the fMRI field is to use a rather stringent cluster defining threshold, e.g.\ p = 0.001 (uncorrected for multiple comparisons), together with an arbitrary cluster extent threshold of 10 voxels (equal to a volume of 80 mm$^3$ for a voxel size of 2 x 2 x 2 mm)~\cite{carp2,lieberman}. Such an approach is ad-hoc, in the sense that one does not know what the (corrected) p-value is for the combined procedure. For this reason, 80,000 additional analyses (1,000 analyses repeated for four levels of smoothing, four pretended activity paradigms and five different software tools) were performed to investigate the validity of this common thresholding approach. The additional analyses were only performed for the Beijing data, for a two sample t-test using a total of 20 subjects. The resulting familywise error rates, i.e.\ how likely it is to detect at least one cluster of 10 voxels in the entire brain when the null hypothesis is true, are given in Figure~\ref{fig:fwe_adhoc_cluster}. Note that the expected degree of false positives is unknown, since the thresholding approach is ad-hoc. Also note that a cluster extent of 3 voxels was used for AFNI, as the AFNI software by default resamples the data to a voxel size of 3 x 3 x 3 mm (a cluster of 3 voxels is for AFNI thereby equal to a volume of 81 mm$^3$). The probability of finding at least one cluster with a volume of 80 mm$^3$, after an initial voxel-wise threshold of p = 0.001, is 50\% - 90\% for all functions except FLAME1 in FSL. The common cluster thresholding approach thus corresponds to a corrected p-value of 0.5 - 0.9. Using an initial voxel-wise threshold of p = 0.005~\cite{lieberman} is of course even more problematic.

\subsection{How close are the statistical test values compared to the theoretical null distributions?}

\begin{figure*}
\center
\subfigure[]{
\includegraphics[scale=0.415]{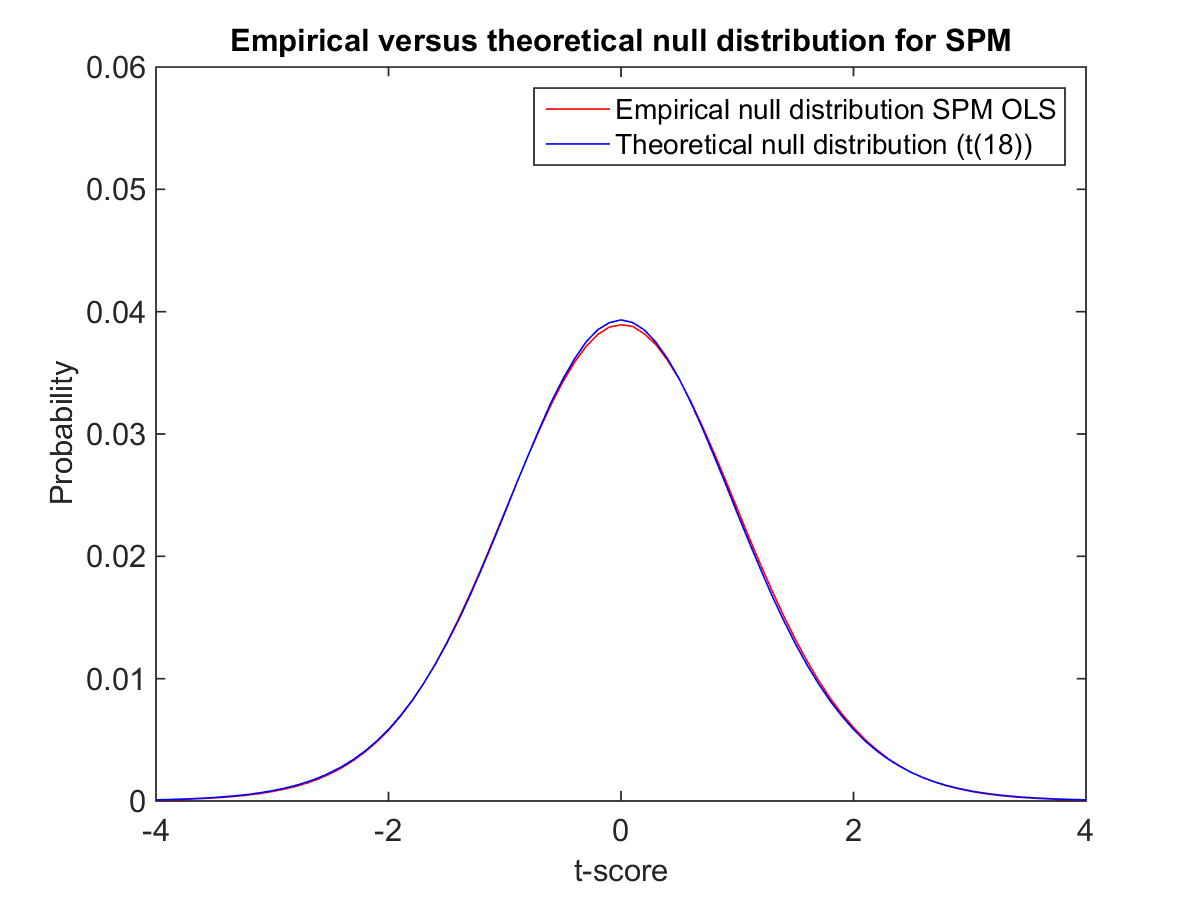}
}
\subfigure[]{
\includegraphics[scale=0.415]{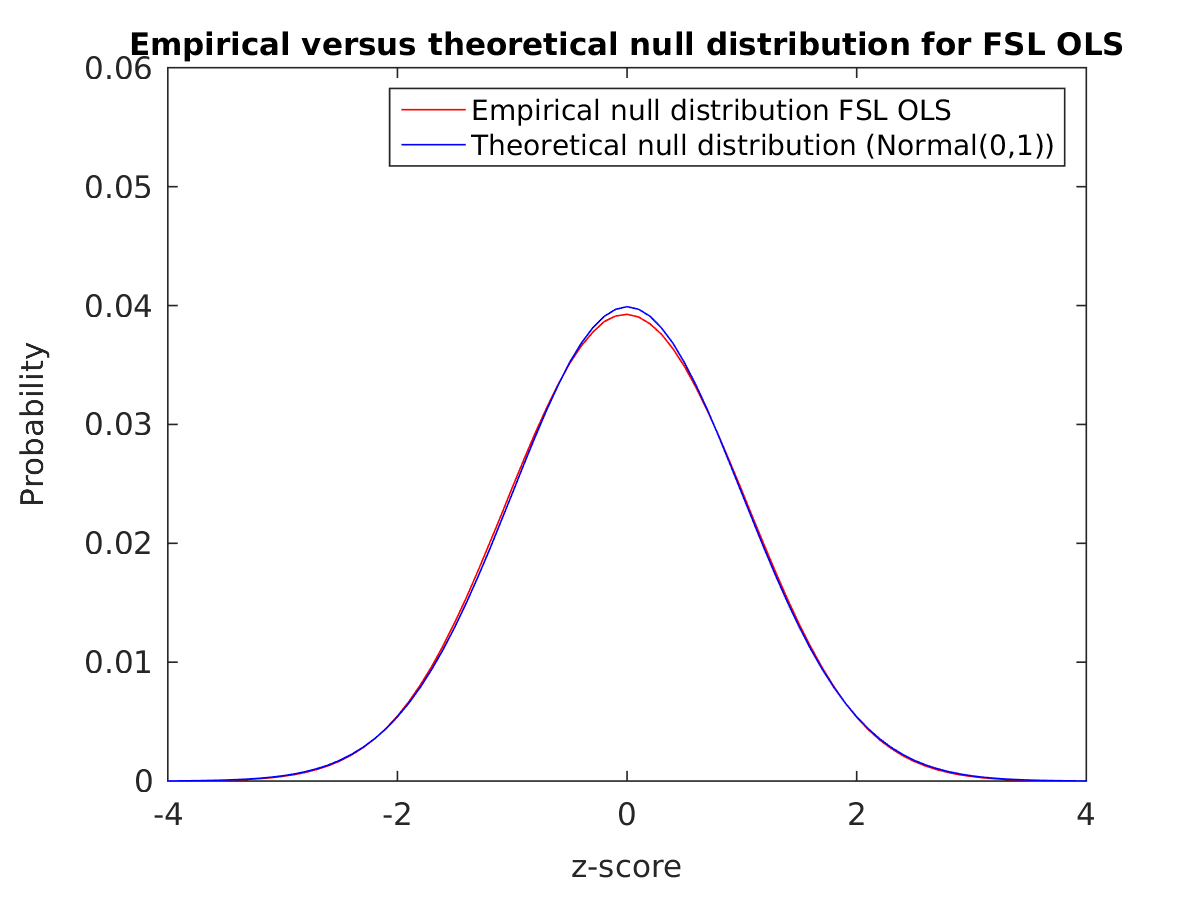}
\includegraphics[scale=0.415]{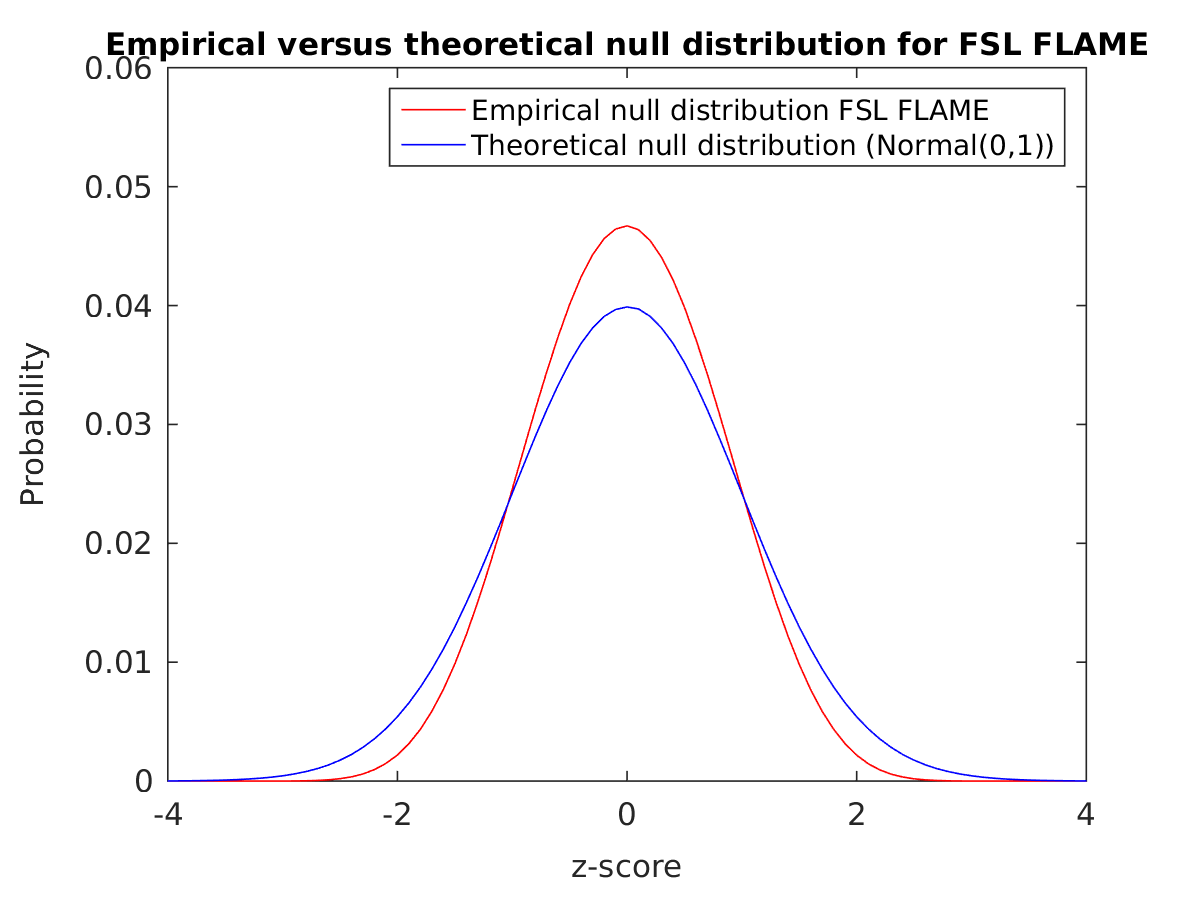}
}
\subfigure[]{
\includegraphics[scale=0.415]{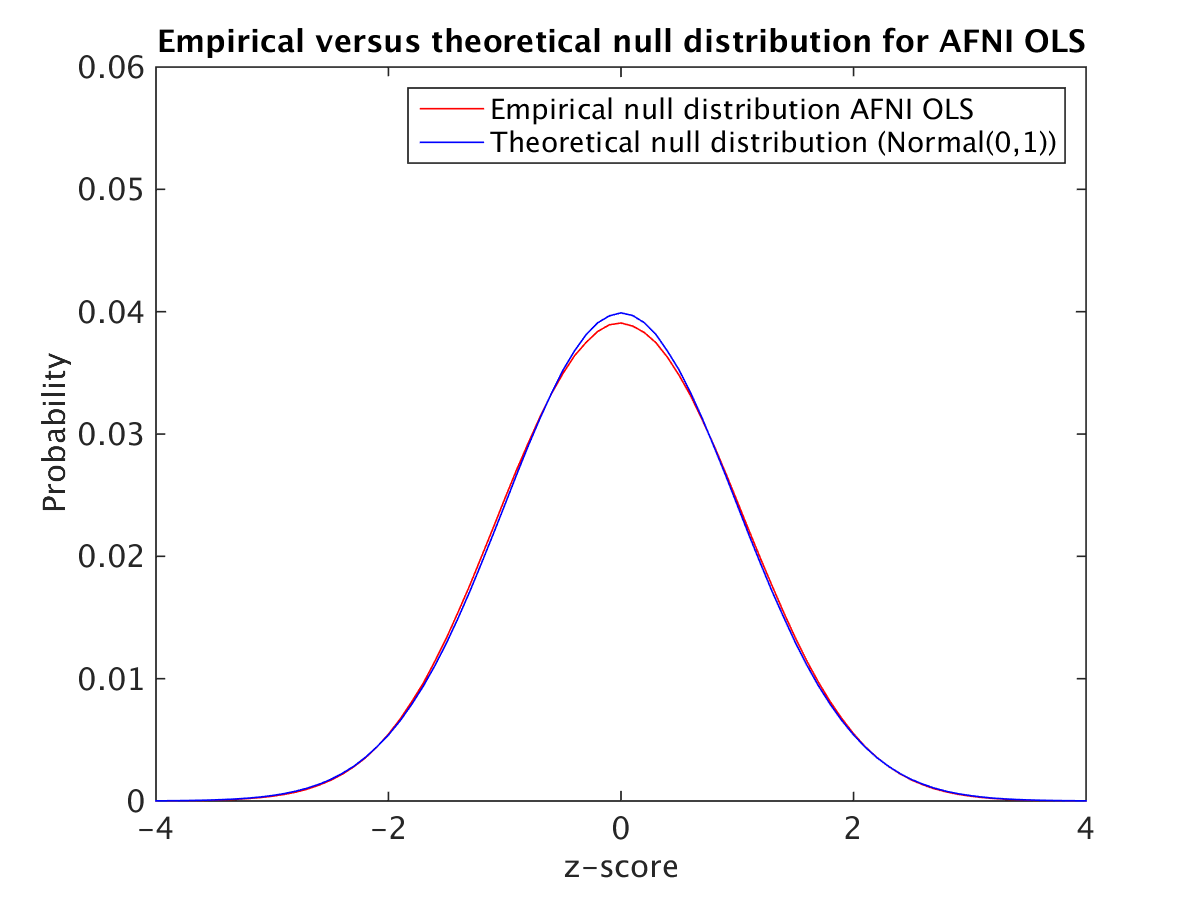}
\includegraphics[scale=0.415]{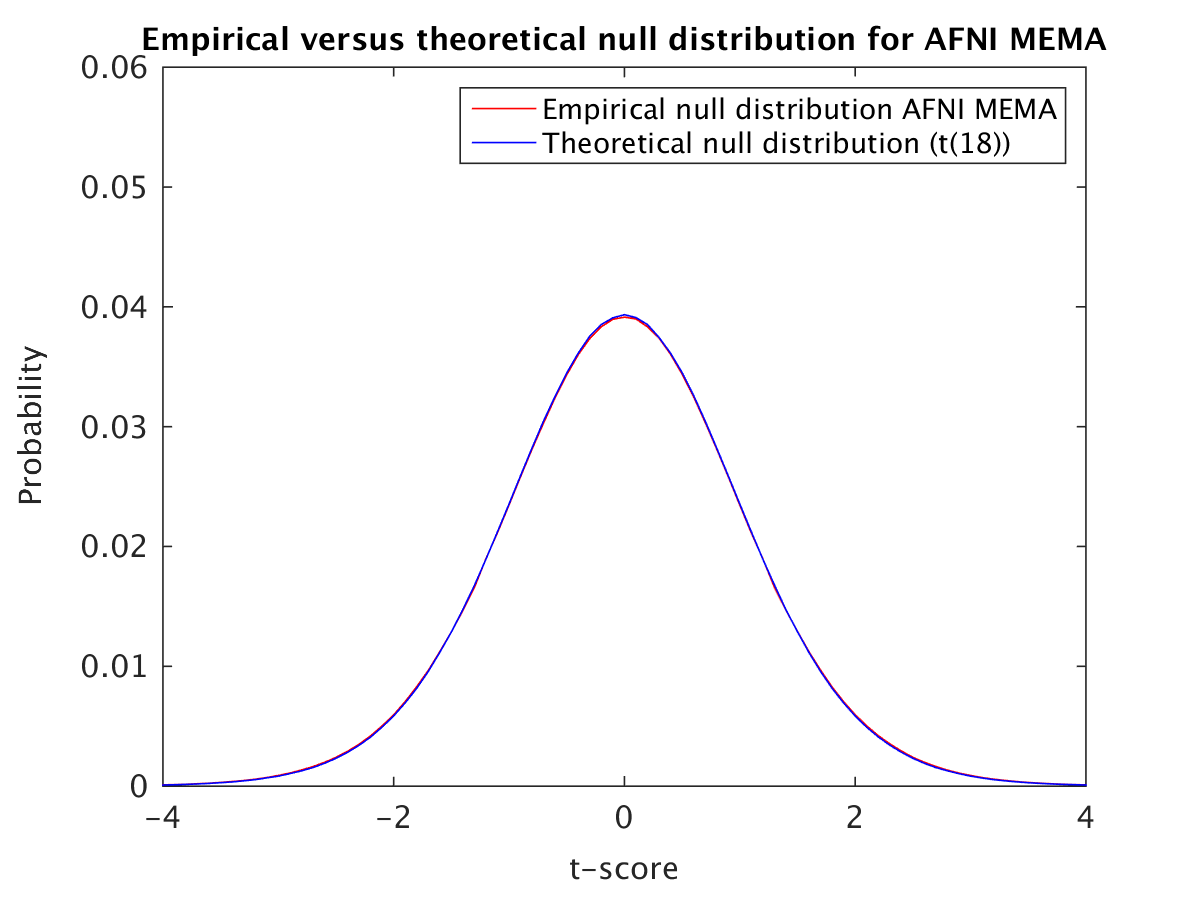}
}
\caption{\emph{Empirical versus theoretical null distributions for a) SPM, b) FSL and c) AFNI. The empirical null distributions were estimated by pooling test values over all brain voxels for 1000 random group comparisons. The test values were drawn from two sample t-tests (10 subjects per group) using the Beijing data (analyzed with the E2 paradigm and 6 mm smoothing). Note that the empirical null distribution for FLAME1 in FSL has a much lower variance compared (0.67) to a normal distribution with unit variance. For this reason, the familywise error rates are much lower for FLAME in FSL, compared to the other functions. }}
\label{fig:empirical_theoretical_null}
\end{figure*}

As a first step to understand the inaccuracies in the parametric methods, the test statistic values (t- or z-scores, as generated by each package) were compared to their theoretical null distributions. Figure~\ref{fig:empirical_theoretical_null} shows the histogram of all brain voxels for 1000 group analyses (see caption for details). The empirical and theoretical nulls are well-matched, except for FSL's FLAME1 which has lower variance ($\hat\sigma^2=0.67$) than the theoretical null. This is the proximal cause of the highly conservative results from FSL FLAME. The mixed effects variance is composed of intra- and inter-subject variance ($\sigma_{WTN}^2, \sigma_{BTW}^2$, respectively), and while other software packages do not separately estimate each (SPM, FSL OLS), FLAME estimates each and constrains $\sigma_{BTW}^2$ to be positive. In this null data, the true effect in each subject is zero and thus the true $\sigma_{BTW}^2 = 0$. Thus unless FLAME's $\hat\sigma_{BWT}^2$ is correctly estimated to be 0, it can only be positively biased, and in fact this point was raised by the original authors~\cite{woolrich2004}. Additionally, FLAME1 (the currently recommended option) uses a conservative degrees of freedom estimate in lieu of a more accurate but slow degrees of freedom estimation used in FLAME2's BIDET. These two factors are the most likely explanation for the observed conservativeness.

\subsection{What does the spatial auto correlation function look like?}

\begin{figure*}
\center
\subfigure[]{
\includegraphics[scale=0.405]{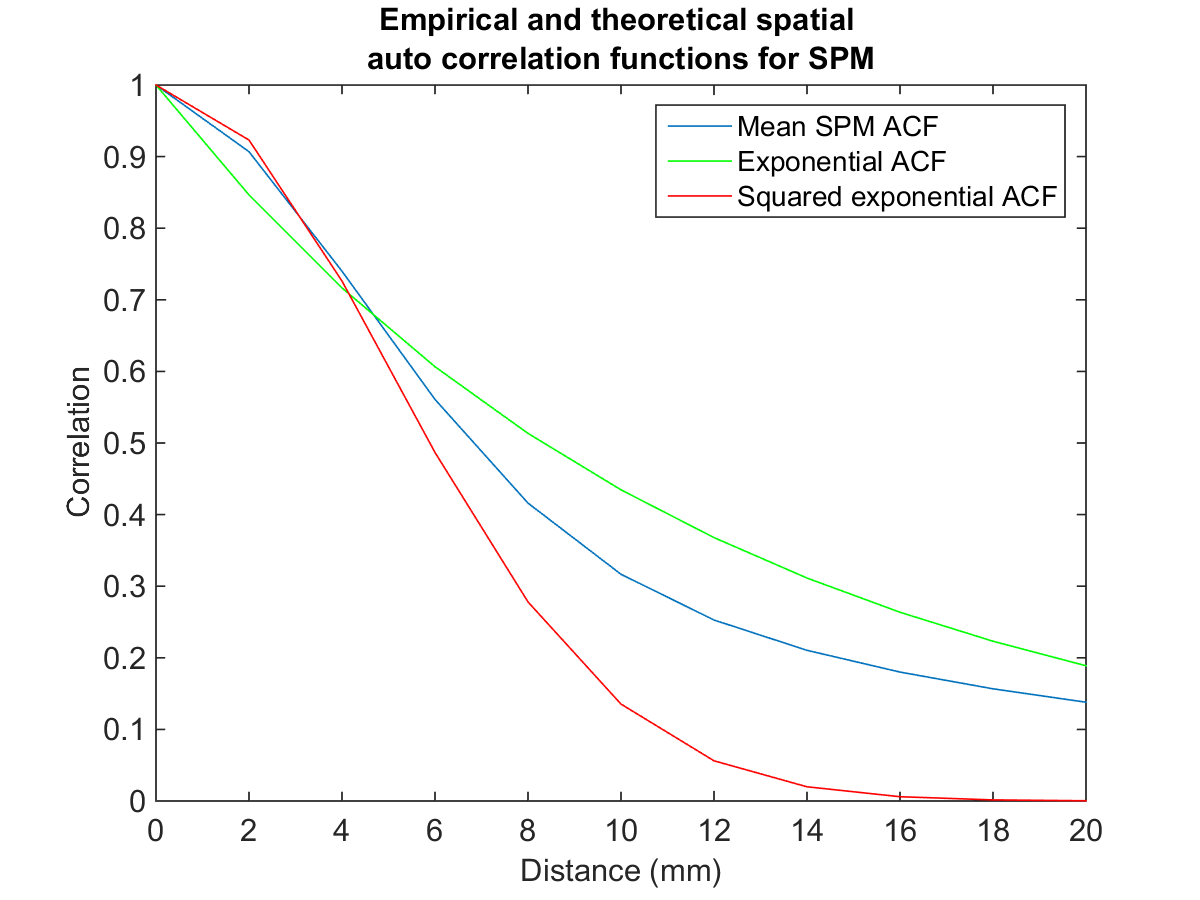}
}
\subfigure[]{
\includegraphics[scale=0.405]{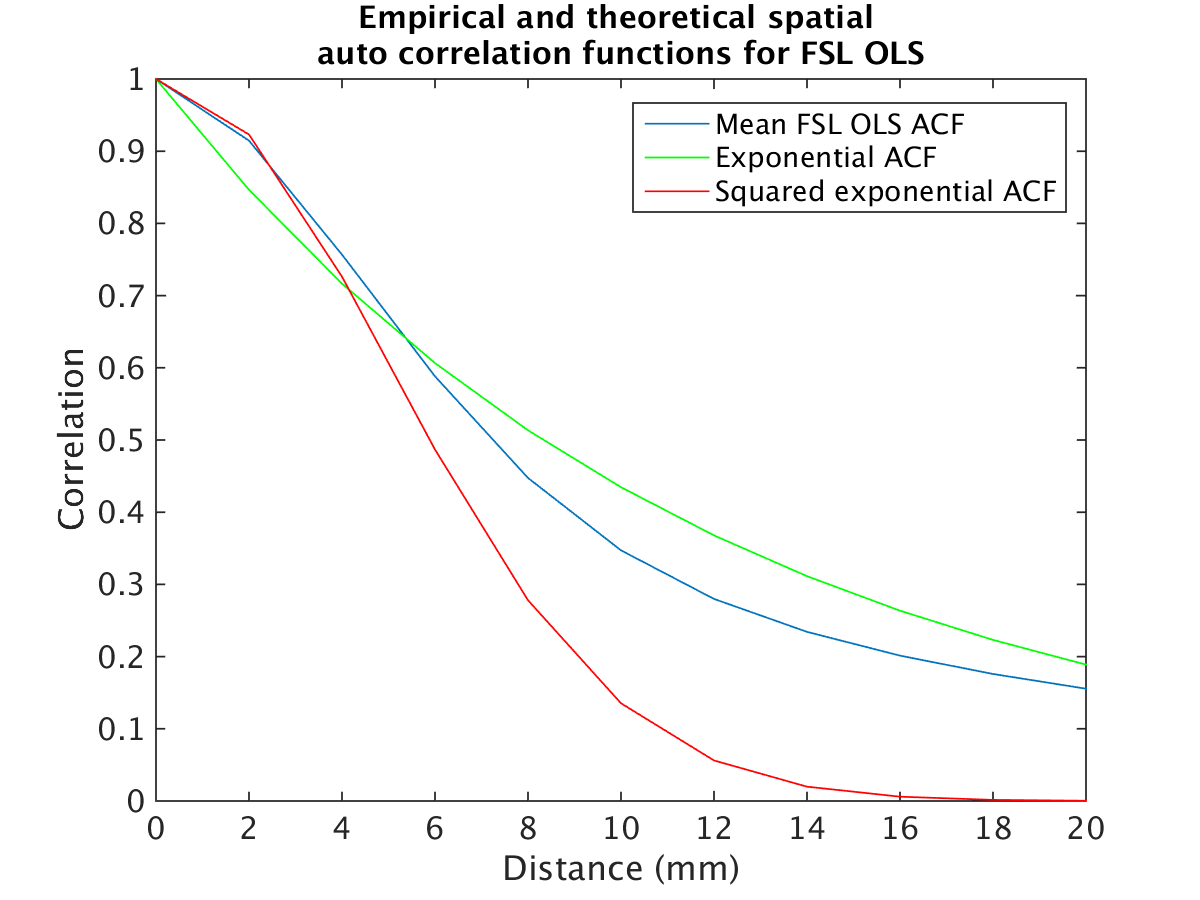}
\includegraphics[scale=0.405]{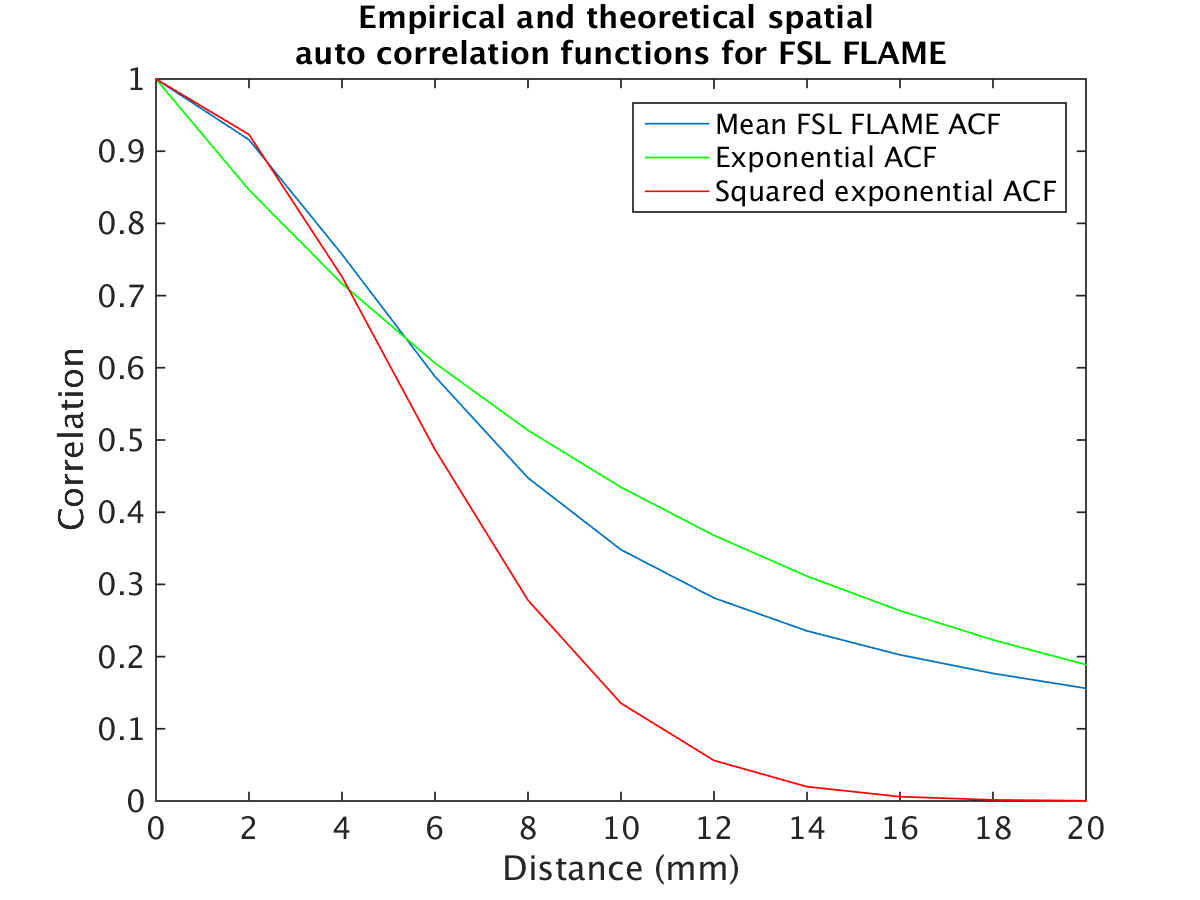}
}
\subfigure[]{
\includegraphics[scale=0.405]{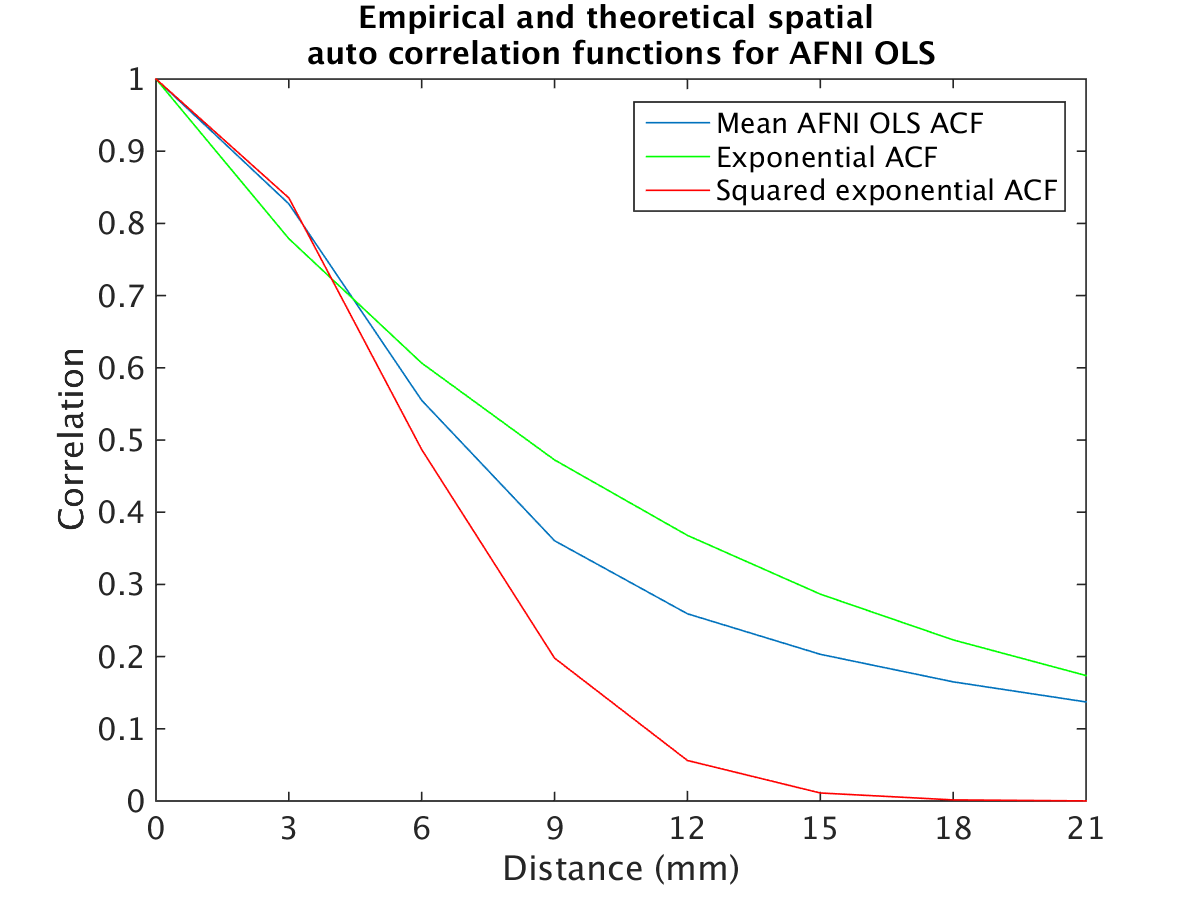}
\includegraphics[scale=0.405]{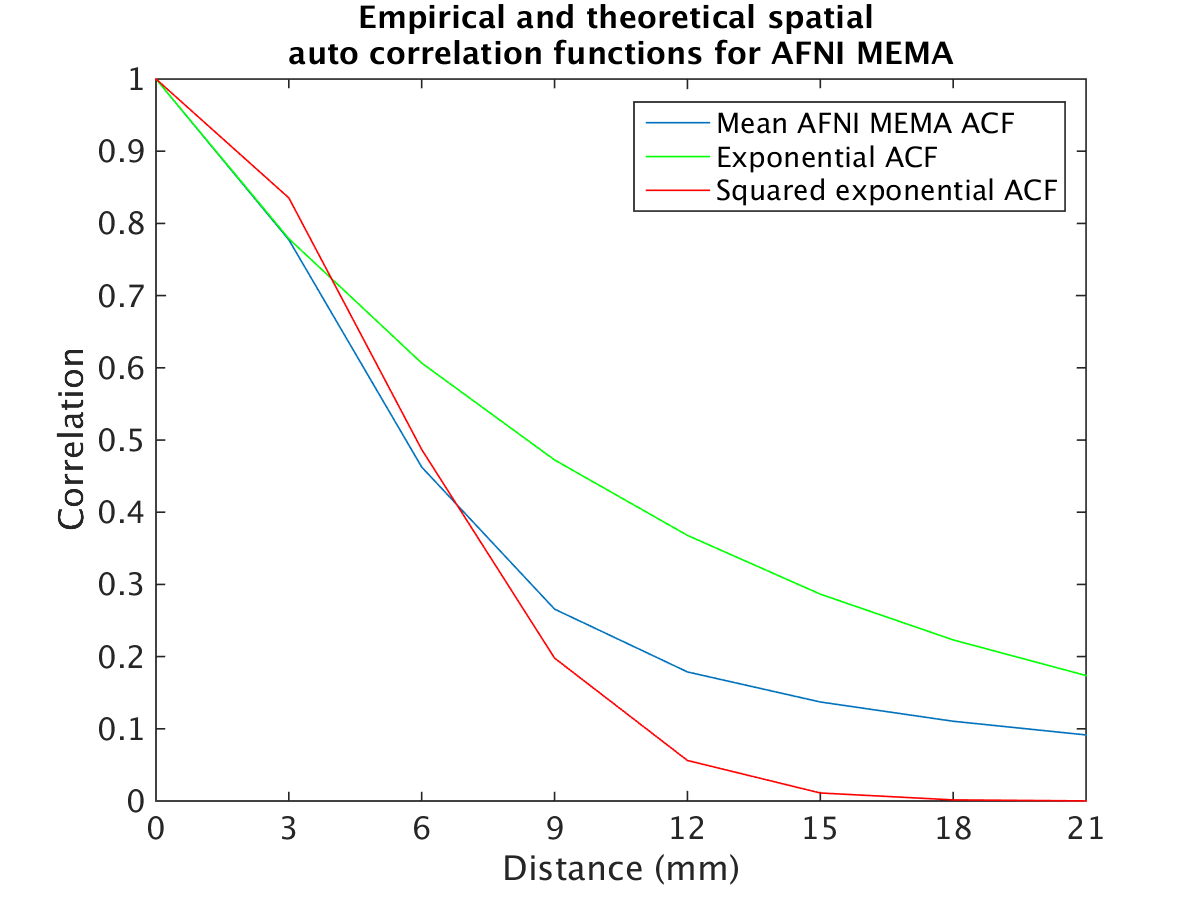}
}
\caption{\emph{Empirical versus theoretical spatial auto correlation functions (SACFs) for a) SPM, b) FSL and c) AFNI. The SACFs were estimated and averaged using 1000 group difference maps.  The difference maps were generated from two sample t-tests (10 subjects per group) using the Beijing data (analyzed with the E2 paradigm and 6 mm smoothing). Note that the empirical SACFs have a much longer tail compared to the theoretical squared exponential SACF, thereby violating one of the required assumptions for parametric cluster-wise inference using Gaussian random field theory. Both SPM and FSL resample the fMRI data to a resolution of 2 mm, while AFNI instead uses a resolution of 3 mm. For this reason, the SACFs are sampled differently for AFNI.}}
\label{fig:empirical_theoretical_sacf}
\end{figure*}

SPM and FSL depend on Gaussian random field theory (RFT) for FWE-corrected voxel-wise and cluster-wise inference. However, RFT cluster-wise inference depends on two additional assumptions. The first assumption is that the spatial smoothness of the fMRI signal is constant over the brain, and the second assumption is that the spatial auto correlation function has a specific shape (a squared exponential)~\cite{hayasaka}. To investigate the second assumption, the spatial auto correlation function was estimated and averaged using 1000 group difference maps. For each group difference map and each distance (1 - 20 mm), the spatial auto correlation was estimated along x, y and z, by calculating the correlation between the original difference map and a shifted version of the difference map. The final estimate is an average of the spatial auto correlation along x, y and z. The empirical spatial auto correlation functions are given in Figure~\ref{fig:empirical_theoretical_sacf}. A reference squared exponential is also included, it is proportional to a Gaussian density with $\sigma=5$ mm, corresponding to an intrinsic smoothness of 8.3 mm (FWHM). The empirical spatial auto correlation function is clearly far from a squared exponential; it has a much longer tail. This may explain why the parametric methods work rather well for a high cluster defining threshold (resulting in small clusters) and not as well for a low cluster defining threshold (resulting in large clusters).

\subsection{How do the smoothness estimates and the cluster extent thresholds differ between SPM, FSL and AFNI?}

\begin{figure}
\center
\includegraphics[scale=0.45]{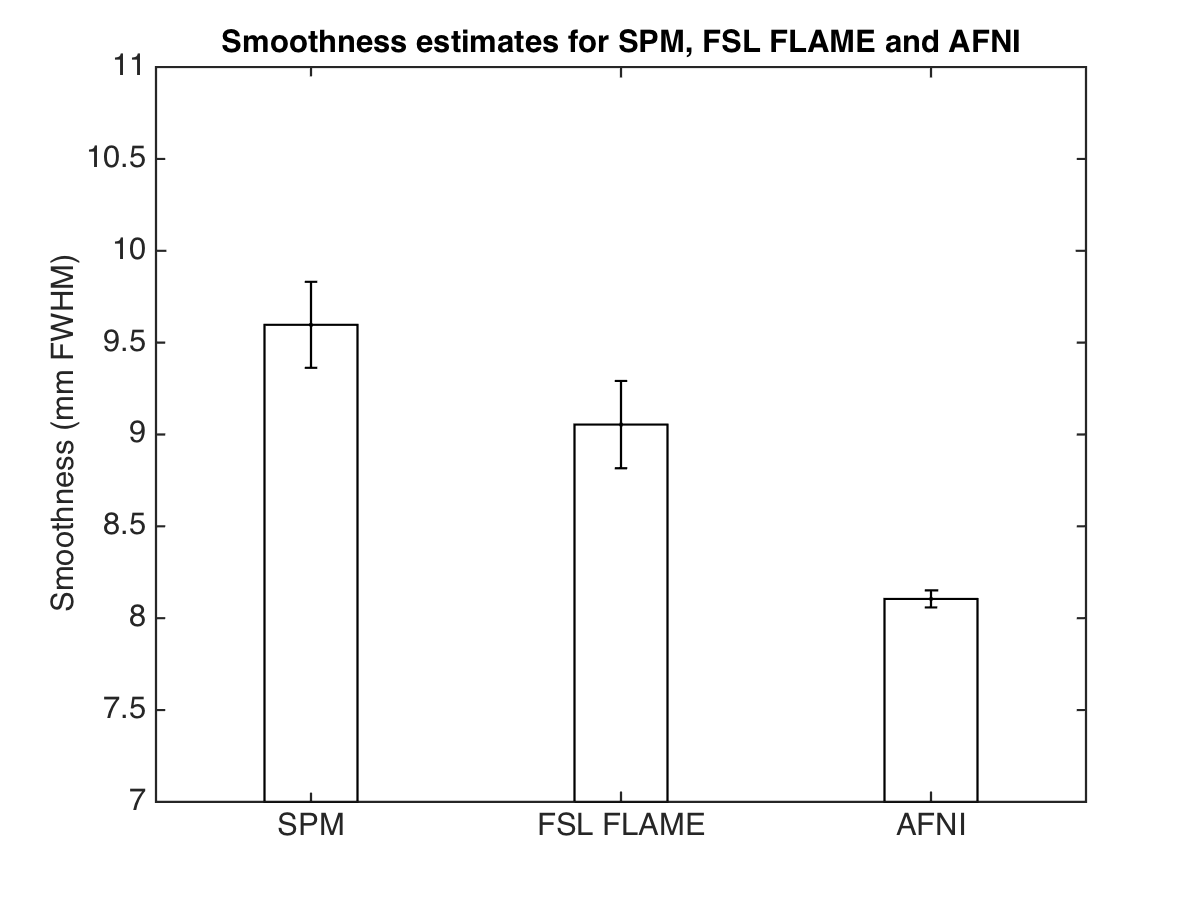}
\caption{\emph{Group smoothness estimates (mm full width at half maximum) for SPM, FSL FLAME and AFNI. The smoothness estimates originate from two sample t-tests (10 subjects per group) using the Beijing data (analyzed with the E2 paradigm and 6 mm smoothing). Note that AFNI estimates the group smoothness differently compared to SPM and FSL. Also note that AFNI uses higher order interpolation for motion correction and spatial normalization, which leads to a lower smoothness compared to more common linear interpolation.}}
\label{fig:smoothness}
\end{figure}

\begin{figure}
\center
\includegraphics[scale=0.45]{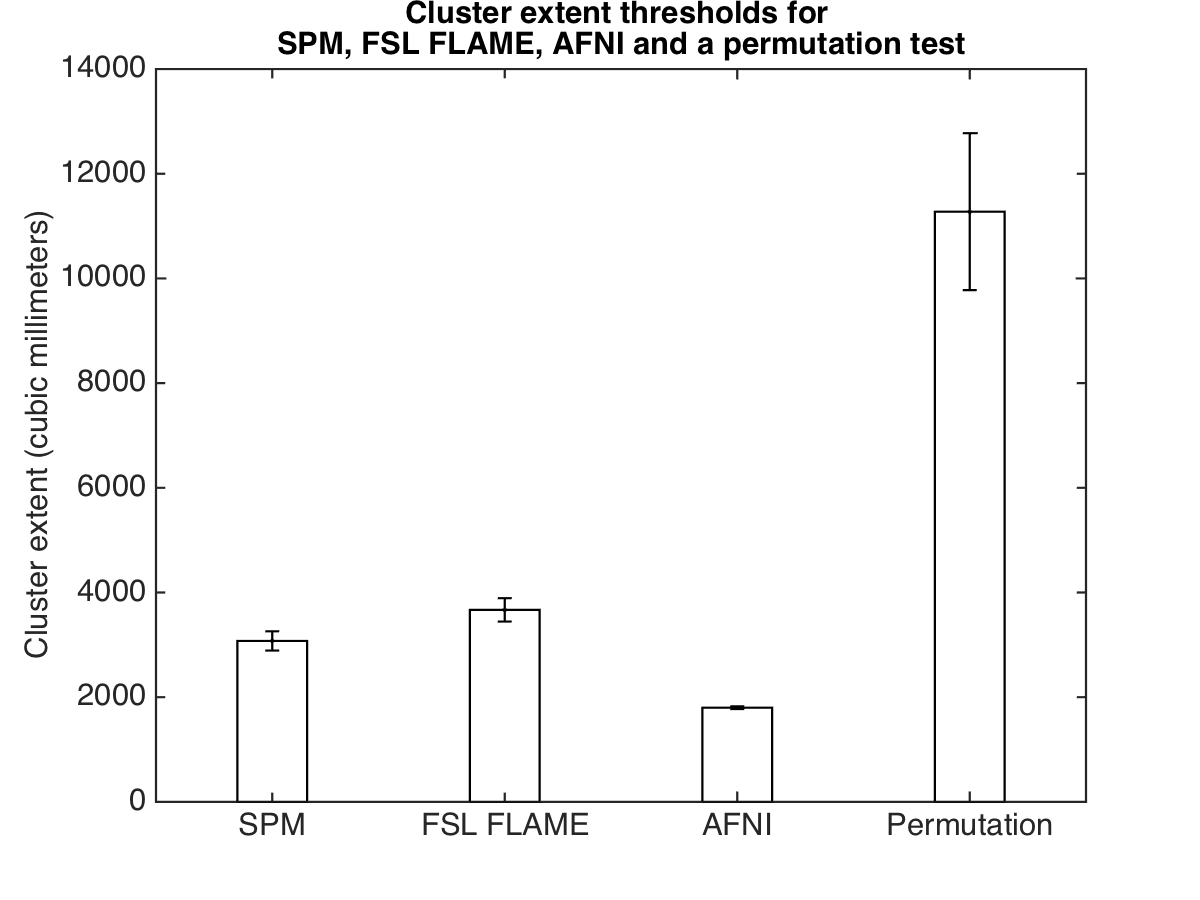}
\caption{\emph{Cluster extent thresholds (in cubic millimeters) for SPM, FSL FLAME, AFNI and a permutation test, for a cluster defining threshold of p = 0.01 and a familywise cluster error rate of p = 0.05. The thresholds originate from two sample t-tests (10 subjects per group) using the Beijing data (analyzed with the E2 paradigm and 6 mm smoothing). Note that the permutation threshold can only be directly compared with the threshold from the FSL software, as first level results from FSL were used for the non-parametric analyses.}}
\label{fig:cluster_thresholds}
\end{figure}

For both voxel and cluster-wise inference, the smoothness of the group difference map or group activation map needs to be estimated to calculate the RESEL (resolution element) count, a key parameter for the RFT based p-values. Group model smoothness estimates for SPM, FSL FLAME and AFNI, for 1000 group comparisons, are given in Figure~\ref{fig:smoothness} (see Figure caption for details). As the preprocessing was unique to each package, it is difficult to make absolute comparisons between intrinsic noise smoothness found with each tool. However, the notably lower smoothness estimates from AFNI could be due to AFNI's use of first level (intrasubject) residuals instead of second level (intersubject) residuals like SPM and FSL. A direct comparison can be made between cluster size thresholds of FSL FLAME and a non-parametric test conducted on the FSL-preprocessed data (see Figure~\ref{fig:cluster_thresholds}), showing the more stringent cluster thresholds the non-parametric method uses to control the familywise error rate. 

\subsection{Where are the false clusters located in the brain?}

\begin{figure*}
\center
\subfigure[]{
\includegraphics[scale=0.39]{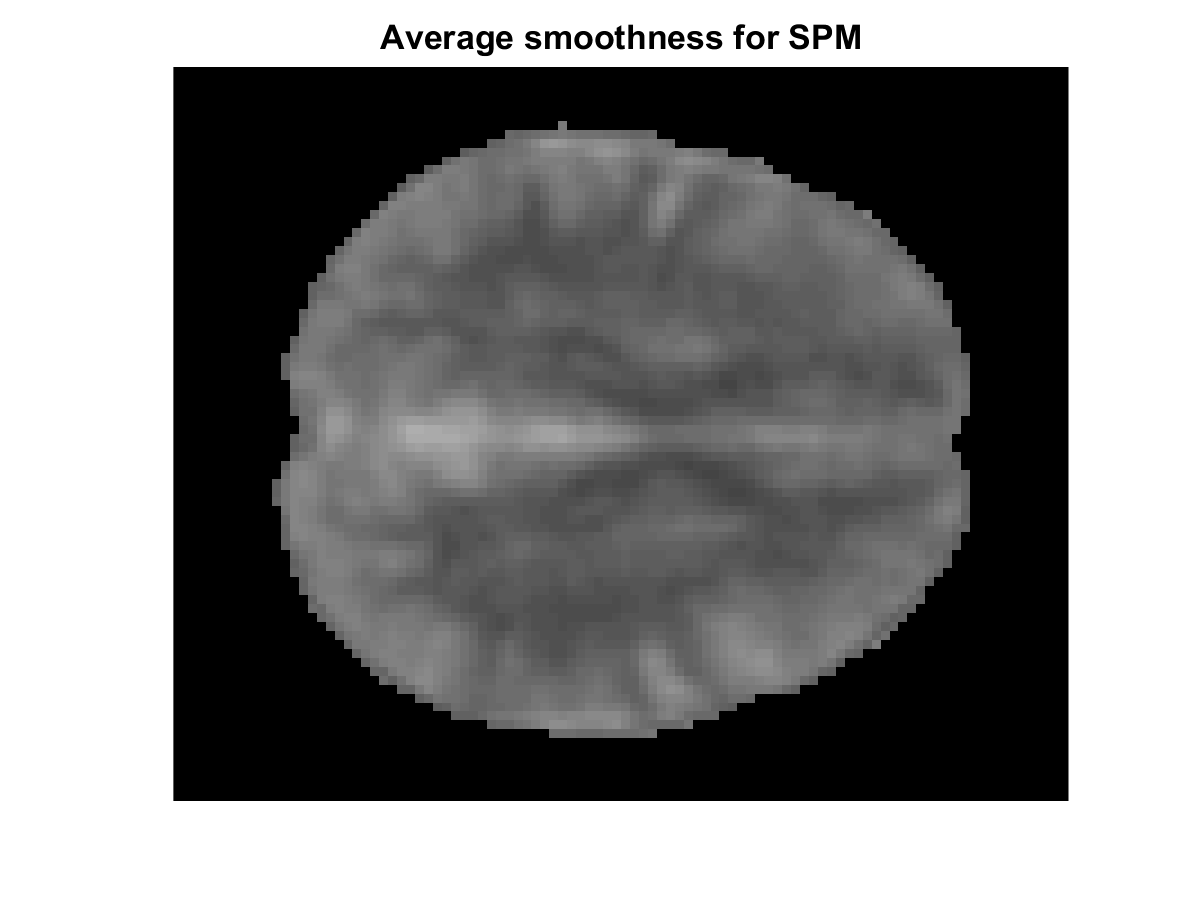}
}
\subfigure[]{
\includegraphics[scale=0.39]{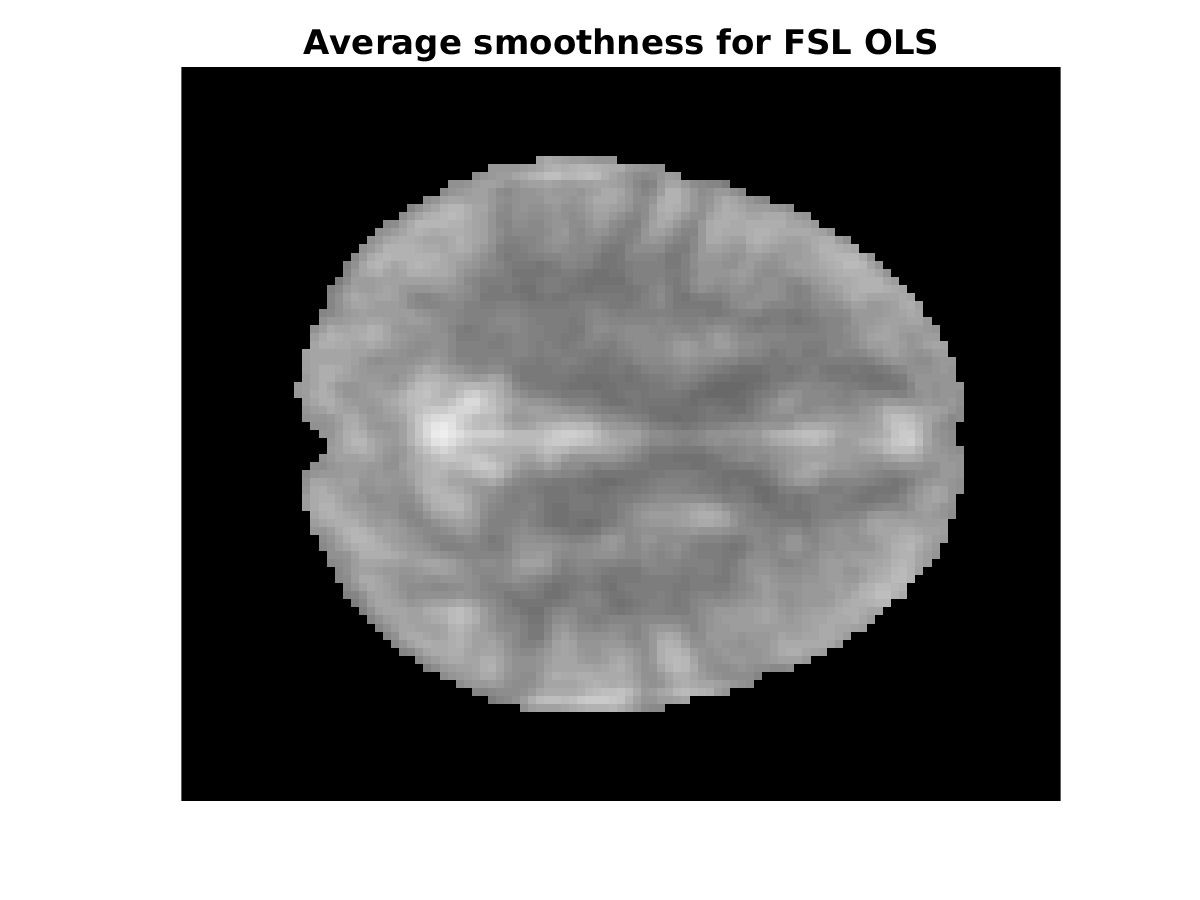}
\includegraphics[scale=0.39]{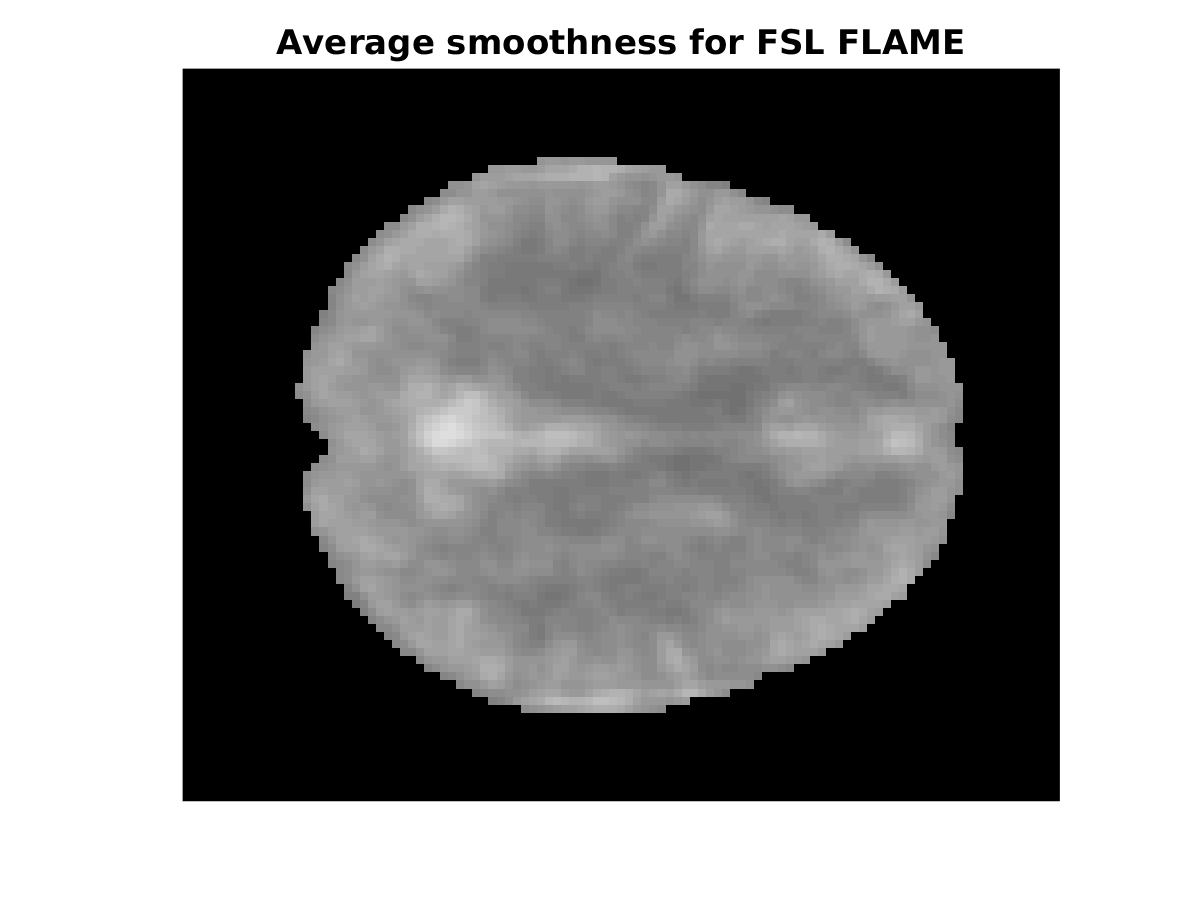}
}
\subfigure[]{
\includegraphics[scale=0.39]{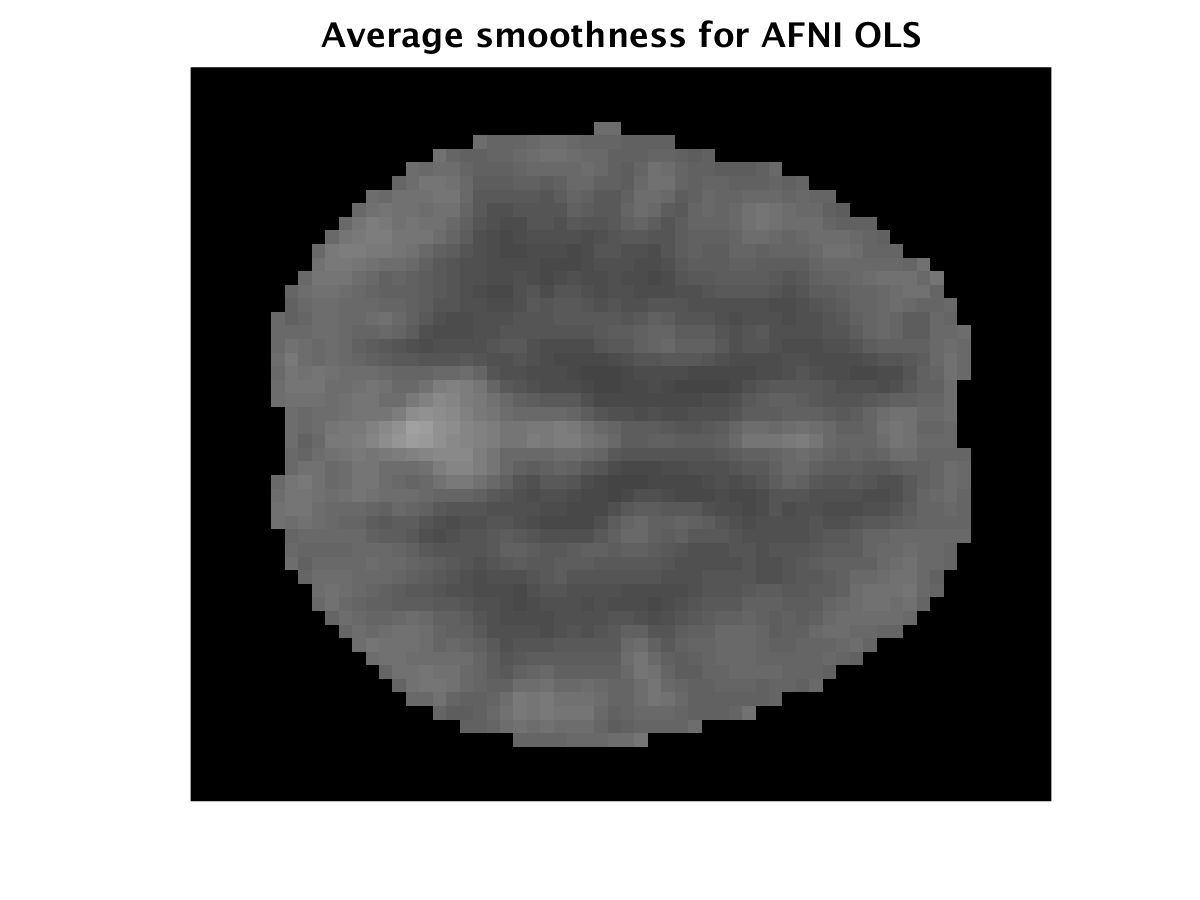}
\includegraphics[scale=0.39]{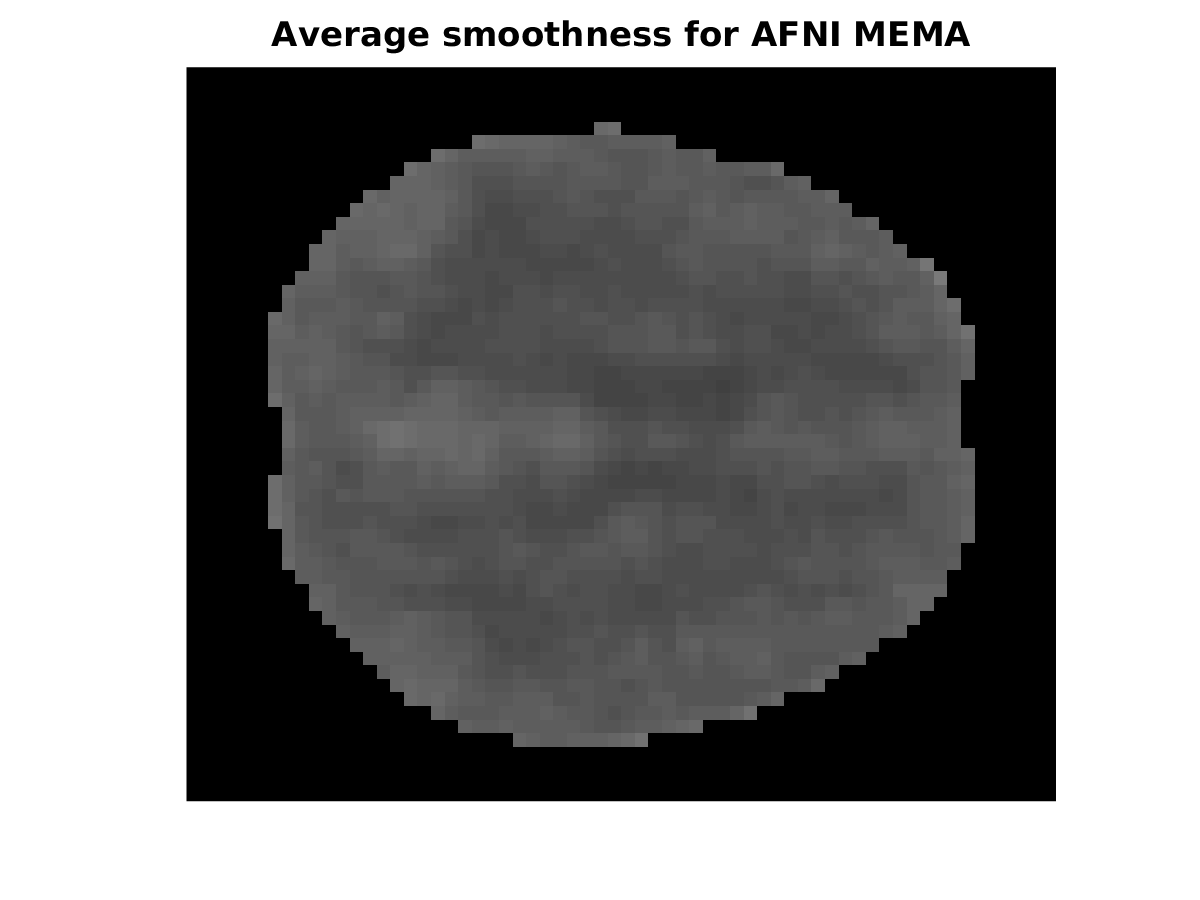}
}
\caption{\emph{Maps of inverse spatial gradient magnitude, reflecting the spatial smoothness of the statistic images, for a) SPM, b) FSL and c) AFNI. The smoothness was estimated from 10,000 difference maps generated from two sample t-tests (10 subjects per group) using the Beijing data (analyzed with the E2 paradigm and 6 mm smoothing). It is clear that the smoothness varies spatially; one of the required assumptions for parametric cluster-wise inference using Gaussian random field theory is thereby violated. Note that the bright areas (high smoothness) match the spatial maps of the false clusters; it is more likely to find a large cluster for areas with a high smoothness. The AFNI software generally results in group difference maps with a lower smoothness compared to SPM and FSL. A possible explanation is that AFNI uses higher order interpolation for motion correction and spatial normalization, which leads to a lower smoothness compared to more common linear interpolation. Also note the reduced smoothness for the iterative methods (FSL FLAME \& AFNI 3dMEMA) and their corresponding non-iterative methods (FSL OLS and AFNI OLS, respectively); the voxel-by-voxel estimation of between subject variance in the iterative methods reduces the smoothness slightly.}}
\label{fig:spatial_smoothness}
\end{figure*}

To investigate if the false clusters appear randomly in the brain, all significant clusters (p $<$ 0.05, corrected) were saved as binary maps and summed together, see Figure~\ref{fig:spatial_smoothness}. These maps of voxel-wise cluster frequency show the areas more and less likely to be marked as significant in a cluster-wise analysis (see Figure caption for details). Posterior cingulate was the most likely area to be covered by a cluster, while white matter was least likely. 

To investigate the assumption of a stationary spatial smoothness, three gradient filters (oriented along x, y and z) were applied to each group difference map, and the average gradient magnitude $\left( \sqrt{(\nabla x)^2 + (\nabla y)^2 + (\nabla z)^2} \right)$ (roughness) was calculated over all analyses (while actual analyses use the residuals, as the null hypothesis is true we can equivalently use the statistic maps). The smoothness was finally obtained as the inverse roughness, and is given in Figure~\ref{fig:spatial_smoothness}. Reductions in smoothness will mean that random clusters are smaller in size (reducing false positive rate) but also that there are more clusters (increasing false positive rate); how these two factors balance out are hard to predict. 

Clearly, the false clusters appear in spatial patterns that match the degree of smoothness in the group difference maps. For SPM, FSL OLS and AFNI OLS, the degree of smoothness is correlated with the brain tissue type; the smoothness is generally higher for gray brain tissue compared to white brain tissue. This effect has previously been observed for VBM data~\cite{silver}.

\begin{figure*}
\center
\subfigure[]{
\includegraphics[scale=0.4]{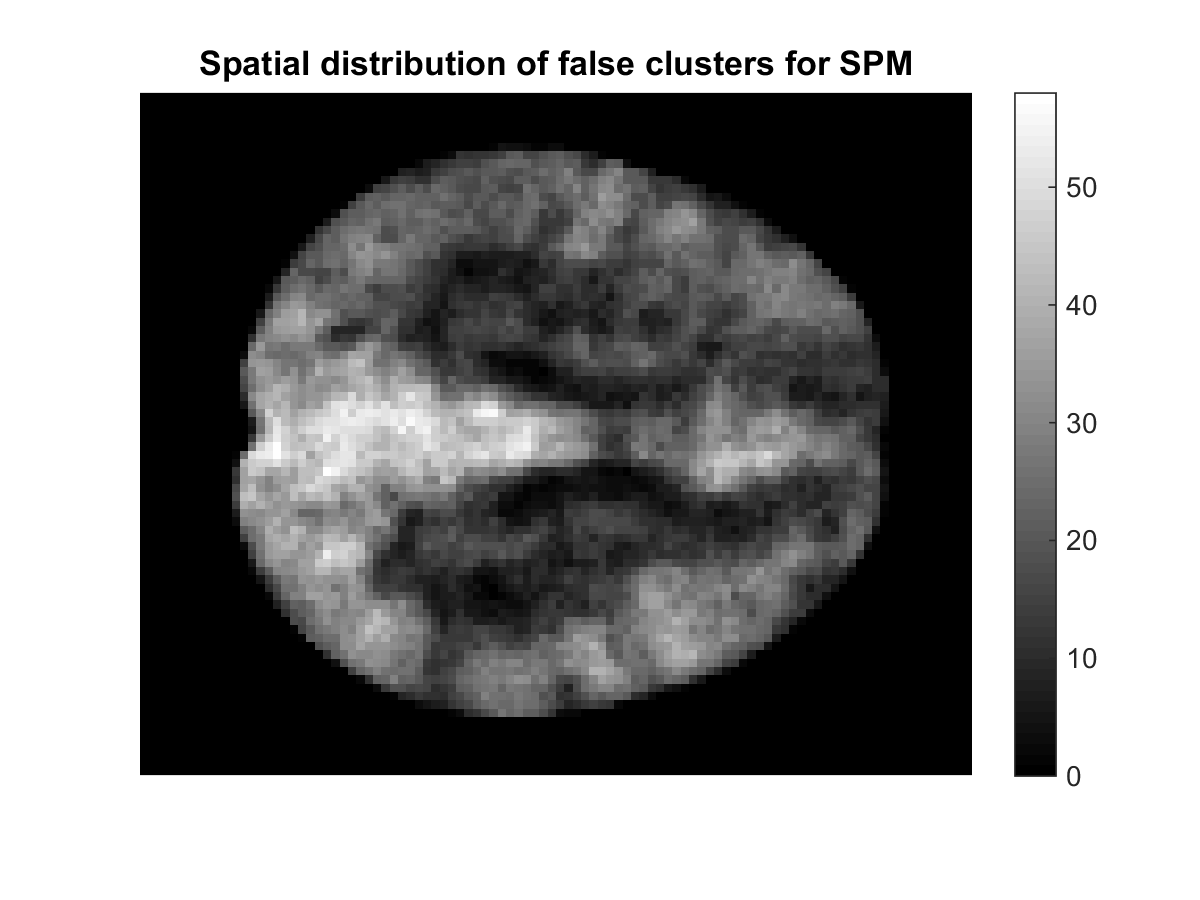}
}
\subfigure[]{
\includegraphics[scale=0.4]{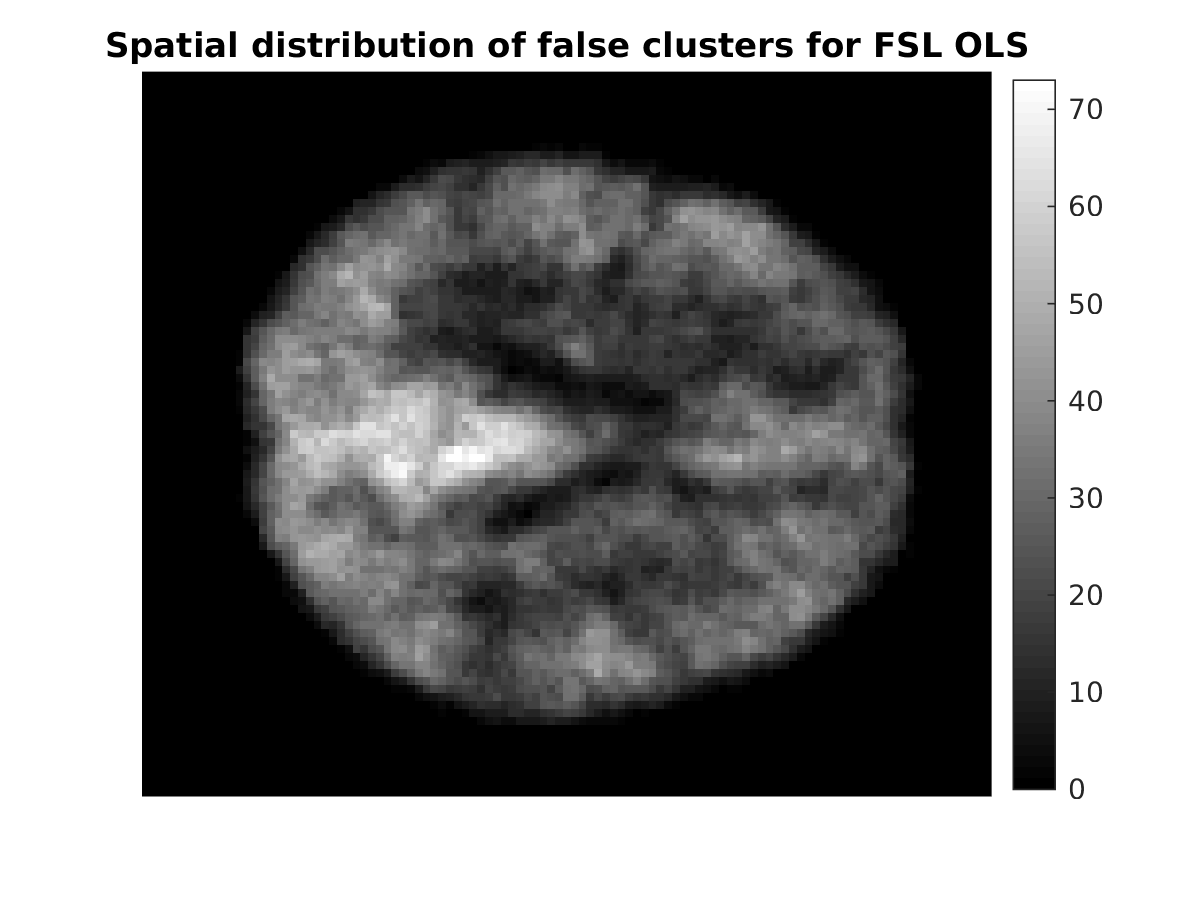}
\includegraphics[scale=0.4]{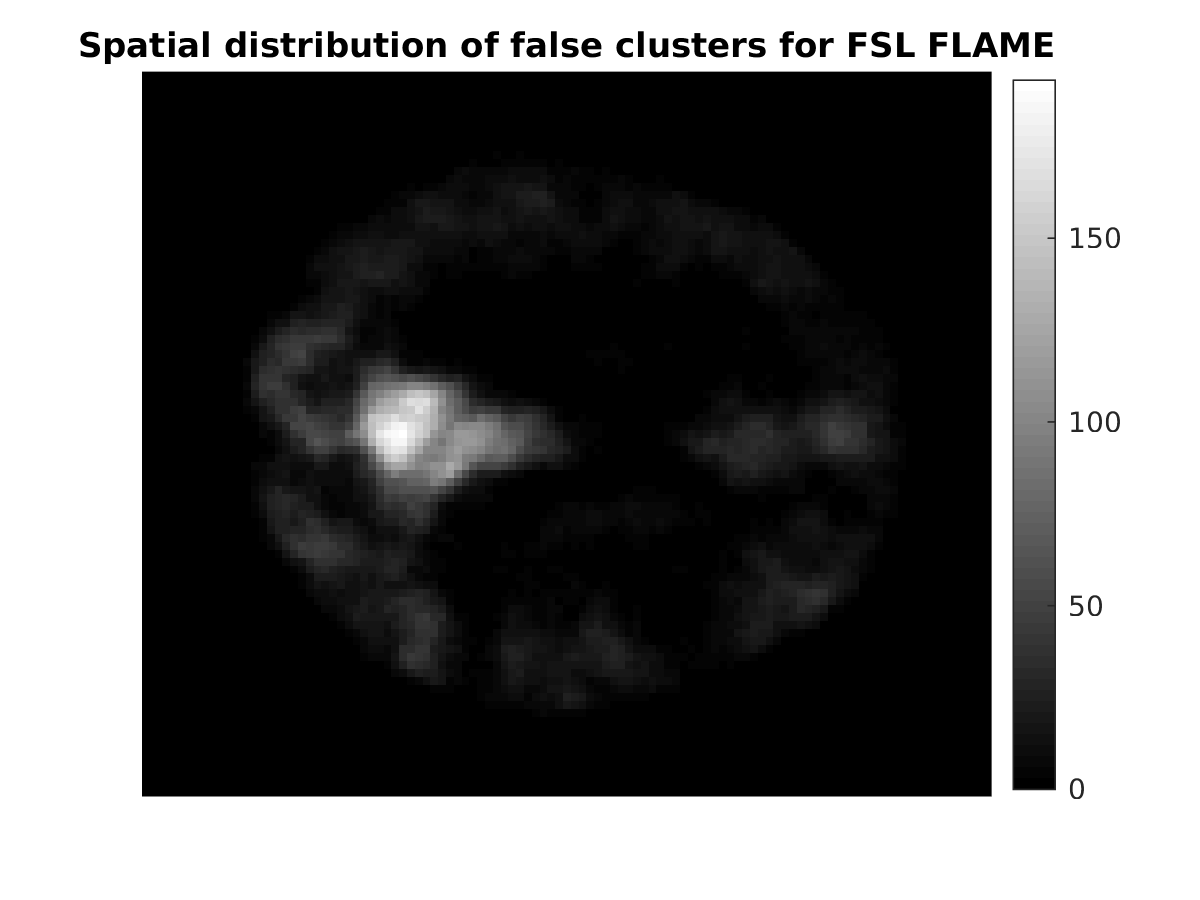}
}
\subfigure[]{
\includegraphics[scale=0.4]{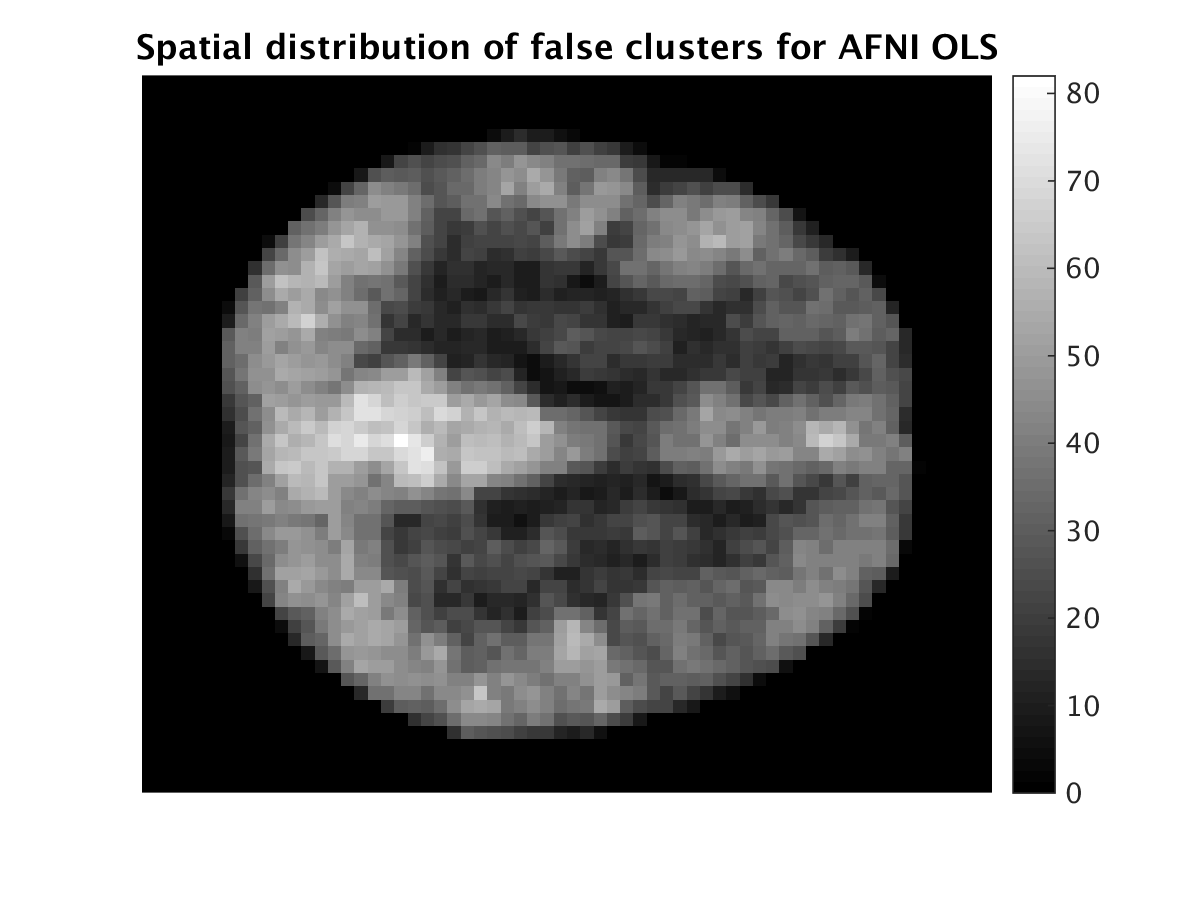}
\includegraphics[scale=0.4]{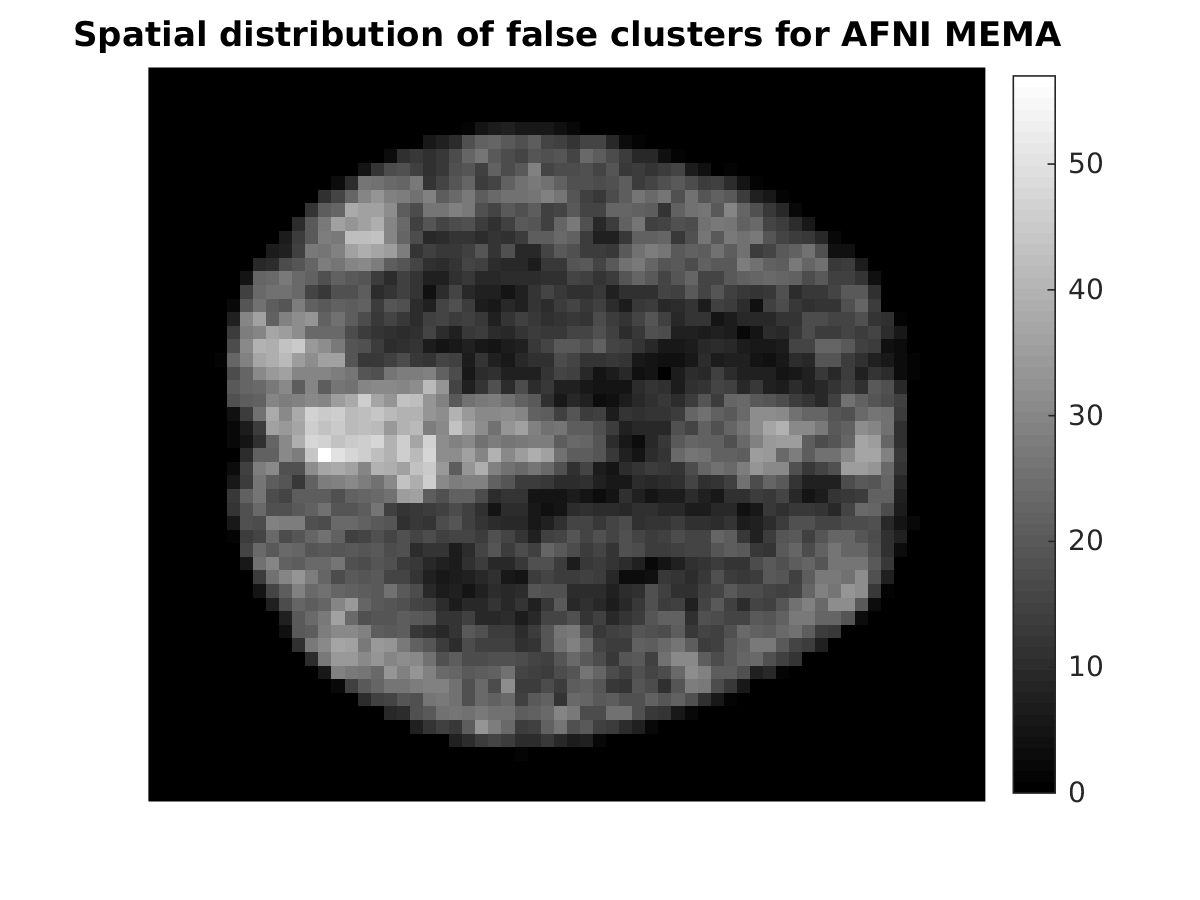}
}
\caption{\emph{The maps show voxel-wise incidence of clusters. Image intensity is the number of times, out of 10,000 random analyses (200,000 for FSL FLAME, to account for fewer clusters per analysis), a cluster occured at a given voxel (CDT p = 0.01), for a) SPM, b) FSL and c) AFNI. Each analysis is a two sample t-test (10 subjects per group) using the Beijing data, analyzed with the E2 paradigm and 6 mm smoothing. The bright spot in the posterior cingulate corresponds to a region of high smoothness, and suggests non-stationarity as a possible contributing factor.}}
\label{fig:false_clusters}
\end{figure*}

\subsection{What is the difference between parametric and non-parametric cluster-wise inference for a typical group study?}

\begin{figure}
\center
\includegraphics[scale=0.45]{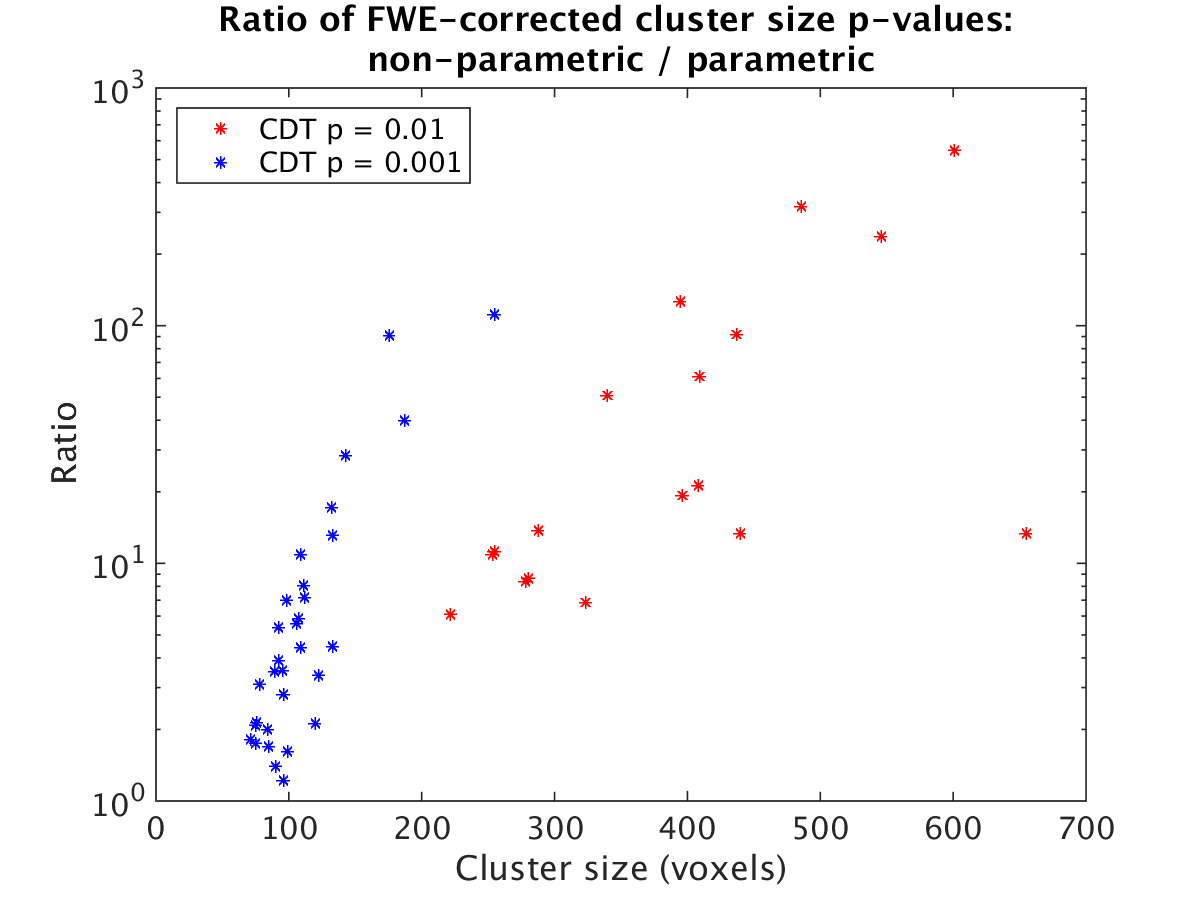}
\caption{\emph{Ratio of non-parametric to parametric FWE corrected p-values for cluster size inference on 4 task (non-null) fMRI datasets, for parametric FWE p-values $0.05\geq p \geq 10^{-4}$. Results for two CDT are shown, p = 0.01 and p = 0.001, and larger ratios indicate parametric p-values being smaller (more significant) than non-parametric p-values (note the logarithmic scale on the y-axis). Clusters with a parametric FWE p-value more significant than $10^{-4}$ are excluded because a permutation test with 5000 permutations can only resolve p-values down to 0.0002, and such p-values would generate large ratios inherently. These results suggest cluster size inference with a CDT of p = 0.01 has FWE inflated by 2 to almost 3 orders of magnitude, and a CDT of p = 0.001 has FWE significance inflated by up to 2 orders of magnitude.}}
\label{fig:ratios}
\end{figure}

All of the analyses to this point have been based on resting state fMRI data, where the null hypothesis should be true. We now use task data to address the practical question of ``How will my FWE-corrected cluster p-values change?'' if a user were to switch from a parametric to a non-parametric method. We use four task data sets (rhyme judgment, mixed gambles~\cite{mixedgambles}, living-nonliving decision with plain or mirror-reversed text, word and object processing~\cite{wordobject}) downloaded from the OpenfMRI~\cite{openfmri} homepage. The data sets were analyzed using a parametric (the OLS option in FSL's FEAT) and a non-parametric method (the randomise function in FSL). The only difference between these two methods is that FSL FEAT-OLS relies on Gaussian random field theory to calculate the corrected cluster p-values, while randomise instead uses the data itself. The resulting cluster p-values are given in Table~\ref{table:typicalstudies1} (cluster defining threshold of p = 0.01) and Tables~\ref{table:typicalstudies2} -~\ref{table:typicalstudies3} (cluster defining threshold of p = 0.001). Figure~\ref{fig:ratios} summarizes these results, plotting the ratio of FWE-corrected p-values, non-parametric to parametric, against cluster size. Given the previous null data evaluations showing valid non-parametric and invalid parametric cluster size inference, we take ratios larger than 1.0 as evidence of inflated (biased) significance in the parametric inferences.

For a cluster defining threshold of p = 0.01 and a cluster size of 400 voxels, the non-parametric cluster p-value is approximately 10 - 100 times larger compared to the parametric p-value. For a cluster defining threshold of p = 0.001 and a cluster size of 100 voxels, the non-parametric cluster p-value is approximately 1.25 - 10 times larger compared to the parametric p-value. For contrast 1 of the word and object processing task data set (Table~\ref{table:typicalstudies1}), one cluster has a parametric p-value of 0.0182 and a non-parametric p-value of 0.249. This matches the empirically estimated familywise error rate of FSL OLS, according to Figure~\ref{fig:fwe_cluster_onesample_40_subjects}. These findings indicate that the problems exist also for task based fMRI data, and not only for resting state data. 

\cleardoublepage
\newpage

\section{Discussion}
\label{sec:discussion}

Our results clearly show that the parametric statistical methods used for group fMRI analysis with the packages SPM, FSL and AFNI can produce FWE-corrected cluster p-values that are erroneous, being spuriously low and inflating statistical significance. This calls into question the validity of countless published fMRI studies based on parametric cluster-wise inference. It is important to stress that we have focused on inferences corrected for multiple comparisons in each group analysis, yet some 40\% of a sample of 241 recent fMRI papers did not report correcting for multiple comparisons~\cite{carp2}, meaning that many group results in the fMRI literature suffer even worse false positive rates than found here~\cite{ioannidis}. A possible explanation for the lack of multiple comparison correction is that the correction methods are believed to be too conservative, resulting in familywise error rates far below the expected 5\%. However, we have found that correction methods based on parametric assumptions can actually be very liberal for cluster-wise inference, yielding familywise error rates of up to 60\%.  

Compared to our previous work~\cite{eklund_ni}, the results presented here are more important for three reasons. First, the current study considers group analyses, while our previous study looked at single subject analyses. Group analyses are much more common in the fMRI field, and are essential for drawing conclusions that should generalize to a population. Second, we here investigate the validity of the three most common fMRI software packages~\cite{carp2}, while we only considered SPM in our previous study. Third, the non-parametric permutation test gives valid results for all two sample t-tests, and for 60\% of the one sample t-tests. This may be due to violations of the stronger assumptions of the one-sided permutation test which assumes symmetrically distributed errors, in contrast to a two-sample permutation test that only assumes exchangeable errors~\cite{nichols2002}. In our previous study, the permutation test only performed well in some cases (mainly because single subject fMRI data contain temporal auto correlation which needs to be removed prior to permuting the volumes). For group analyses, the brain activity of each subject can be seen as independent, and thus justifies the permutation's exchangeability assumption.

\subsection{Should resting state data be used to test statistical assumptions?}

One possible criticism is that resting state fMRI data does not truly compromise null data, as it may be affected by consistent trends or transients, for example, at the start of the session. If this was the case, the excess false positives would appear only in certain paradigms and, in particular, least likely in the randomized event-related (E2) design. Rather, the inflated false positives were observed across all experiment types with parametric cluster size inference, implying this effect cannot be responsible. 

\subsection{Why is cluster-wise inference more problematic than voxel-wise?}

It is clear that the parametric statistical methods work well, if conservatively, for voxel-wise inference, but not for cluster-wise inference. We note that other authors have found random field theory cluster-wise inference to be invalid in certain settings under stationarity~\cite{hayasaka,woo} and non-stationarity~\cite{hayasaka2,silver}.  This present work, however, is the most comprehensive to explore the typical parameters used in task fMRI for a variety of software tools. Our results are also corroborated by similar experiments for structural brain analysis (voxel based morphometry, VBM)~\cite{scarpazza,vbm,silver,scarpazza2,meyer}, showing that cluster based p-values are more sensitive to the statistical assumptions. For voxel-wise inference, our results are consistent with a previous comparison between parametric and non-parametric methods for fMRI, showing that a non-parametric permutation test results in lower significance thresholds~\cite{silver,nichols_comparing}. 

Both SPM and FSL rely on RFT to correct for multiple comparisons. For voxel-wise inference, RFT is based on the assumption that the activity map is sufficiently smooth (roughly, at least 3 voxel FWHM~\cite{hayasaka}), and that the spatial auto correlation function (SACF) is twice-differentiable at the origin. Further, RFT is only accurate for sufficient large statistic values (precisely, a statistic value used as a threshold would produce at most one peak under the null hypothesis on average). For cluster-wise inference, RFT additionally assumes a Gaussian shape of the SACF (i.e.\ a squared exponential covariance function), and that the spatial smoothness is constant over the brain. The cluster defining threshold (CDT) must also be sufficiently large, but to accomodate a different sets of approximations. First, it must be large enough to ensure that handles or voids in the clusters do not occur on average (which is much lower than needed for voxel-wise RFT accuracy), ensuring that the expected Euler characteristic is a good approximation to the (null) expected number of clusters. Second, the CDT must be large enough for the (null) distribution of cluster size to be accurate. Note that both SPM and FSL apply cluster size results for Gaussian images to T images after Gaussianizing the CDT, even though there are T image results available~\cite{cao}. Hayasaka and Nichols~\cite{hayasaka} found that the proper T results did not dramatically improve the performance of the RFT cluster size inferences. 

The 3dClustSim function in AFNI also assumes a constant spatial smoothness and a Gaussian form of the SACF (since a Gaussian smoothing is applied to each generated noise volume). The Monte Carlo approach should be accurate for any CDT as the other assumptions hold. As the familywise error rates are far above the expected 5\% for cluster-wise inference, but not for voxel-wise inference, one or more of the Gaussian SACF, the stationary SACF, or the sufficiently large CDT assumptions must be invalid. Cluster-wise inference was discouraged for VBM already 15 years ago~\cite{vbm}, due to non-stationary spatial smoothness in the statistical maps. The non-stationary smoothness can be modeled~\cite{hayasaka2}, but parametric methods still give invalid results for VBM with low smoothing and CDT~\cite{silver}. In the fMRI field, however, the assumption of a stationary spatial smoothness has not really been investigated. 

Figure~\ref{fig:empirical_theoretical_sacf} shows that the SACF is far from a squared exponential. The empirical SACFs are close to a squared exponential for small distances, but the auto correlation is higher than expected for large distances. This could be the reason why the parametric methods work rather well for a high cluster defining threshold (p = 0.001), and not at all for a low threshold (p = 0.01). A low threshold gives large clusters with a large radius, for which the tail of the SACF is quite important. For a high threshold, resulting in rather small clusters with a small radius, the tail is not as important. Also, it could simply be that the high-threshold assumption is not satisfied for a CDT of p = 0.01. Figure~\ref{fig:spatial_smoothness} shows that the spatial smoothness is not constant in the brain, but varies spatially. Note that the bright areas match the spatial distribution of false clusters in Figure~\ref{fig:false_clusters}; it is more likely to find a large cluster for a high smoothness. The permutation test does not assume a specific shape of the SACF, nor does it assume a constant spatial smoothness, nor require a high CDT. For these reasons, the permutation test provides valid results, for two sample t-tests, for both voxel and cluster-wise inference.

\subsection{Why does AFNI's Monte Carlo approach, with fewer parametric assumptions, not perform better?}

As can be observed in Figures~\ref{fig:fwe_cluster_twosample_20_subjects},~\ref{fig:fwe_cluster_onesample_20_subjects},~\ref{fig:fwe_cluster_twosample_40_subjects} and ~\ref{fig:fwe_cluster_onesample_40_subjects}, AFNI results in familywise error rates that are high even for a cluster defining threshold of p = 0.001. There are two main factors that explain these results. 

Firstly, AFNI estimates the spatial group smoothness differently compared to SPM and FSL. AFNI averages smoothness estimates from the first level analysis, whereas SPM and FSL estimate the group smoothness using the group residuals from the general linear model~\cite{kiebel}. The group smoothness used by 3dClustSim may for this reason be too low (compared to SPM and FSL, see Figure~\ref{fig:smoothness}); the variation of smoothness over subjects is not considered.  

Secondly, a 15 year old bug was found in 3dClustSim while testing the three software packages (the bug was fixed by the AFNI group as of May 2015\footnote{http://afni.nimh.nih.gov/pub/dist/doc/program\_help/3dClustSim.html}, during preparation of this manuscript). The effect of the bug was an underestimation of how likely it is to find a cluster of a certain size (in other words, the p-values reported by 3dClustSim were too low). The main idea behind the 3dClustSim function is to generate Gaussian noise with unit variance, and then smooth it using a Gaussian lowpass filter with a size corresponding to the estimated group smoothness. This procedure is repeated a large number of times, to obtain an estimate of how common different cluster sizes are for Gaussian noise. The smoothed noise is rescaled back to unit variance, and 3dClustSim performs the rescaling by first estimating the variance of the smoothed noise. Due to edge effects caused by the smoothing operation the boundary of the volume is attenuated, which has two effects. First, the estimated variance used for standardization will be biased down, increasing the variance of the simulated images\footnote{The variance of independent unit variance noise after convolution is equal to the sum of squares of the smoothing kernel; this result is not used by 3dClustSim, which instead uses the empirical variance over the image to standardize the images.}. Second, the attenuation will reduce the chance that clusters will ever occur near the boundary, effectively reducing the search volume and under estimating the severity of the multiple testing problem. 

Together, the lower group smoothness and the bug in 3dClustSim resulted in cluster extent thresholds that are much lower compared to SPM and FSL, see Figure~\ref{fig:cluster_thresholds}, which resulted in particularly high familywise error rates. Note that the cluster extent thresholds for SPM, FSL and AFNI match the degree of false positives according to Figure~\ref{fig:fwe_cluster_twosample_20_subjects} (d). AFNI has the lowest cluster extent thresholds, and therefore results in a higher familywise error rate compared to SPM and FSL. FSL has higher extent thresholds compared to SPM, and the familywise error rates are therefore slightly lower.

The familywise error rates for AFNI will be lower with the fixed 3dClustSim function, especially for high levels of smoothing (for which the bug in 3dClustSim is more noticeable). As an example of the difference between the old and the new 3dClustSim function, the new function gives a cluster extent threshold that is 15\% higher compared to the old function (for a smoothness of 8 mm). These findings are rather alarming, as 3dClustSim is one of the most popular choices for multiple comparison correction~\cite{carp2}.

\subsection{Which parameters affect the familywise error rate for cluster-wise inference?}

According to Figures~\ref{fig:fwe_cluster_twosample_20_subjects},~\ref{fig:fwe_cluster_onesample_20_subjects},~\ref{fig:fwe_cluster_twosample_40_subjects} and ~\ref{fig:fwe_cluster_onesample_40_subjects}, the cluster defining threshold is the most important parameter for SPM, FSL and AFNI; using a more liberal threshold increases the degree of false positives. This result is consistent with previous work~\cite{silver,woo,eklund_isbi}. However, the permutation test is completely unaffected by changes of this parameter. According to a recent review looking at 484 fMRI studies~\cite{woo}, the used cluster defining threshold varies greatly between the three software packages (mainly due to different default settings). For SPM, p = 0.001 is the default and most common threshold (used in about 70\% of the studies), followed by 20\% for p = 0.005 and 5\% for p = 0.01. For FSL, p = 0.01 is the default and most common choice (65\% of the studies), followed by 20\% for p = 0.001 and 10\% greater than p = 0.01. The AFNI software does not have a default setting for the 3dClustSim function, but a threshold of p = 0.005 seems to be the most common option (used in 40\% of the studies), followed by 25\% for p = 0.001 and 15\% for p = 0.01.

The amount of smoothing has a rather large impact on the degree of false positives, especially for FSL OLS. The results from the permutation test, on the other hand, do not depend on this parameter. The original fMRI data has an intrinsic SACF, which is mixed with the SACF of the smoothing kernel. The combined SACF will more closely resemble a squared exponential for high levels of smoothing, simply because the smoothing operation forces the data to have a more Gaussian SACF. The permutation test does not assume a specific form of the SACF, and therefore performs well for any degree of smoothing. It should be stressed that AFNI uses 4 mm as the default amount of smoothing, while FSL uses 5 mm and SPM uses 8 mm. Both FSL and AFNI OLS show interaction between the amount of smoothing and the fMRI data. For all software packages, the amount of smoothing has a larger effect on the block based activity paradigms, compared to the event related ones. These two effects are consistent with our previous work~\cite{eklund_ni}. SPM, FSL OLS and AFNI also show interaction between the amount of smoothing and the cluster defining threshold. 

Just as for our previous study~\cite{eklund_ni}, the used (pretended) activity paradigm significantly affects the degree of false positives. This was unexpected, but it means that problems that arise due to temporally correlated noise actually propagate from the single subject analysis to the group analysis. SPM, FSL and AFNI 3dMEMA show interaction between the fMRI data and the activity paradigm. This can be explained by the fact that the Beijing data sets were collected with a repetition time of 2 seconds, while the Cambridge data sets were collected with a repetition time of 3 seconds. The temporal auto correlation between two consecutive volumes increases with the sampling rate, and in our previous study about single subject fMRI analysis~\cite{eklund_ni}, the sampling rate was found to be the most important factor for the degree of false positives. Block based designs are more sensitive to temporal auto correlations resembling power spectra with an 1/f appearance (f being frequency), as their power spectra are concentrated at low frequencies. The permutation test is unaffected by the used activity paradigm for two sample t-tests, but slightly affected for one sample t-tests. An unexpected result is that all software packages show significant interaction between the activity paradigm and the number of subjects. 

All software packages are significantly affected by the analysis type; the familywise error rates are generally lower for a two-sample t-test compared to a one sample t-test. This effect can be explained by the fact that a test value that represents a difference (e.g. the difference in brain activation between two groups) can more easily be approximated with a normal distribution, compared to a test value that does not represent a difference. As can be seen in Figures~\ref{fig:fwe_cluster_onesample_20_subjects},~\ref{fig:fwe_voxel_20_subjects},~\ref{fig:fwe_cluster_onesample_40_subjects} and~\ref{fig:fwe_voxel_40_subjects}, the one-sample t-tests are problematic even for the permutation test. For both FSL OLS and AFNI, there is strong interaction between the analysis type and the fMRI data. A possible explanation is that the one sample t-tests are more problematic for fMRI data collected with a higher sampling rate, as such data have stronger temporal auto correlation between two consecutive volumes~\cite{eklund_ni}.

Both FSL FLAME1 and AFNI OLS give a higher degree of false positives when the number of subjects increases. This is counter intuitive, as Gaussian random field theory normally works better for higher degrees of freedom. All software packages except SPM also show interaction between the fMRI data and the number of subjects.

\subsection{The future of fMRI}

It is not realistic to redo 28,000 fMRI studies, or to re-analyze all the data. Considering that it is now possible to evaluate common statistical methods using real fMRI data, the fMRI community should, in our opinion, focus on validation of existing methods (rather than developing new methods based on questionable assumptions). A plethora of excellent methods are available in the statistics field, but are seldom used in the neuroimaging community. A non-parametric permutation test, for example, is based on a small number of assumptions, and has here been proven to yield more accurate results than parametric methods. The main drawback of a permutation test is the increase in computational complexity, as the group analysis needs to be repeated 1,000 - 10,000 times. The increase in processing time is no longer a problem; an ordinary desktop computer can run a permutation test for neuroimaging data in less than a minute~\cite{broccoli,eklund_gpu}. A single desktop computer, with a powerful graphics card, was used in this study to run all the 384,000 permutation tests (with 1,000 permutations each) in about 15 days (the processing time would be 10 - 30 years if the function randomise in FSL was used instead).

In addition to unreliable statistical methods, the neuroimaging field also suffers from studies having low statistical power~\cite{button,durnez}. One possible way to increase the statistical power is to use locally multivariate statistical methods~\cite{friman,eklund_cca,aberg,jin,etzel}, which do not analyze the data one voxel at a time. Multivariate statistical methods can, however, result in more complicated null distributions, making it harder to obtain p-values. More advanced clustering techniques, such as cluster mass inference~\cite{clustermass} or threshold free cluster enhancement~\cite{tfce}, can also result in a higher statistical power, but it is often hard to derive a theoretical null distribution. A permutation test can estimate the null distribution of any test statistic, and can thus increase both the accuracy and the statistical power of fMRI studies.

\section*{Acknowledgment}

This research was supported by the neuroeconomic research initiative at Link\"{o}ping university, and by the Swedish research council (grant 2013-5229 'statistical analysis of fMRI data'). This study would not be possible without the recent data sharing initiatives in the neuroimaging field. We therefore thank the Neuroimaging Informatics Tools and Resources Clearinghouse (NITRC) and all the researchers that have contributed with resting state data to the 1000 functional connectomes project. We would also like to thank Russ Poldrack and his colleagues for starting the OpenfMRI project (supported by NSF grant OCI-1131441) and all the researchers that have shared their task based data. The Nvidia corporation, who donated the Tesla K40 graphics card used to run all the permutation tests, is also acknowledged.

\bibliographystyle{IEEEbib}
\bibliography{references}

\begin{thebibliography}{10}

\bibitem{ogawa}
{Ogawa, S. et al.},
\newblock ``{Intrinsic signal changes accompanying sensory stimulation:
  functional brain mapping with magnetic resonance imaging},''
\newblock {\em PNAS}, vol. 89, pp. 5951--5955, 1992.

\bibitem{fMRI}
N.K. Logothetis,
\newblock ``What we can do and what we cannot do with {fMRI},''
\newblock {\em Nature}, vol. 453, pp. 869--878, 2008.

\bibitem{biswal}
B.~Biswal, F.~Zerrin Yetkin, V.~M. Haughton, and J.~S. Hyde,
\newblock ``{Functional connectivity in the motor cortex of resting human brain
  using echo-planar {MRI}},''
\newblock {\em Magnetic resonance in medicine}, vol. 34, pp. 537--541, 1995.

\bibitem{dynamic}
{Hutchison, R.M. et al.},
\newblock ``Dynamic functional connectivity: Promise, issues, and
  interpretations,''
\newblock {\em NeuroImage}, vol. 80, pp. 360 -- 378, 2013.

\bibitem{welvaert}
M.~Welvaert and Y.~Rosseel,
\newblock ``A review of {fMRI} simulation studies,''
\newblock {\em {PLoS ONE}}, vol. 9, pp. e101953, 2014.

\bibitem{biswal2}
{Biswal, B. et al.},
\newblock ``Toward discovery science of human brain function,''
\newblock {\em PNAS}, vol. 107, pp. 4734--4739, 2010.

\bibitem{essen}
{Van Essen, D. et al.},
\newblock ``{The WU-Minn Human Connectome Project: An overview},''
\newblock {\em NeuroImage}, vol. 80, pp. 62--79, 2013.

\bibitem{poldrack}
R.~Poldrack and K.~Gorgolewski,
\newblock ``Making big data open: data sharing in neuroimaging,''
\newblock {\em Nature Neuroscience}, vol. 17, pp. 1510--1517, 2014.

\bibitem{openfmri}
{Poldrack, R. et al.},
\newblock ``Toward open sharing of task-based {fMRI} data: the {OpenfMRI}
  project,''
\newblock {\em Frontiers in Neuroinformatics}, vol. 7, no. 12, 2013.

\bibitem{adni1}
{Mueller, S.G. et al.},
\newblock ``The {Alzheimer's} disease neuroimaging initiative,''
\newblock {\em Neuroimaging Clinics of North America}, vol. 15, no. 4, pp. 869
  -- 877, 2005.

\bibitem{adni2}
{Jack, C.R. et al.},
\newblock ``The {Alzheimer's} disease neuroimaging initiative {(ADNI): MRI}
  methods,''
\newblock {\em Journal of Magnetic Resonance Imaging}, vol. 27, no. 4, pp.
  685--691, 2008.

\bibitem{poline}
{Poline, J.B. et al.},
\newblock ``Data sharing in neuroimaging research,''
\newblock {\em Frontiers in Neuroinformatics}, vol. 6, no. 9, 2012.

\bibitem{scarpazza}
C.~Scarpazza, G.~Sartori, M.~de~Simone, and A.~Mechelli,
\newblock ``When the single matters more than the group: very high false
  positive rates in single case voxel based morphometry,''
\newblock {\em NeuroImage}, vol. 70, pp. 175--188, 2013.

\bibitem{vbm}
J.~Ashburner and K.~Friston,
\newblock ``Voxel-based morphometry - the methods,''
\newblock {\em NeuroImage}, vol. 11, pp. 805--821, 2000.

\bibitem{silver}
M.~Silver, G.~Montana, and T.~Nichols,
\newblock ``False positives in neuroimaging genetics using voxel-based
  morphometry data,''
\newblock {\em NeuroImage}, vol. 54, pp. 992--1000, 2011.

\bibitem{eklund_ni}
A.~Eklund, M.~Andersson, C.~Josephson, M.~Johannesson, and H.~Knutsson,
\newblock ``Does parametric {fMRI} analysis with {SPM} yield valid results? -
  {An} empirical study of 1484 rest datasets,''
\newblock {\em NeuroImage}, vol. 61, pp. 565--578, 2012.

\bibitem{friston}
K.~Friston, J.~Ashburner, S.~Kiebel, T.~Nichols, and W.~Penny,
\newblock {\em Statistical Parametric Mapping: the Analysis of Functional Brain
  Images},
\newblock Elsevier/Academic Press, 2007.

\bibitem{ashburner}
J.~Ashburner,
\newblock ``{SPM}: a history,''
\newblock {\em NeuroImage}, vol. 62, pp. 791--800, 2012.

\bibitem{fsl}
M.~Jenkinson, C.~Beckmann, T.~Behrens, M.~Woolrich, and S.~Smith,
\newblock ``{FSL},''
\newblock {\em NeuroImage}, vol. 62, pp. 782--790, 2012.

\bibitem{afni}
R.~W. Cox,
\newblock ``{AFNI}: Software for analysis and visualization of functional
  magnetic resonance neuroimages,''
\newblock {\em Computers and Biomedical Research}, vol. 29, pp. 162--173, 1996.

\bibitem{scarpazza2}
C.~Scarpazza, S.~Tognin, S.~Frisciata, G.~Sartori, and A.~Mechelli,
\newblock ``False positive rates in voxel-based morphometry studies of the
  human brain: Should we be worried?,''
\newblock {\em Neuroscience \& Biobehavioral Reviews}, vol. 52, pp. 49--55,
  2015.

\bibitem{friston_cluster}
K.~J. Friston, K.~J. Worsley, R.~S.~J. Frackowiak, J.~C. Mazziotta, and A.~C.
  Evans,
\newblock ``Assessing the significance of focal activations using their spatial
  extent,''
\newblock {\em Human Brain Mapping}, vol. 1, pp. 210--220, 1994.

\bibitem{forman_cluster}
{Forman, S. D. et al.},
\newblock ``Improved assessment of significant activation in functional
  magnetic resonance imaging {(fMRI)}: Use of a cluster-size threshold,''
\newblock {\em Magnetic resonance in medicine}, vol. 33, pp. 636--647, 1995.

\bibitem{woo}
C.~Woo, A.~Krishnan, and T.~Wager,
\newblock ``Cluster-extent based thresholding in {fMRI} analyses: {Pitfalls}
  and recommendations,''
\newblock {\em NeuroImage}, vol. 91, pp. 412 -- 419, 2014.

\bibitem{nichols2002}
T.E. Nichols and A.P. Holmes,
\newblock ``Nonparametric permutation tests for functional neuroimaging: a
  primer with examples,''
\newblock {\em Human brain mapping}, vol. 15, pp. 1--25, 2002.

\bibitem{winkler}
A.~Winkler, G.~Ridgway, M.~Webster, S.~Smith, and T.~Nichols,
\newblock ``Permutation inference for the general linear model,''
\newblock {\em NeuroImage}, vol. 92, pp. 381--397, 2014.

\bibitem{carp2}
J.~Carp,
\newblock ``The secret lives of experiments: Methods reporting in the {fMRI}
  literature,''
\newblock {\em NeuroImage}, vol. 63, pp. 289--300, 2012.

\bibitem{broccoli}
A.~Eklund, P.~Dufort, M.~Villani, and S.~LaConte,
\newblock ``{BROCCOLI: Software for fast fMRI analysis on many-core CPUs and
  GPUs},''
\newblock {\em Frontiers in Neuroinformatics}, vol. 8:24, 2014.

\bibitem{lieberman}
M.D. Lieberman and W.A. Cunningham,
\newblock ``{Type I and Type II error concerns in fMRI research: re-balancing
  the scale},''
\newblock {\em Social cognitive and affective neuroscience}, vol. 4, pp.
  423--428, 2009.

\bibitem{woolrich2004}
M.~Woolrich, T.~Behrens, C.~Beckmann, M.~Jenkinson, and S.~Smith,
\newblock ``Multilevel linear modelling for {FMRI} group analysis using
  {Bayesian} inference,''
\newblock {\em NeuroImage}, vol. 21, pp. 1732--1747, 2004.

\bibitem{hayasaka}
S.~Hayasaka and T.E. Nichols,
\newblock ``Validating cluster size inference: random field and permutation
  methods,''
\newblock {\em NeuroImage}, vol. 20, pp. 2343--2356, 2003.

\bibitem{mixedgambles}
S.M. Tom, C.R. Fox, C.~Trepel, and R.A. Poldrack,
\newblock ``The neural basis of loss aversion in decision-making under risk,''
\newblock {\em Science}, vol. 315, pp. 515--518, 2007.

\bibitem{wordobject}
K.J. Duncan, C.~Pattamadilok, I.~Knierim, and J.T. Devlin,
\newblock ``Consistency and variability in functional localisers,''
\newblock {\em NeuroImage}, vol. 46, pp. 1018--1026, 2009.

\bibitem{ioannidis}
J.P.~A. Ioannidis,
\newblock ``{Why most published research findings are false},''
\newblock {\em PLOS Medicine}, vol. 2, pp. e124, 2005.

\bibitem{hayasaka2}
S.~Hayasakaa, K.~L. Phanb, I.~Liberzonc, K.~J. Worsley, and T.~E. Nichols,
\newblock ``Nonstationary cluster-size inference with random field and
  permutation methods,''
\newblock {\em NeuroImage}, vol. 22, pp. 676--687, 2004.

\bibitem{meyer}
{Meyer-Lindenberg, A. et al.},
\newblock ``False positives in imaging genetics,''
\newblock {\em NeuroImage}, vol. 40, pp. 655--661, 2008.

\bibitem{nichols_comparing}
T.~Nichols,
\newblock ``Controlling the familywise error rate in functional neuroimaging: a
  comparative review,''
\newblock {\em Statistical methods in medical research}, vol. 12, pp. 419--446,
  2003.

\bibitem{cao}
J.~Cao and K.~Worsley,
\newblock ``Applications of random fields in human brain mapping,''
\newblock in {\em Spatial Statistics: Methodological Aspects and Applications},
  M.~Moore, Ed., pp. 169--182. Springer, New York, 2001.

\bibitem{kiebel}
S.J. Kiebel, J.-B. Poline, K.J. Friston, A.P. Holmes, and K.J. Worsley,
\newblock ``Robust smoothness estimation in statistical parametric maps using
  standardized residuals from the general linear model,''
\newblock {\em NeuroImage}, vol. 10, pp. 756--766, 1999.

\bibitem{eklund_isbi}
A.~Eklund, T.~Nichols, M.~Andersson, and H.~Knutsson,
\newblock ``Empirically investigating the statistical validity of {SPM}, {FSL}
  and {AFNI} for single subject {fMRI} analysis,''
\newblock in {\em {IEEE} International symposium on biomedical imaging
  ({ISBI})}, 2015, pp. 1376--1380.

\bibitem{eklund_gpu}
A.~Eklund, P.~Dufort, D.~Forsberg, and S.~LaConte,
\newblock ``Medical image processing on the {GPU} - {Past}, present and
  future,''
\newblock {\em Medical Image Analysis}, vol. 17, pp. 1073--1094, 2013.

\bibitem{button}
{Button, K.S. et al.},
\newblock ``Power failure: why small sample size undermines the reliability of
  neuroscience,''
\newblock {\em Nature Reviews Neuroscience}, vol. 14, pp. 365--376, 2013.

\bibitem{durnez}
J.~Durnez, B.~Moerkerke, and T.~Nichols,
\newblock ``{Post-hoc power estimation for topological inference in {fMRI}},''
\newblock {\em NeuroImage}, vol. 84, pp. 45--64, 2014.

\bibitem{friman}
O.~Friman, M.~Borga, P.~Lundberg, and H.~Knutsson,
\newblock ``Adaptive analysis of {fMRI} data,''
\newblock {\em NeuroImage}, vol. 19, pp. 837--845, 2003.

\bibitem{eklund_cca}
A.~Eklund, M.~Andersson, and H.~Knutsson,
\newblock ``Fast random permutation tests enable objective evaluation of
  methods for single-subject {fMRI} analysis,''
\newblock {\em International Journal of Biomedical Imaging, Article ID 627947},
  vol. 2011, 2011.

\bibitem{aberg}
M.~Bj\"{o}rnsdotter, K.~Rylander, and J.~Wessberg,
\newblock ``A {Monte Carlo} method for locally multivariate brain mapping,''
\newblock {\em NeuroImage}, vol. 56, pp. 508--516, 2011.

\bibitem{jin}
M.~Jin, R.~Nandy, T.~Curran, and D.~Cordes,
\newblock ``Extending local canonical correlation analysis to handle general
  linear contrasts for {fMRI} data,''
\newblock {\em International Journal of Biomedical Imaging, Article {ID}
  574971}, vol. 2012, 2012.

\bibitem{etzel}
J.A. Etzel, J.M. Zacks, and T.S. Braver,
\newblock ``Searchlight analysis: Promise, pitfalls, and potential,''
\newblock {\em NeuroImage}, vol. 78, pp. 261 -- 269, 2013.

\bibitem{clustermass}
H.~Zhang, T.E. Nichols, and T.D. Johnson,
\newblock ``Cluster mass inference via random field theory,''
\newblock {\em NeuroImage}, vol. 44, pp. 51 -- 61, 2009.

\bibitem{tfce}
S.M. Smith and T.E. Nichols,
\newblock ``Threshold-free cluster enhancement: Addressing problems of
  smoothing, threshold dependence and localisation in cluster inference,''
\newblock {\em NeuroImage}, vol. 44, pp. 83--98, 2009.

\end{thebibliography}

\clearpage
\newpage

\section{Methods}
\label{sec:methods}

\subsection{Resting state fMRI data}

Resting state fMRI data from 396 healthy controls were downloaded from the homepage of the 1000 functional connectomes project~\cite{biswal2} \\ (http://fcon\_1000.projects.nitrc.org/fcpClassic/FcpTable.html). The Beijing and the Cambridge data sets were selected for their large sample sizes (198 subjects each) and their narrow age ranges (21.2 $\pm$ 1.8 and 21.0 $\pm$ 2.3 years, respectively). The Beijing data were collected with a repetition time (TR) of 2 seconds and consist of 225 time points per subject, the spatial resolution is 3.125 x 3.125 x 3.6 mm$^3$. The Cambridge data were collected with a TR of 3 seconds and consist of 119 time points per subject, the spatial resolution is 3 x 3 x 3 mm$^3$. For each subject there is one $T_1$-weighted anatomical volume which can be used for normalization to a brain template. According to the motion plots from FSL, no subject moved more than 1 mm in any direction. According to motion plots from AFNI, one Cambridge subject and three Beijing subjects moved slightly more than 1 mm. The fMRI data have not been corrected for geometric distortions, and no field maps are available for this purpose.

Since all the subjects are healthy and of similar age, it should be impossible to find any significant brain activity differences between two randomly generated subgroups (analyses were performed separately for the Beijing data and the Cambridge data). The same approach has previously been used to test the validity of parametric statistics for voxel based morphometry~\cite{vbm,scarpazza2}. As the subjects have not performed any specific task in the MR scanner, it should also be impossible to find significant group activations. The data can thus be used to test both two-sample t-tests (group differences) and one sample t-tests (group activations). 

\subsubsection{Random group generation}

Each random group was created by first applying a random permutation to a list containing all the 198 subject numbers. To create two random groups of 20 subjects each, the first 20 permuted subject numbers were put into group 1, and the following 20 permuted subject numbers were put into group 2. According to the n choose k formula $\frac{n!}{k!(n-k)!}$ it is possible to create approximately $1.31 \cdot 10^{42}$ such random group divisions (n = 198 and k = 40). The analyses will not be independent, but the estimate of the familywise false positive rate will still be unbiased. A total of 1000 random analyses were used to estimate the familywise false positive rate, giving a 95\% confidence interval of 3.65\% - 6.35\% for an expected false positive rate of 5\%. To make a fair comparison between the different software packages, the same 1000 permutations were used for all software packages and all parameter settings. 

\subsubsection{Code availability}

Parametric group analyses were performed using SPM 8 (http://www.fil.ion.ucl.ac.uk/spm/software/spm8/), FSL 5.0.7 (http://fsl.fmrib.ox.ac.uk/fsldownloads/) and AFNI \\(http://afni.nimh.nih.gov/afni/download/afni/releases, compiled August 13 2014, version 2011\_12\_21\_1014). FSL can perform non-parametric group analyses using the function randomise, but we here used our BROCCOLI software~\cite{broccoli} (https://github.com/wanderine/BROCCOLI) to lower the processing time. All the processing scripts are freely available \\ (https://github.com/wanderine/ParametricMultisubjectfMRI) to show all the processing settings and to facilitate replication of the results. Since all the software packages and all the fMRI data are also freely available, anyone can replicate the results in this paper.

\subsubsection{First level analyses}

A first processing script was used for each software package to perform first level analyses for each subject, resulting in brain activation maps in a standard brain space (Montreal Neurological Institute (MNI) for SPM and FSL, and Talairach for AFNI). All first level analyses involved normalization to a brain template, motion correction and different amounts of smoothing (4, 6, 8 and 10 mm full width at half maximum). Slice timing correction was not performed, as the slice timing information is not available in the fMRI data sets. A general linear model (GLM) was finally applied to the preprocessed fMRI data, using different regressors for activity (B1, B2, E1, E2). The estimated head motion parameters were used as additional regressors in the design matrix, for all packages, to further reduce effects of head motion.

First level analyses were for SPM performed using a Matlab batch script, mainly created using the SPM manual. The spatial normalization was done as a two step procedure, where the mean fMRI volume was first aligned to the anatomical volume (using the function 'Coregister' with default settings). The anatomical volume was aligned to MNI space using the function 'Segment' (with default settings), and the two transforms were finally combined to transform the fMRI data to MNI space at 2 mm isotropic resolution (using the function 'Normalise: Write'). Spatial smoothing was finally applied to the spatially normalized fMRI data. The first level models were then fit in the atlas space, i.e. not in the subject space.

For FSL, first level analyses were setup through the FEAT GUI. The spatial normalization to the brain template (MNI152\_T1\_2mm\_brain.nii.gz) was performed as a two step linear registration using the function FLIRT (which is the default option). One fMRI volume was aligned to the anatomical volume using the BBR (boundary based registration) option in FLIRT (default). The anatomical volume was aligned to MNI space using a linear registration with 12 degrees of freedom (default), and the two transforms were finally combined. The first level models were fit in the subject space (after spatial smoothing), and the contrasts and their variances were then transformed to the atlas space. 

First level analyses in AFNI were performed using the standardized processing script afni\_proc.py, which creates a tcsh script which contains all the calls to different AFNI functions. The spatial normalization was performed as a two step procedure. One fMRI volume was first linearly aligned to the anatomical volume, using the script align\_epi\_anat.py. The anatomical volume was then linearly aligned to the brain template (TT\_N27+tlrc) using the script @auto\_tlrc. The transformations from the spatial normalization and the motion correction were finally applied using a single interpolation, resulting in normalized fMRI data in an isotropic resolution of 3 mm. Spatial smoothing was applied to the spatially normalized fMRI data, and the first level models were then fit in the atlas space (i.e. not in the subject space). 

Default drift modelling or highpass filtering options were used in each of SPM, FSL and AFNI. A discrete cosine transform with cutoff of 128 seconds was used for SPM, while highpass filters with different cutoffs where used for FSL (20 seconds for activity paradigm B1, 60 seconds for B2 and 100 seconds for E1 and E2), matching the defaults used by the FEAT GUI, and AFNI's Legende polynomial order is 4 and 3 for the Beijing and the Cambridge data, respectively (based on total scan duration). Temporal correlations were further corrected for with a global AR(1) model in SPM, an arbitrary temporal auto correlation function regularized with a Tukey taper and adaptive spatial smoothing in FSL and a voxel-wise ARMA(1,1) model in AFNI.

\subsubsection{Group analyses}

A second processing script was used for each software package to perform random effect group analyses, using the results from the first level analyses. For SPM, group analyses were only performed with the resulting beta weights from the first level analyses, using ordinary least squares (OLS) regression over subjects. For FSL, group analyses were performed both using FLAME1 (which is the default option) and OLS. The FLAME1 function uses both the beta weight and the corresponding variance of each subject, subsequently estimating a between subject variance. For AFNI, group analyses were performed using the functions 3dttest++ (OLS, using beta estimates from the function 3dDeconvolve which assumes independent errors) and 3dMEMA (which is similar to FLAME1 in FSL, using beta and variance estimates from the function 3dREMLfit which uses a voxel-wise ARMA(1,1) model of the errors). 

For the non-parametric analyses in BROCCOLI, first level results from FSL were used and OLS regression was performed in each permutation. The largest test value across the entire brain was saved in each permutation, to empirically form the null distribution of the maximum test statistic (which is required to correct for multiple comparisons). For cluster-wise inference, the cluster defining threshold was first applied and the size of the largest cluster was then saved in each permutation. A permutation test cannot be used for testing group activations (one sample t-tests), as the mean brain activity is invariant to permutations of the subjects. An alternative is to instead use random sign flipping of the subjects, justified by an assumption of symmetrically distributed errors, which is the solution that FSL and BROCCOLI use for one sample t-tests. Each non-parametric group analysis was performed using 1000 permutations or sign flips, giving a total of 192 million permutations and 192 million sign flips for all group analyses (hence the need to lower the processing time).

Voxel-wise FWE-corrected p-values from SPM and FSL were obtained based on their respective implementations of random field theory, while AFNI FWE p-values were obtained with a Bonferroni correction for the number of voxels (AFNI does not provide any specific program for voxel-wise FWE p-values). For the non-parametric analyses, FWE-corrected p-values were calculated with the empirical null distribution of the voxel-wise maximum statistic, computed as the proportion of the null distribution being larger than a particular statistic value. 

Cluster-wise FWE-corrected p-values from SPM and FSL were likewise obtained based on their implementations of random field theory. AFNI estimates FWE p-values with a simulation based procedure, 3dClustSim. SPM and FSL estimate smoothness from the residuals of the group level analysis (used for both voxel-wise and cluster-wise inference), while AFNI uses the average of the first level analyses' smoothness estimates. For the non-parametric analyses, FWE-corrected p-values were calculated as the proportion of cluster sizes in the empirically estimated null distribution being larger than each cluster in the group difference map or group activation map. 

Each group analysis was considered to give a significant result if any cluster or voxel had a FWE-corrected p-value $p < 0.05$.

\subsection{Task based fMRI data}

Task based fMRI data were downloaded from the homepage of the OpenfMRI project~\cite{openfmri} (http://openfmri.org), to investigate how cluster based p-values differ between parametric and non-parametric group analyses. Each task dataset contains fMRI data, anatomical data and timing information for each subject. The data sets were only analyzed with FSL, using 5 mm of smoothing (the default option). Motion regressors were used in all cases, to further suppress effects of head motion. Group analyses were performed using the parametric OLS option (i.e. not the default FLAME1 option) and the non-parametric randomise function.

\subsubsection{Rhyme judgment}

The rhyme judgment dataset is available at \\ http://openfmri.org/dataset/ds000003. The 13 subjects were presented with pairs of either words or pseudowords, and made rhyming judgments for each pair. The fMRI data were collected with a repetition time of 2 seconds and consist of 160 time points per subject, the spatial resolution is 3.125 x 3.125 x 4 mm$^3$. The data were analyzed with two regressors; one for words and one for pseudo words. A total of four contrasts were applied; words, pseudowords, words - pseudowords, pseudowords - words. For a cluster defining threshold of p = 0.01, a t-threshold of 2.65 was used. For a cluster defining threshold of p = 0.001, a t-threshold of 3.95 was used. 

\subsubsection{Mixed-gambles task}

The mixed-gambles task dataset is available at \\ http://openfmri.org/dataset/ds000005. The 16 subjects were presented with mixed (gain/loss) gambles, and decided whether they would accept each gamble.  No outcomes of these gambles were presented during scanning, but after the scan three gambles were selected at random and played for real money. The fMRI data were collected using a 3 T Siemens Allegra scanner. A repetition time of 2 seconds was used and a total of 240 volumes were collected for each run, the spatial resolution is 3.125 x 3.125 x 4 mm$^3$. The dataset contains three runs per subject, but only the first run was used in our analysis. The data were analyzed using four regressors; task, parametric gain, parametric loss and distance from indifference. A total of four contrasts were applied; parametric gain, - parametric gain, parametric loss, - parametric loss. For a cluster defining threshold of p = 0.01, a t-threshold of 2.57 was used. For a cluster defining threshold of p = 0.001, a t-threshold of 3.75 was used.
 
\subsubsection{Living-nonliving decision with plain or mirror-reversed text}

The living-nonliving decision task dataset is available at \\ http://openfmri.org/dataset/ds000006a. The 14 subjects made living-nonliving decisions on items presented in either plain or mirror-reversed text. The fMRI data were collected using a 3 T Siemens Allegra scanner. A repetition time of 2 seconds was used and a total of 205 volumes were collected for each run, the spatial resolution is 3.125 x 3.125 x 5 mm$^3$. The dataset contains six runs per subject, but only the first run was used in our analysis. The data were analyzed using five regressors; mirror-switched, mirror-repeat, plain-switched, plain-repeat and junk. A total of four contrasts were applied; mirrored versus plain (1,1,-1,-1,0), switched versus non-switched (1,-1,1,-1,0), switched versus non-switched mirrored only (1,-1,0,0,0) and switched versus non-switched plain only (0,0,1,-1,0). For a cluster defining threshold of p = 0.01, a t-threshold of 2.615 was used. For a cluster defining threshold of p = 0.001, a t-threshold of = 3.87 was used.
 
\subsubsection{Word and object processing}

The word and object processing task dataset is available at \\ http://openfmri.org/dataset/ds000107. The 49 subjects performed a visual one-back task with four categories of items: written words, objects, scrambled objects and consonant letter strings. The fMRI data were collected using a 1.5 T Siemens scanner. A repetition time of 3 seconds was used and a total of 165 volumes were collected for each run, the spatial resolution is 3 x 3 x 3 mm$^3$. The dataset contains two runs per subject, but only the first run was used in our analysis. The data were analyzed using four regressors; words, objects, scrambled objects, consonant strings. A total of six contrasts were applied; words, objects, scrambled objects, consonant strings, objects versus scrambled objects (0,1,-1,0) and words versus consonant strings (1,0,0,-1). For a cluster defining threshold of p = 0.01, a t-threshold of 2.38 was used. For a cluster defining threshold of p = 0.001, a t-threshold of 3.28 was used.

\clearpage
\newpage

\begin{table*}[htb]
\scriptsize
\caption{Cluster p-values (corrected for multiple comparisons) for FSL OLS and a permutation test, for typical fMRI studies available on the OpenfMRI homepage. A cluster defining threshold of p = 0.01 (z = 2.3) was used. Note that the resolution of the permutation p-values is 0.0002, since 5000 permutations (or sign flips) were used. A star denotes that the parametric p-value is below 0.2, while the non-parametric p-value is not.}
\begin{center}
\begin{tabular}{|c|c|c|c|c|}
\hline 
\textbf{\normalsize OpenfMRI data set}  & \textbf{\normalsize Subjects} & \textbf{\normalsize Cluster size (voxels)}  & \textbf{\normalsize FSL OLS p-value }  & \textbf{\normalsize Perm OLS p-value }   \\
\hline

\normalsize Rhyme judgment, contrast 1 & \normalsize 13 & \normalsize 53877   & \normalsize 0 & \normalsize 0.0002 \\ 

\normalsize Rhyme judgment, contrast 2 & \normalsize 13 & \normalsize 27484  & \normalsize 0 & \normalsize 0.0002  \\  
\normalsize                            & \normalsize  & \normalsize 14682  & \normalsize 1.37 $ \cdot $ 10$^{-40}$ & \normalsize 0.002 \\ 
\normalsize                            & \normalsize  & \normalsize 3467  & \normalsize 1.16 $ \cdot $ 10$^{-14}$ &  \normalsize 0.024 \\ 

\normalsize Rhyme judgment, contrast 3 & \normalsize 13 & \normalsize 799   & \normalsize 4.8 $ \cdot $ 10$^{-5}$ & \normalsize 0.097  \\ 
\normalsize                            & \normalsize      & \normalsize 408   & \normalsize 0.0103 & \normalsize 0.219 * \\ 

\normalsize Rhyme judgment, contrast 4 & \normalsize 13 & \normalsize No surviving clusters   & &  \\

\hline

\normalsize Mixed gambles, contrast 1 & \normalsize 16 & \normalsize 13284   & \normalsize 1.36 $ \cdot $ 10$^{-36}$ & \normalsize 0.005 \\ 
\normalsize                           & \normalsize  & \normalsize 440   & \normalsize 0.0152 & \normalsize 0.202 * \\ 

\normalsize Mixed gambles, contrast 2 & \normalsize 16 & \normalsize No surviving clusters   &  &  \\

\normalsize Mixed gambles, contrast 3 & \normalsize 16 & \normalsize No surviving clusters   & &  \\

\normalsize Mixed gambles, contrast 4 & \normalsize 16 & \normalsize 655  &\normalsize 0.00888 & \normalsize 0.118 \\ 

\hline

\normalsize Living-nonliving decision, contrast 1 & \normalsize 14 & \normalsize 8612   & \normalsize3.46 $ \cdot $ 10$^{-33}$ &\normalsize 0.001 \\ 
\normalsize  & \normalsize  & \normalsize 7577   & \normalsize 2.37 $ \cdot $ 10$^{-30}$ &\normalsize 0.002 \\ 
\normalsize  & \normalsize  & \normalsize 5920   & \normalsize 1.6 $ \cdot $ 10$^{-25}$ &\normalsize 0.003 \\ 
\normalsize  & \normalsize  & \normalsize 1439   & \normalsize 4.88 $ \cdot $ 10$^{-9}$ & \normalsize 0.035 \\ 
\normalsize  & \normalsize  & \normalsize 601    & \normalsize 0.000213 & \normalsize 0.116  \\ 

\normalsize Living-nonliving decision, contrast 2 & \normalsize 14 & \normalsize 751  &\normalsize 2.75 $ \cdot $ 10$^{-5}$ &\normalsize 0.08  \\ 
\normalsize  & \normalsize  & \normalsize 669  & \normalsize 8.71 $ \cdot $ 10$^{-5}$ &\normalsize 0.096 \\ 
\normalsize  & \normalsize  & \normalsize 546  &\normalsize  0.000541 &\normalsize 0.128  \\ 

\normalsize Living-nonliving decision, contrast 3 & \normalsize 14 & \normalsize 396  &\normalsize 0.00889 &\normalsize   0.172  \\ 
\normalsize  & \normalsize  & \normalsize 323  & \normalsize 0.0302  &\normalsize  0.207 * \\

\normalsize Living-nonliving decision, contrast 4 & \normalsize 14 & \normalsize No surviving clusters & &    \\ 

\hline

\normalsize Word and object processing, contrast 1 & \normalsize 49 & \normalsize 7397  &\normalsize 2.3 $ \cdot $ 10$^{-32}$ &\normalsize 0.001 \\ 
\normalsize  & \normalsize  & \normalsize  6586 &\normalsize 7.57 $ \cdot $ 10$^{-30}$ &\normalsize 0.001 \\ 
\normalsize  & \normalsize  & \normalsize  6232 &\normalsize 1.02 $ \cdot $ 10$^{-28}$ &\normalsize 0.002 \\ 
\normalsize  & \normalsize  & \normalsize  2834 &\normalsize 2.11 $ \cdot $ 10$^{-16}$ &\normalsize 0.01 \\ 
\normalsize  & \normalsize  & \normalsize  486 &\normalsize 0.00044 &\normalsize 0.139  \\ 
\normalsize  & \normalsize  & \normalsize  288 &\normalsize 0.0182 &\normalsize 0.249 * \\ 

\normalsize Word and object processing, contrast 2 & \normalsize 49 & \normalsize 7062  &\normalsize 9.77 $ \cdot $ 10$^{-30}$ &\normalsize 0.001 \\ 
\normalsize  & \normalsize  & \normalsize  6158 &\normalsize 5.08 $ \cdot $ 10$^{-27}$ &\normalsize 0.002 \\ 
\normalsize  & \normalsize  & \normalsize  5529 &\normalsize 4.73 $ \cdot $ 10$^{-25}$ &\normalsize 0.003 \\ 
\normalsize  & \normalsize  & \normalsize  1853 &\normalsize 2.27 $ \cdot $ 10$^{-11}$ &\normalsize 0.025 \\ 
\normalsize  & \normalsize  & \normalsize  1523 &\normalsize 8.3 $ \cdot $ 10$^{-10}$ &\normalsize 0.035 \\ 
\normalsize  & \normalsize  & \normalsize  1465 &\normalsize 1.6 $ \cdot $ 10$^{-9}$ &\normalsize 0.037 \\ 
\normalsize  & \normalsize  & \normalsize  1382 &\normalsize 4.18 $ \cdot $ 10$^{-9}$ &\normalsize 0.04 \\ 
\normalsize  & \normalsize  & \normalsize  437 &\normalsize 0.00174 &\normalsize 0.159  \\ 
\normalsize  & \normalsize  & \normalsize  409 &\normalsize 0.00283 &\normalsize 0.173  \\ 

\normalsize Word and object processing, contrast 3 & \normalsize 49 & \normalsize 42205  &\normalsize 0 &\normalsize 0.0002 \\ 
\normalsize  & \normalsize  & \normalsize  998 &\normalsize 1.79 $ \cdot $ 10$^{-7}$ &\normalsize 0.054  \\ 

\normalsize Word and object processing, contrast 4 & \normalsize 49 & \normalsize 32404  &\normalsize 0 &\normalsize 0.0002 \\ 
\normalsize  & \normalsize  & \normalsize  12837 &\normalsize 2.8 $ \cdot $ 10$^{-45}$ &\normalsize 0.001 \\ 
\normalsize  & \normalsize  & \normalsize  280 &\normalsize 0.0287 &\normalsize 0.248 * \\ 
\normalsize  & \normalsize  & \normalsize  278 &\normalsize 0.0299 &\normalsize 0.251 * \\ 

\normalsize Word and object processing, contrast 5 & \normalsize 49 & \normalsize 2118  &\normalsize 3.05 $ \cdot $ 10$^{-14}$ &\normalsize  0.017\\
\normalsize  & \normalsize  & \normalsize  881 &\normalsize 2.98 $ \cdot $ 10$^{-7}$ &\normalsize  0.051 \\
\normalsize  & \normalsize  & \normalsize  395 &\normalsize 0.00115 &\normalsize 0.146   \\
\normalsize  & \normalsize  & \normalsize  340 &\normalsize 0.00354 &\normalsize  0.18  \\
\normalsize  & \normalsize  & \normalsize  255 &\normalsize 0.0226 &\normalsize  0.253 * \\
\normalsize  & \normalsize  & \normalsize  253 &\normalsize 0.0237 &\normalsize  0.258 *\\
\normalsize  & \normalsize  & \normalsize  222 &\normalsize 0.0486 &\normalsize  0.297 *\\

\normalsize Word and object processing, contrast 6 & \normalsize 49 & \normalsize 27767  &\normalsize 0 &\normalsize 0.0002 \\ 
\normalsize  & \normalsize  & \normalsize  6183 &\normalsize 9.65 $ \cdot $ 10$^{-29}$ &\normalsize 0.005  \\ 

\hline
\end{tabular}
\end{center}
\label{table:typicalstudies1}
\end{table*}

\begin{table*}[htb]
\scriptsize
\caption{Cluster p-values (corrected for multiple comparisons) for FSL OLS and a permutation test, for typical fMRI studies available on the OpenfMRI homepage. A cluster defining threshold of p = 0.001 (z = 3.1) was used. Note that the resolution of the permutation p-values is 0.0002, since 5000 permutations (or sign flips) were used. A star denotes that the parametric p-value is below 0.05, while the non-parametric p-value is not.}
\begin{center}
\begin{tabular}{|c|c|c|c|c|}
\hline 
\textbf{\normalsize OpenfMRI data set}  & \textbf{\normalsize Subjects} & \textbf{\normalsize Cluster size (voxels)}  & \textbf{\normalsize FSL OLS p-value}  & \textbf{\normalsize Perm OLS p-value}   \\
\hline
\normalsize Rhyme judgment, contrast 1 & \normalsize 13 & \normalsize  13877 &\normalsize 0           &\normalsize 0.0002 \\ 
\normalsize                           & \normalsize      & \normalsize  4859 &\normalsize 1.18 $ \cdot $  10$^{-38}$ &\normalsize 0.001 \\ 
\normalsize                           & \normalsize      & \normalsize  2273 &\normalsize 5.44 $ \cdot $  10$^{-23}$ &\normalsize 0.002 \\ 
\normalsize                           & \normalsize      & \normalsize  2039 &\normalsize 2.49 $ \cdot $  10$^{-21}$ &\normalsize 0.002 \\ 
\normalsize                           & \normalsize      & \normalsize  1081 &\normalsize 1.04 $ \cdot $  10$^{-13}$ &\normalsize 0.005 \\ 
\normalsize                           & \normalsize      & \normalsize  473 &\normalsize 1.19 $ \cdot $  10$^{-7}$ &\normalsize 0.008 \\ 
\normalsize                           & \normalsize      & \normalsize  306 &\normalsize 1.78 $ \cdot $  10$^{-5}$ &\normalsize 0.011 \\ 
\normalsize                           & \normalsize      & \normalsize  133 &\normalsize 0.00806 &\normalsize 0.038 \\ 
\normalsize                           & \normalsize      & \normalsize  122 &\normalsize 0.0127  &\normalsize 0.042 \\ 
\normalsize                           & \normalsize      & \normalsize  99 &\normalsize 0.0347 &\normalsize 0.055 * \\ 

\normalsize  Rhyme judgment, contrast 2 & \normalsize 13 & \normalsize  14470 &\normalsize 0 &\normalsize 0.0002  \\ 
\normalsize                           & \normalsize     & \normalsize  3074 &\normalsize 5.43 $ \cdot $  10$^{-27}$ &\normalsize 0.001 \\ 
\normalsize                           & \normalsize     & \normalsize 1868 &\normalsize 3.7 $ \cdot $  10$^{-19}$ &\normalsize 0.002 \\ 
\normalsize                           & \normalsize     & \normalsize 1558 &\normalsize 6.83 $ \cdot $  10$^{-17}$ &\normalsize 0.002 \\ 
\normalsize                           & \normalsize     & \normalsize  874 &\normalsize 2.95 $ \cdot $  10$^{-11}$ &\normalsize 0.004 \\ 
\normalsize                           & \normalsize     & \normalsize  422 &\normalsize 1.19 $ \cdot $  10$^{-6}$ &\normalsize 0.008 \\ 
\normalsize                           & \normalsize     & \normalsize  255 &\normalsize 0.000153 &\normalsize 0.014 \\ 
\normalsize                           & \normalsize     & \normalsize  96 &\normalsize 0.0498 &\normalsize  0.06 *\\ 

\normalsize  Rhyme judgment, contrast 3 & \normalsize 13 & \normalsize  No surviving clusters &  &   \\
\normalsize  Rhyme judgment, contrast 4 & \normalsize 13 & \normalsize  No surviving clusters &  & \\

\hline

\normalsize Mixed gambles, contrast 1 & \normalsize 16 & \normalsize 766   &\normalsize 7.01 $ \cdot $ 10$^{-10}$ &\normalsize 0.001  \\ 
\normalsize  & \normalsize  & \normalsize 120   &\normalsize 0.0237 &\normalsize  0.053 *\\ 
\normalsize Mixed gambles, contrast 2 & \normalsize 16 & \normalsize No surviving clusters   &  &  \\
\normalsize Mixed gambles, contrast 3 & \normalsize 16 & \normalsize No surviving clusters   & &  \\
\normalsize Mixed gambles, contrast 4 & \normalsize 16 & \normalsize No surviving clusters &  &  \\

\hline

\normalsize Living-nonliving decision, contrast 1 & \normalsize 14 & \normalsize 3310  &\normalsize 1.69 $ \cdot $ 10$^{-33}$ &\normalsize 0.0002  \\
\normalsize  & \normalsize & \normalsize 1901  &\normalsize 6.95 $ \cdot $ 10$^{-23}$ &\normalsize  0.0002 \\
\normalsize  & \normalsize & \normalsize 761  &\normalsize 4.8 $ \cdot $ 10$^{-12}$ &\normalsize  0.001 \\
\normalsize  & \normalsize & \normalsize 569  &\normalsize 8.96 $ \cdot $ 10$^{-10}$ &\normalsize  0.002\\
\normalsize  & \normalsize & \normalsize 417  &\normalsize 5.96 $ \cdot $ 10$^{-8}$ &\normalsize  0.002\\
\normalsize  & \normalsize & \normalsize 187  &\normalsize 0.000326 &\normalsize  0.013 \\
\normalsize  & \normalsize & \normalsize 109  &\normalsize 0.0109 &\normalsize  0.038 \\
\normalsize  & \normalsize & \normalsize 96  &\normalsize 0.021 &\normalsize  0.047 \\
\normalsize  & \normalsize & \normalsize 85  &\normalsize 0.0373 &\normalsize 0.06 *\\
\normalsize  & \normalsize & \normalsize 84  &\normalsize 0.0394 &\normalsize 0.061 *\\ 

\normalsize Living-nonliving decision, contrast 2 & \normalsize 14 & \normalsize 90  &\normalsize  0.0301 &\normalsize 0.046 \\ 

\normalsize Living-nonliving decision, contrast 3 & \normalsize 14 & \normalsize No surviving clusters \normalsize   &   &   \\
\normalsize Living-nonliving decision, contrast 4 & \normalsize 14 & \normalsize No surviving clusters \normalsize   &  &  \\

\hline
\end{tabular}
\end{center}
\label{table:typicalstudies2}
\end{table*} 

\begin{table*}[htb]
\scriptsize
\caption{Cluster p-values (corrected for multiple comparisons) for FSL OLS and a permutation test, for typical fMRI studies available on the OpenfMRI homepage. A cluster defining threshold of p = 0.001 (z = 3.1) was used. Note that the resolution of the permutation p-values is 0.0002, since 5000 permutations (or sign flips) were used. A star denotes that the parametric p-value is below 0.05, while the non-parametric p-value is not.}
\begin{center}
\begin{tabular}{|c|c|c|c|c|}
\hline
\textbf{\normalsize OpenfMRI data set}  & \textbf{\normalsize Subjects} & \textbf{\normalsize Cluster size (voxels)}  & \textbf{\normalsize FSL OLS p-value}  & \textbf{\normalsize Perm OLS p-value}   \\
\hline
\normalsize Word and object processing, contrast 1 & \normalsize 49 & \normalsize 4644  &\normalsize 1.4 $ \cdot $ 10$^{-45}$ &\normalsize 0.0002  \\
\normalsize  & \normalsize & \normalsize 4017  &\normalsize 2.35 $ \cdot $ 10$^{-41}$ &\normalsize  0.0002 \\
\normalsize  & \normalsize & \normalsize 2615  &\normalsize 7.21 $ \cdot $ 10$^{-31}$ &\normalsize  0.0002 \\
\normalsize  & \normalsize & \normalsize 828  &\normalsize 6.38 $ \cdot $ 10$^{-14}$ &\normalsize  0.001 \\
\normalsize  & \normalsize & \normalsize 765  &\normalsize 3.64 $ \cdot $ 10$^{-13}$ &\normalsize  0.001 \\
\normalsize  & \normalsize & \normalsize 543  &\normalsize 2.57 $ \cdot $ 10$^{-10}$ &\normalsize  0.003 \\
\normalsize  & \normalsize & \normalsize 306  &\normalsize 8.34 $ \cdot $ 10$^{-7}$ &\normalsize 0.006  \\
\normalsize  & \normalsize & \normalsize 292  &\normalsize 1.49 $ \cdot $ 10$^{-6}$ &\normalsize  0.006 \\
\normalsize  & \normalsize & \normalsize 176  &\normalsize 0.000187 &\normalsize  0.017 \\ 

\normalsize Word and object processing, contrast 2 & \normalsize 49 & \normalsize 5000  &\normalsize 1.4 $ \cdot $ 10$^{-45}$ &\normalsize 0.0002  \\
\normalsize  & \normalsize & \normalsize 3902  &\normalsize 1.61 $ \cdot $ 10$^{-38}$ &\normalsize  0.0002 \\
\normalsize  & \normalsize & \normalsize 1540  &\normalsize 2.22 $ \cdot $ 10$^{-20}$ &\normalsize  0.0002 \\
\normalsize  & \normalsize & \normalsize 1199  &\normalsize 3.92 $ \cdot $ 10$^{-17}$ &\normalsize  0.0002 \\
\normalsize  & \normalsize & \normalsize 1035  &\normalsize 1.84 $ \cdot $ 10$^{-15}$ &\normalsize  0.0002 \\
\normalsize  & \normalsize & \normalsize 989  &\normalsize 5.6 $ \cdot $ 10$^{-15}$ &\normalsize  0.001 \\
\normalsize  & \normalsize & \normalsize 759  &\normalsize 1.96 $ \cdot $ 10$^{-12}$ &\normalsize  0.001 \\
\normalsize  & \normalsize & \normalsize 699  &\normalsize 9.92 $ \cdot $ 10$^{-12}$ &\normalsize  0.001 \\
\normalsize  & \normalsize & \normalsize 497  &\normalsize 3.42 $ \cdot $ 10$^{-9}$ &\normalsize  0.003 \\
\normalsize  & \normalsize & \normalsize 413  &\normalsize 5.96 $ \cdot $ 10$^{-8}$ &\normalsize  0.004 \\
\normalsize  & \normalsize & \normalsize 133  &\normalsize 0.00222 &\normalsize  0.029 \\
\normalsize  & \normalsize & \normalsize 95  &\normalsize 0.0149 &\normalsize  0.053 *\\

\normalsize Word and object processing, contrast 3 & \normalsize 49 & \normalsize 27735  &\normalsize 0 &\normalsize 0.0002  \\
\normalsize  & \normalsize & \normalsize 1312  &\normalsize 2.62 $ \cdot $ 10$^{-19}$ &\normalsize  0.001\\
\normalsize  & \normalsize & \normalsize 1264  &\normalsize 8.15 $ \cdot $ 10$^{-19}$ &\normalsize  0.001\\
\normalsize  & \normalsize & \normalsize 789  &\normalsize 1.57 $ \cdot $ 10$^{-13}$ &\normalsize  0.002\\
\normalsize  & \normalsize & \normalsize 525  &\normalsize 3.99 $ \cdot $ 10$^{-10}$ &\normalsize  0.004\\
\normalsize  & \normalsize & \normalsize 415  &\normalsize 1.53 $ \cdot $ 10$^{-8}$ &\normalsize  0.005\\
\normalsize  & \normalsize & \normalsize 209  &\normalsize 4.05 $ \cdot $ 10$^{-5}$ &\normalsize  0.013\\
\normalsize  & \normalsize & \normalsize 143  &\normalsize 0.000845 &\normalsize  0.024 \\

\normalsize Word and object processing, contrast 4 & \normalsize 49 & \normalsize 24890  &\normalsize 0 &\normalsize 0.0002  \\
\normalsize  & \normalsize & \normalsize 3525  &\normalsize 2.79 $ \cdot $ 10$^{-36}$ &\normalsize  0.0002\\
\normalsize  & \normalsize & \normalsize 1678  &\normalsize 7.98 $ \cdot $ 10$^{-22}$ &\normalsize  0.0002\\
\normalsize  & \normalsize & \normalsize 1492  &\normalsize 4.01 $ \cdot $ 10$^{-20}$ &\normalsize  0.001\\
\normalsize  & \normalsize & \normalsize 996  &\normalsize 3.42 $ \cdot $ 10$^{-15}$ &\normalsize  0.002\\
\normalsize  & \normalsize & \normalsize 845  &\normalsize 1.55 $ \cdot $ 10$^{-13}$ &\normalsize  0.003\\
\normalsize  & \normalsize & \normalsize 346  &\normalsize 4.17 $ \cdot $ 10$^{-7}$ &\normalsize  0.008\\
\normalsize  & \normalsize & \normalsize 112  &\normalsize 0.00582 &\normalsize  0.042\\
\normalsize  & \normalsize & \normalsize 107  &\normalsize 0.00751 &\normalsize  0.044\\
\normalsize  & \normalsize & \normalsize 106  &\normalsize 0.00791 &\normalsize  0.044\\
\normalsize  & \normalsize & \normalsize 75  &\normalsize 0.0425 &\normalsize  0.074 *\\

\normalsize Word and object processing, contrast 5 & \normalsize 49 & \normalsize 373  &\normalsize 2.03 $ \cdot $ 10$^{-8}$ &\normalsize 0.005  \\
\normalsize  & \normalsize & \normalsize 282  &\normalsize 7.75 $ \cdot $ 10$^{-7}$ &\normalsize 0.006 \\
\normalsize  & \normalsize & \normalsize 109  &\normalsize 0.00302 &\normalsize  0.033 \\
\normalsize  & \normalsize & \normalsize 98  &\normalsize 0.00574 &\normalsize  0.04 \\
\normalsize  & \normalsize & \normalsize 92  &\normalsize 0.00821 &\normalsize  0.044\\
\normalsize  & \normalsize & \normalsize 78  &\normalsize 0.0195 &\normalsize  0.06 *\\

\normalsize Word and object processing, contrast 6 & \normalsize 49 & \normalsize 11134  &\normalsize 0 &\normalsize 0.0002   \\
\normalsize  & \normalsize & \normalsize 3466  &\normalsize 1.21 $ \cdot $ 10$^{-37}$ &\normalsize  0.001\\
\normalsize  & \normalsize & \normalsize 1630  &\normalsize 1.87 $ \cdot $ 10$^{-22}$ &\normalsize  0.001 \\
\normalsize  & \normalsize & \normalsize 609  &\normalsize 2.86 $ \cdot $ 10$^{-11}$ &\normalsize  0.003\\
\normalsize  & \normalsize & \normalsize 475  &\normalsize 1.98 $ \cdot $ 10$^{-9}$ &\normalsize  0.004\\
\normalsize  & \normalsize & \normalsize 270  &\normalsize 3.16 $ \cdot $ 10$^{-6}$ &\normalsize  0.009\\
\normalsize  & \normalsize & \normalsize 132  &\normalsize 0.00145 &\normalsize  0.025\\
\normalsize  & \normalsize & \normalsize 111  &\normalsize 0.00433 &\normalsize  0.035\\
\normalsize  & \normalsize & \normalsize 92  &\normalsize 0.0123 &\normalsize  0.048 \\
\normalsize  & \normalsize & \normalsize 89  &\normalsize 0.0146 &\normalsize  0.051 *\\
\normalsize  & \normalsize & \normalsize 76  &\normalsize 0.0313 &\normalsize  0.067 *\\
\normalsize  & \normalsize & \normalsize 75  &\normalsize 0.0332 &\normalsize  0.069 *\\

\hline
\end{tabular}
\end{center}
\label{table:typicalstudies3}
\end{table*} 

\end{document}